\documentclass[11pt]{report}
\usepackage[dissertation,final,copyright,]{USCthesis}
\usepackage[]{amsfonts, bm, amsthm, graphicx, amsmath}

\newcommand{\ra}{\rangle}
\newcommand{\la}{\langle}
\newcommand{\tr}{{\rm Tr}}
\newcommand{\be}{\begin{equation}}
\newcommand{\ee}{\end{equation}}
\newcommand{\ber}{\begin{eqnarray}}
\newcommand{\eer}{\end{eqnarray}}
\def\identity{\leavevmode\hbox{\small1\kern-3.2pt\normalsize1}}

\newenvironment{example}[1][Example]{\begin{trivlist}
\item[\hskip \labelsep {\bfseries #1}]}{\end{trivlist}}

\def\thebibliography#1{\chapter*{References} 
    \addcontentsline{toc}{chapter}{References} 
    \list 
    {[\arabic{enumi}]}{\settowidth\labelwidth{[#1]}\leftmargin 
\labelwidth 
    \advance\leftmargin\labelsep 
    \usecounter{enumi}} \def\newblock{\hskip .11em plus .33em minus -. 
07em} 
    \sloppy \sfcode`\.=1000\relax}

\begin{document}
\newtheorem{theorem}{Theorem}[chapter]
\newtheorem{corollary}[theorem]{Corollary}
\newtheorem{conjecture}[theorem]{Conjecture}
\newtheorem{lemma}[theorem]{Lemma}
\newtheorem{proposition}[theorem]{Proposition}
\newtheorem{remark}[theorem]{Remark}
\newtheorem{definition}[theorem]{Definition}
\newlength{\myVSpace}
\setlength{\myVSpace}{1ex}
\newcommand\xstrut{\raisebox{-.5\myVSpace}{\rule{0pt}{\myVSpace}}}
\title{SYMMETRY IN QUANTUM WALKS}
\author{Hari Krovi}
\majorfield{ELECTRICAL ENGINEERING} 
\committee{T.~A.~Brun & (Chairperson)\\*
                     D.~A.~Lidar &\\*
                     S.~Haas & (Outside Member)}
\submitdate{August 2007}
\copyrightyear{2007}                  
\begin{preface}
\prefacesection{Dedication}
\vspace{0.5in}
\begin{center}
{\bf To my parents,} \\
\vspace{0.1in}
{\bf Kamala and Ganapathy Rao Krovi}
\end{center}
\newpage
\prefacesection{Acknowledgments}
I am greatly indebted to Todd Brun for being an excellent advisor and mentor. Todd gave a lot of his time especially during the initial stages of my PhD when it is so easy to get lost. I am especially grateful for his guidance and the valuable insights he has provided at various points which have led us to a lot of interesting results.

I would like to thank Daniel Lidar for his advice, for the useful discussions we had on various topics and the project that I did with him. Thanks are due to Igor Devetak for his time and help on the project that I did with him. Thanks go to Stephan Haas, P. Vijay Kumar, Urbashi Mitra and Antonio Ortega for their help and interest in my graduate studies.

Special thanks to Diane Demetras and Tim Boston for their help with many administrative issues which have bothered me many times during my PhD. Thanks to Milly Montenegro, Mayumi Thrasher and Gerrelyn Ramos for making CSI a great place to work.

A big thank-you to Pinni, Babaiah, Ranu, Archana, Harsha and my friends for all their love and support and for making me feel like I do the coolest thing on the planet.
\tableofcontents
\listoffigures
\prefacesection{Abstract}
A discrete-time quantum walk on a graph is the repeated application of a unitary evolution operator to a Hilbert space corresponding to the graph. Hitting times for discrete quantum walks on graphs give an average
time before the walk reaches an ending condition.  We derive an expression
for hitting time using superoperators, and numerically evaluate it for the
walk on the hypercube for various coins and decoherence models. The hitting time for a classical random walk on a connected graph will always be finite. We show that, by contrast, quantum walks can have infinite hitting times for some initial states. We seek criteria to determine if a given walk on a graph will have infinite hitting times, and find a sufficient condition, which for discrete time quantum walks is that the degeneracy of the evolution operator be greater than the degree of the graph.  The phenomenon of infinite hitting times is in general a consequence of the symmetry of the graph and its automorphism group. 

Symmetries of a graph, given by its automorphism group, can be inherited by the evolution operator. Using the irreducible representations of the automorphism group, we derive conditions such that quantum walks defined on this graph must have infinite hitting times for some initial states. Symmetry can cause the walk to also be confined to a subspace of the original Hilbert space for cartain initial states. We show that a quantum walk confined to the subspace corresponding to this symmetry group can be seen as a different quantum walk on a smaller {\it quotient} graph. We give an explicit construction of the quotient graph for any subgroup $H$ of the automorphism group. The automorphisms of the quotient graph which are inherited from the original graph are the original automorphism group modulo the subgroup $H$ used to construct it. We conjecture that the existence of a small quotient graph with finite hitting times is necessary for a walk to exhibit a quantum speed-up. Finally, we use symmetry and the theory of decoherence-free subspaces to determine when the subspace of the quotient graph is a decoherence-free subspace of the dynamics.
\end{preface}

\chapter{Introduction and preliminaries}
\section{Quantum computing}
In the course of the 20th century, quantum mechanics was established as a fundamental theory of physics. Every physical theory is built (or needs to be built) within the framework of quantum mechanics. But quantum mechanics can seem very counter-intuitive even to an expert. Superposition, interference and entanglement are some of the main and bizarre aspects of quantum mechanics. These aspects can lead to many spectacular and often puzzling effects in physical systems. One of the main goals of the field of quantum computing and quantum information is to use these non-classical aspects to build more powerful computers, communication devices and cryptographic systems.

The invention of the transistor by Bardeen, Brattain and Shockley has revolutionized computer hardware. Computers have since been becoming more powerful as more and more transistors are made to fit on an integrated circuit (IC). In 1965, Gordon Moore characterized this growth in his famous Moore's law which states that the number of transistors on an IC doubles in number roughly every two years. This means that the power of the computer doubles for the same cost roughly every two years. This law has held sway for many years and seems to be valid even today. However, experts predict that its rule will have run its time sometime in the next twenty years. As electronic devices become smaller they will eventually run into fundamental difficulties of size due to quantum effects. Conventional fabrication techniques are going to have a problem when these quantum effects come into play. New technologies will have to replace them to make progress. Quantum computing is a paradigm which can provide a solution to this problem. This is another reason why research in quantum computing and quantum information is critical.

The original idea of a quantum computer can be traced back to simulation of quantum systems. Any physical system adheres to the laws of quantum mechanics and hence it is inherently a quantum system. Thus, understanding quantum systems is essential to physics. But any reasonable and large-scale model of such a system is hard to simulate on a classical computer since it requires exponential resources. In the 1980s, it  was realized that an appropriate system which might efficiently simulate any quantum system is one which is based on quantum mechanics. The essential idea of a such a quantum computer was suggested by Manin \cite{Manin}, Benioff \cite{Benioff} and Feynman \cite{Feynman}. This idea was taken to the next level by Deutsch, who introduced the notion of a quantum computer as a universal computing machine \cite{Deu1,Deu2}. An early application of these concepts was in quantum cryptography \cite{BB84}, where a cryptographic protocol was proposed which has since been shown to be unconditionally secure \cite{PreSho}. The first instance of a computational problem that could be solved faster by a quantum computer than by a classical computer was given by Deutsch and Josza \cite{DeuJoz}. The next problem which has a faster quantum algorithm than a classical one, was given by Simon \cite{Simon}. Though these algorithms demonstrated the power of quantum computing, the problems they solve are rather artificial in nature. The discovery of quantum algorithms for prime factoring and discrete-logarithm by Shor \cite{Sho94} was the first example where quantum algorithms work faster than the best known classical algorithms for problems of great practical importance. Since then there has been a surge of interest in quantum computing and quantum information which led to the development of more areas, such as quantum Shannon theory, which was built from the original ideas of quantum teleportation and superdense coding \cite{Bennett} and quantum error correction.

One of the most important goals of quantum computing is the design of fast algorithms for computational problems. A quantum  algorithm for database search which works faster than any classical algorithm was given by Grover \cite{Gro96}. The algorithms of Grover \cite{Gro96} and Shor \cite{Sho94} are among the famous examples of quantum algorithms which have a speed-up over classical algorithms. These two algorithms are very different in structure:  Grover's algorithm exploits an invariant two dimensional subspace within the search space, while Shor's algorithm exploits the properties of the quantum Fourier transform (QFT). Efficient algorithms for a class of problems called the hidden subgroup problem (HSP) (factoring and discrete-logarithm belong to this class) use the QFT \cite{Lom04}. The QFT is useful for the Abelian version of the HSP and for some non-Abelian groups \cite{RotBet}. But for other non-Abelian groups, most notably, the symmetric group and the associated HSP-- the graph isomorhism problem, the power of the QFT seems to be limited. Grover's algorithm, although very useful in many search problems, gives only a quadratic speed up, and hence a straightforward application of this algorithm is not very efficient for the HSP. This is because it ignores structure in the problem which can be used to obtain a speed up. New concepts and tools might be needed to design algorithms to solve these problems. In this context, quantum walks might provide such tools. A goal of this thesis is to explore the properties of quantum walks which might be useful in designing fast algorithms.

We begin by giving a brief introduction to the basic concepts and definitions of quantum computing which are used in this thesis. To get a more detailed explanation refer to the book by Nielsen and Chuang \cite{NielsenChuang}.
\section{The rules of quantum mechanics}
{\bf State space.} An isolated quantum system can be described by a quantum state which is a vector in a complex vector space endowed with an inner product (i.e., a Hilbert space). Given an orthonormal basis $\{|i\ra\}$ $i\in S$, for this Hilbert space, the state of the system $|\alpha\ra$ can be written as $|\alpha\ra = \sum_i a_i |i\ra$, such that $\sum_i |a_i|^2=1$. $\la\alpha |$ denotes the complex conjugate of the state and $\la\alpha|\beta\ra$ denotes the inner product between two states $|\alpha\ra$ and $|\beta\ra$.

{\bf Unitary evolution.} The evolution of a closed system can be described by a unitary transformation. In other words, the state of the system at two different times can be related by a unitary operator. If $|\alpha\ra$ and $|\beta\ra$ are the states at times $t_1$ and $t_2$ respectively, then
\begin{equation}
|\beta\ra=\hat{U}|\alpha\ra ,
\end{equation}
where $\hat{U}$ is unitary i.e., $\hat{U}\hat{U}^{\dag}=\hat{I}$, $\hat{U}^{\dag}$ is the transpose complex conjugate of $\hat{U}$. Note that the norm of the state does not change after the evolution, i.e., $|\la\beta |\beta\ra|=|\la\alpha |\hat{U}^\dag\hat{U}|\alpha\ra|=|\la\alpha |\alpha\ra|$.

{\bf Measurement.} Generalized quantum measurements are described by a set of measurement operators $\{M_i\}$ which can act on the Hilbert space of the system and satisfy the {\it completeness relation} $\sum_i M_i^{\dag} M_i=I$. The index $i$ denotes the measurement outcome and when a measurement is performed any one of these outcomes can occur with a certain probability. Given a state of the system, say $|\alpha\ra$ before the measurement, an outcome $k$ occurs with a probability $\la\alpha|M_k^{\dag}M_k|\alpha\ra$ and if this outcome $k$ occurs, then the resulting state of the system is,
\begin{equation}
\frac{M_k|\alpha\ra}{\sqrt{\la\alpha|M_k^{\dag}M_k|\alpha\ra}} .
\end{equation}
Projective measurements are a special case of generalized measurements in which the measurement operators are {\it projectors}. An operator $P$ is a projector if $P^2=P$ and $P=P^{\dag}$. Thus, in such measurements, the completeness relation becomes $\sum_i P_i=I$, where $P_i$ are the measurement operators. The operators $P_i$ also satisfy orthogonality relations $P_i P_j=\delta_{i j}$.

{\bf Density operators.} Suppose the exact quantum state is not known, but rather it is known that the state belongs to a set $|\psi_i\ra$ where $i$ indexes the set. If each $|\psi_i\ra$ could be the state with a probability $p_i$, then the density operator (also known as the density matrix) for the system is defined as
\begin{equation}
\rho=\sum_i p_i |\psi_i\ra\la\psi_i | .
\end{equation}
If the state of a quantum system is known exactly ($|\psi\ra$, say), it is called a {\it pure} state. The density operator corresponding to a pure state $|\psi\ra$ is $\rho=|\psi\ra\la\psi |$. If the state is not pure, it is called {\it mixed}. The unitary evolution and measurement postulates can be written in the density operator picture in the following way. Given that the density matrices of a closed quantum system at times $t_1$ and $t_2$ are $\rho_1$ and $\rho_2$ respectively, this evolution can be related by a unitary operator as
\begin{equation}
\rho_2=\hat{U}\rho_1\hat{U}^{\dag} .
\end{equation}
Now, given a measurement described by measurement operators $\{M_i\}$ (such that $\sum_i M_i^\dag M_i=I$), the outcome $k$ occurs with a probability $p(k)=\tr (M_k^\dag M_k \rho)$ and the state of the system is then $M_k\rho M_k^\dag /\tr (M_k^\dag M_k \rho)$. For a projective measurement with operators $\{P_i\}$, an outcome $k$ occurs with probability $p(k)=\tr (P_k\rho)$ and the resulting state is $P_k \rho P_k/\tr (P_k\rho)$. Density operators are characterized by the following two properties.
\begin{enumerate}
	\item The trace of a density matrix is unity.
	\item A density matrix is a positive matrix.
\end{enumerate}

\section{Outline}
The rest of this thesis is organized as follows. In Chapter 2, we define classical random walks and then give the definitions of quantum walks. Then we define Cayley graphs and give a detailed description of the hypercube as an example of a Cayley graph.

In Chapter 3, we define a notion of hitting time and provide a formula for the hitting time of a quantum walk on a graph. We then give the results of simulations for various cases with and without decoherence for the hitting time on the hypercube. We also provide evidence of infinite hitting times and then derive rigorously the conditions under which a quantum walk has infinite hitting times.

In Chapter 4, we relate the notion of infinite hitting times to symmetry of the graph and give examples of graphs with sufficient symmetry to have infinite hitting times. We also provide the necessary definitions and results in the representation theory of finite groups used in the analysis.

In Chapter 5, we use symmetry again to explain fast hitting times. We develop the notion of a quotient graph and use it to show that quantum walks on a graph with symmetry are actually walks on a smaller quotient graph. Quantum walks with symmetry have invariant subspaces which can be used for algorithm design.

In Chapter 6, we make use of symmetry to find conditions on the decoherence such that the subspace of a quotient graph will lie in a decoherence-free subspaces of the dynamics.

\chapter{Random and quantum walks on graphs}
Quantum walks were formulated in studies involving the dynamics of quantum diffusion \cite{Feynman_lects}, but the analysis of quantum walks for use in quantum algorithms was first done by Farhi and Gutmann \cite{FG98}. The motivation for the development of quantum walks is three-fold. First, algorithm design in classical computer science has benefitted enormously with the advent of randomized algorithms. The best known algorithms for some important problems like 3-SAT (3-satisfiability) are based on random walks \cite{MotRag95}. A natural question is whether quantum walks would perform better for these problems. Second, as mentioned above existing quantum algorithms for the so called hidden subgroup problem, based on the quantum Fourier transform, like Shor's algorithm and related algorithms \cite{Sho94}, do not seem to be effective for certain non-Abelian problems like the graph isomorphism problem. Therefore, there is a need for a new class of algorithms to tackle these problems, and quantum walk based algorithms may provide a new approach. Finally, a large number of problems in computer science can be reformulated as graph related problems and having a quantum formalism which is specific for graphs may be useful in designing algorithms for these problems.

Quantum walks have been already been used in quantum algorithms which have a speed-up over the corresponding classical algorithms. Childs et al. showed in \cite{CCDFGS03} that a continuous quantum walk would find a certain {\it final} node exponentially faster than a classical random walk on the so-called ``glued trees" graph. The classical walk tends to take exponentially longer due to the large number of vertices in the middle of the graph, where as a quantum walk would traverse it in a superposition of paths and move toward the final vertex much faster. Shenvi et al. \cite{SKW03} showed that a quantum walk on an unsorted database (represented as a hypercube) has a quadratic speed up over a classical algorithm. The algorithm begins in a superposition of all the vertices and coin states and proceeds by applying the ``Grover" coin at every vertex except the final vertex where it applies another coin ($-I$, which they point out in the paper, is an arbitrary choice since other coins worked just as well in numerical simulations.) This makes the walk converge to the final vertex in $O(\sqrt{N})$ steps, where $N$ is the number of vertices. Quantum search for a marked item in $N$ items arranged on a $d$-dimensional $N^{1/d}\times N^{1/d}\times\dots N^{1/d}$ grid was analyzed using a continuous-time walk by Childs and Goldstone in \cite{ChildsGoldstone} and using a discrete-time walk by Ambainis et al. in \cite{AKR05}. This is one of the few places where discrete walks are better. An algorithm based on a continuous-time walk worked in $O(\sqrt{N})$ time steps for $d\geq 5$, in $O(\sqrt{N}\log N)$ for $d=4$ and had no speed up in $d=2,3$ whereas a discrete walk took $O(\sqrt{N})$ for $d\geq 3$ and $O(\sqrt{N}\log N)$ for $d=2$. Ambainis \cite{Ambainis03} has applied quantum walks to the element distinctness problem i.e., the problem of determining whether $N$ elements in a given set are all distinct or not. This algorithm is based on a discrete-time quantum walk on the Johnson graph. Other algorithms based on quantum walks are an algorithm for matrix product verification \cite{BS06}, triangle finding \cite{MSS05} and group commutativity testing \cite{MN05}.

Several quantities of interest have been defined for quantum walks analogous to classical walks in \cite{AAKV00}, such as mixing time, sampling time, filling time and dispersion time. In \cite{AAKV00}, a lower bound on the mixing time of a discrete quantum walk on the $N$-cycle is found and shown to be at most polynomially faster than that of the classical walk. Hitting time is another important quantity for classical walks on graphs. Two definitions of hitting time are given in \cite{Kem03b} and an upper bound for one of them was found for the walk on a hypercube. A different definition of hitting time is given in \cite{KB05}, where the unitary evolution of the discrete walk is replaced by a measured walk. In such a walk, after the application of the unitary evolution operator, a measurement is performed to see if the particle is in the final vertex or not. In \cite{KB06} the phenomenon of {\it infinite} hitting times is analyzed and it was shown that graphs with sufficient symmetry can have infinite hitting times for certain initial states. This is a purely quantum phenomenon and does not have a classical analogue. Cayley graphs on the symmetric group are shown to be examples of graphs which have this symmetry and hence have an infinite hitting time for certain starting states. The theory of irreducible representations is used to estimate the amount of degeneracy that a given group of symmetries produces. The use of representation theory to explain aspects of quantum walks on certain classes of graphs was also done in \cite{GW}, where the behavior of mixing times of Cayley graphs on the symmetric group is explained based on its irreducible representations.

Quantum walks have applications other than in the design of new algorithms. In \cite{CDEL04, CDDEKL05}, quantum walks on weighted graphs have been used to efficiently transfer quantum states with perfect fidelity. Here and in \cite{F06}, symmetry has been used to demonstrate a class of graphs on which a continuous time quantum walk reduces to the walk on a line (quantum wire) like in the case of the glued-trees graph.

Quantum walks come in two distinct flavors: discrete-time and continuous-time. The main difference between them is that discrete time walks require a ``coin"---which is just any unitary matrix---plus an extra Hilbert space on which the coin acts, while continuous time walks do not need this extra Hilbert space. Aside from this, these two versions are similar to their classical counterparts. Discrete-time quantum walks evolve by the application of a unitary evolution operator at discrete time intervals, and continuous-time walks evolve under a (usually time-independent) Hamiltonian. Unlike the classical case, the extra Hilbert space for discrete-time quantum walks means that one cannot obtain the continuous quantum walk from the discrete walk by taking a limit as the time step goes to zero. Although there is no natural limit to go from the discrete to continuous walks for general graphs, for the quantum walk on the line \cite{Strauch06} offers a treatment of this limit, where it is possible to meaningfully extract the continuous-time walk as a limit of the discrete-time walk. The dynamics of quantum walks of both types has been studied in detail for walks on an infinite line---for the continuous-time case in  Refs.~\cite{CCDFGS03,FG98,CFG02,Konno1,Konno2} and for the discrete-time case in \cite{NV00,BCGJW04,BCA03a,BCA03b,BCA03c}. There has also been considerable work on other regular graphs.  The $N$-cycle is treated in \cite{AAKV00,TFMK}, and the hypercube in \cite{SKW03,MooRus02,Kem03b,KB05,KB06,KB07, KB07b}.  Quantum walks on general undirected graphs are defined in \cite{Ken03, Amb03}, and on directed graphs in \cite{Mon05}. Reviews of quantum walks include an introductory review by Kempe in \cite{Kem03a}, and a review from the perspective of algorithms by Ambainis in \cite{Amb03}. The role of symmetry in quantum walks has been analyzed in \cite{KB05,KB06,KB07}. It has been shown in \cite{KB06} that the evolution operator of a quantum walk inherits symmetries from the automorphisms of the graph and this leads to degeneracy in the operator. This degeneracy can determine subspaces to which the walk remains confined. Thus, the walk never explores some parts of the Hilbert space, leading to the phenomenon of infinite hitting times. In \cite{KB07}, it was shown that due to the symmetry of the graph, certain subspaces which confine the walk have fast hitting times, and moreover these subspaces are the corresponding Hilbert spaces of a different graph - a {\it quotient} graph. Thus, the walk is confined to smaller graph due to symmetry. The exponential speed-up observed for quantum walks in \cite{CCDFGS03,SKW03} on the ``glued-trees" graph and the hypercube respectively is due to the fact that the quantum walk is on a smaller quotient graph.

Decoherence in quantum systems can be broadly defined as any process that destroys quantum coherence. Aharanov \cite{ADZ92} first considered quantum walks with measurements. Decoherence was later considered in quantum walks mainly to demonstrate the classical behavior of such decohering quantum walks. Quantum to classical transition due to decoherence has been considered  for the walk on the line in Refs.~\cite{BCA03a,BCA03b} and for the hypercube in \cite{Alagic05,KT03}. Kendon et al \cite{KS04} provide a treatment of this transition for any quantum walk. More recently, decoherence has been shown to be beneficial for fast mixing behavior of quantum walks. For the continuous-time walk on the hypercube, decohering quantum walks have been shown to mix faster \cite{Ric07, Ric07a}. Kendon \cite{Ken06} gives a review of the work done in this field so far, focusing mainly on decoherence. Studies of decoherence in quantum walks is useful not only to understand the quantum behavior of the walk but also for possible implementations. Any implementation of a quantum walk will involve having to deal with decoherence. Such studies may be useful in determining the kind of decoherence that has little or no effect on the useful properties of the walk. Some implementation schemes proposed so far are in Refs. \cite{Dur,Fedichkin,KS04,Sanders,Ryan}.

\section{Random walks}
A simple random walk on an undirected graph is defined as the repeated application of a stochastic matrix $P$, where $P(i,j)=1/d_i$ if $i$ and $j$ are connected and $0$ otherwise, where $P(i,j)$ is the probability to go from vertex $i$ to vertex $j$ and $d_i$ and $d_j$ are the degrees of these vertices. Such a process is called a {\it Markov chain}. If the graph is connected and non-bipartite then a fundamental property of Markov chains is that the distribution tends to a stationary distribution ($\pi$) which is independent of the initial distribution ($p^0$). If the graph is regular (if every vertex is connected to the same number of other vertices), then this final distribution is uniform over all the vertices. By contrast, quantum walks do not converge to a final distribution since norms of states do not change under a unitary operation and hence the distance between the states describing the system does not converge to zero (\cite{AAKV00}, \cite{Kem03a}). The induced probability distribution does not converge either, but it turns out that the time-averaged probability distribution converges. The time averaged distribution can be defined as \begin{equation}
\bar{P}_T(v|\alpha_0)=\frac{1}{T}\sum_{t=0}^{T-1} P_t (v|\alpha_0) ,
\end{equation}
where $|\alpha_0\ra$ is the initial state, $P_t$ is the instantaneous distribution and $v$ is any vertex of the graph.

The rate of convergence to this final distribution can expressed in terms of many quantities, but the one used commonly is mixing time. Mixing time is defined as
\begin{equation}
M_{\epsilon}=\min\{T|\forall t\geq T,p^0:||p^t-\pi ||\leq \epsilon\} ,
\end{equation}
where $||p-q||=\sum_i|p_i-q_i|$ is the total variation distance between the distributions $p$ and $q$. Mixing time for a classical walk is related to the {\it spectral gap} (the difference between the largest and the second largest eigenvalues of $P$) in the following way:
\begin{equation}
\frac{\lambda_2}{(\lambda_1-\lambda_2)\log 2\epsilon} \leq M_{\epsilon}\leq \frac{1}{(\lambda_1-\lambda_2)}(\max_i\log\pi_i^{-1}+\log\epsilon^{-1}) ,
\end{equation}
where $\lambda_1$ is the largest eigenvalue and $\lambda_2$ is the second largest eigenvalue. If the graph is regular, it turns out that the largest eigenvalue of $P$ (i.e., $\lambda_1$) is $1$. The above relation connects mixing time and the second largest eigenvalue of the stochastic matrix. The classical mixing time has been found for a number of graphs. For the $N$-cycle the mixing time of a simple random walk can be calculated to be O$(N^2\log (1/\epsilon))$. The mixing time for the hypercube turns out to be $O(d\log d\log (1/\epsilon))$, where $d$ is the dimension of the hypercube. The hypercube is an example of a class of graphs called expander graphs for which it has been shown that a classical random walk is rapidly mixing. But quantum walks on the other hand do not have good mixing properties on the hypercube. The discrete-time quantum walk has a mixing time of at least O$(d^{3/2}/\epsilon)$ \cite{MooRus02}, where $d$ again is the dimension of the hypercube. But a quantity that looks promising for quantum walks on the hypercube is hitting time. Hitting time measures the average time it takes for a walk to reach a certain vertex from a given starting vertex. The analysis of hitting times under different situations is one of the aspects of this thesis. In order to motivate the analysis of hitting times of quantum walks on the hypercube, we briefly present the following classical random walk based algorithm.

Classical random walks have been used in many randomized algorithms in computer science. One of the applications of random walks to algorithms is for the 3-satisfiability (3-SAT) problem where the best known classical algorithm is based on the hitting time of a random walk on the hypercube. The 3-SAT problem can be defined as follows. Consider a set of $n$ literals $\{x_1,x_2,\dots ,x_n\}$, each taking a value of 0 or 1, and a set of clauses such that each clause contains a logical OR of only three literals (or their negations). The problem of 3-satisfiability consists of finding an assignment for the literals such that each of the clauses is satisfied (i.e., each has a value 1). For example, $(x_1\vee \neg x_2\vee \neg x_3)\wedge (\neg x_1\vee x_2 \vee x_4)$ is an expression with two clauses, where $\wedge$ and $\vee$ denote the binary operations of AND and OR and $\neg x$ denotes the negation of $x$. $(x_1,x_2,x_3,x_4)=(1,0,0,1)$ is an assignment of the literals that satisfies both the clauses and thus represents a solution. The classical random-walk based algorithm takes O$(1.329...^n {\rm poly} (n))$ steps \cite{Rolf,Shoning}. It consists of the following steps.
\begin{enumerate}
	\item Choose a random initial assignment for the literals.
	\item Repeat $3n$ times
	\begin{enumerate}
	\item If all the clauses are satisfied, then stop.	
	\item If not, pick one unsatisfied clause, choose one literal uniformly at random and flip it.
	\end{enumerate}
\end{enumerate}

This algorithm is a random walk on the hypercube since an initial assignment of the literals represents a vertex on the hypercube and flipping a bit of this vertex represents moving along an edge connected to it. Thus analyzing the behavior of hitting times of quantum walks on the hypercube may give us important clues to design quantum algorithms for the 3-SAT. However, in the above algorithm, the walk is on a directed hypercube since after moving along an edge to a new vertex, the literal that was flipped need not occur in an unsatisfied clause. This means that it cannot be flipped at the new vertex and therefore, there is no path from this vertex to the original vertex. This directed nature of the graph causes problems when we try to make this a quantum algorithm. A quantum walk as with any quantum computation, needs to be unitary. It has been shown in \cite{Mon05} that on a directed graph, it is possible to define a unitary walk if and only if the graph is {\it reversible}. A directed edge of a graph going from vertex $v_i$ to $v_j$ is called reversible if there is a path from $v_j$ to $v_i$. A graph is called reversible if every edge in it is reversible. Thus, given an expression for 3-SAT, one could define a unitary quantum walk if the corresponding directed graph of the problem is reversible. This may be overcome by defining a non-unitary walk, one that involves measurements, if it is not reversible. But the next step, which is to design an algorithm based on this walk, is not clear. There does not exist a quantum walk algorithm for 3-SAT presently, but a quantum algorithm based on Grover search and amplitude amplification which works in O$(1.153...^n {\rm poly}(n))$ steps is presented in \cite{ambainis_3sat}. 

\section{Quantum walks--discrete and continuous}

Quantum walks as noted earlier, are primarily of two types. Depending on the way the evolution operator is defined, they can be either discrete-time or continuous-time quantum walks. These two definitions of quantum walks are not exactly equivalent to the two types of classical random walks. While they are both based on the classical definitions, unlike the classical case the discrete quantum walk does not reduce to the continuous walk when we let the time step between repeated applications of the unitary tend to zero. This is because discrete-time walks need an extra Hilbert space, called the ``coin''  space (from the idea that one flips a coin at each step to determine which way to walk), and taking the limit where the time step goes to zero does not eliminate this Hilbert space. Therefore, the properties of discrete and continuous walks are different. Though there is no obvious reason why one should be preferred, in some cases it has been shown that coins make these walks faster \cite{AKR05}.

\subsection{Discrete-time walks}

A discrete-time quantum walk can broadly be defined as the repeated application of a unitary evolution operator on a Hilbert space whose size depends on the graph. This Hilbert space usually consists of the space of possible positions (i.e., the vertices) together with the space of possible directions in which the particle can move along from each vertex (the coin space). All the concepts which we develop in this thesis such as a formula for hitting time of a dicrete-time quantum walk, infinite hitting times, effect of symmetry and quotient graphs are applicable to any undirected graph. However, in most of our simulations we consider $d$-regular, undirected and $d$-colorable graphs. This is because the structure of quantum walks reduces to something more manageable in this case. We will briefly review the definitions of these graph properties.

\begin{figure}[tbh]
\begin{center}\includegraphics[width=4in]{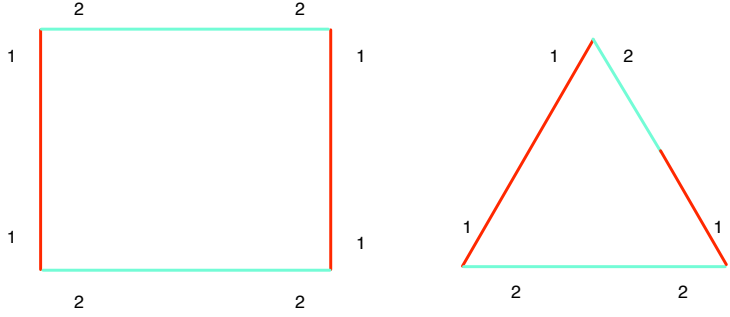}\end{center}
\caption{\label{fig1} Examples of 2-colorable (square) and non-2-colorable (triangle) regular graphs of degree 2, where the colors are numbered 1 and 2.}
\end{figure}

A {\it regular} graph is one where every vertex is connected to the same number $d$ of other vertices.  This number is called the the {\it degree} of the graph.  A graph is {\it undirected} if for every edge between vertices A and B going from A to B, an edge goes from B to A as well.  In this case, we identify the edge from A to B with the edge from B to A, and consider them a single edge.  A regular, undirected graph with $N$ vertices of degree $d$ is considered $d$-{\it colorable} if the edges incident on every vertex can be numbered $1$ through $d$ such that every edge between two vertices has the same number at either end. Not all $d$-regular and undirected graphs can be $d$-colored. A simple example is the triangle graph where $N=3$ and $d=2$.  (See Fig.~\ref{fig1}.) For $d$-regular, undirected and $d$-colored graphs, the Hilbert space of the walk is $\mathcal{H}^p\otimes\mathcal{H}^c$, i.e., the tensor product of the position and direction (or coin) space. The evolution operator $\hat{U}$ is given by $\hat{U}=\hat{S}(\hat{I}\otimes\hat{C})$, where $\hat{S}$ is called the shift matrix and $\hat{C}$ is the coin matrix. The shift matrix encodes the structure of the graph and is very similar to its adjacency matrix. The vertices, numbered $|0\ra$ through $|N-1\ra$, are basis states for the vertex Hilbert space $\mathcal{H}^p$ and the set of all directions from each vertex, numbered $|1\ra$ through $|d\ra$, are basis states for the coin Hilbert space $\mathcal{H}^c$. In this basis, the shift matrix for the graph can be given the explicit form:
\[
\hat{S} = \sum_v \sum_i |v(i),i\ra\la v,i| ,
\]
where $v(i)$ is the vertex connected to $v$ along the edge numbered $i$.

The coin matrix $\hat{C}$ acts only on the coin space, and ``flips'' the directions before the shift matrix is applied. Then $\hat{S}$ moves the particle from its present vertex to the vertex connected to it along the edge indicated by the coin direction. Though $\hat{C}$ can be any unitary matrix, usually coins with some structure are considered. The coins that we used in our previous analysis are the Grover coin $\hat{G}$ and the Discrete Fourier Transform (DFT) coin $\hat{D}$. The matrices for these coins are given by:
\begin{equation}
\hat{G} = 2|\Psi\ra\la\Psi|-I = \begin{pmatrix}
  \frac{2}{d}-1 & \frac{2}{d} & \ldots & \frac{2}{d} \\
  \frac{2}{d} & \frac{2}{d}-1 & \ldots & \frac{2}{d} \\
  \vdots & \vdots & \ddots & \vdots \\
  \frac{2}{d} & \frac{2}{d} & \ldots & \frac{2}{d}-1
\end{pmatrix} ,
\label{Grover_matrix}
\end{equation}
and
\begin{equation}
\hat{D}=\frac{1}{\surd{d}}\begin{pmatrix}
  1 & 1 & 1 & \ldots & 1 \\
  1 & \omega & \omega^2 & \ldots & \omega^{d-1} \\
  \vdots & \vdots & \vdots & \ddots & \vdots \\
  1 & \omega^{d-1} & \omega^{2(d-1)} & \ldots &
  \omega^{(d-1)(d-1)} 
\end{pmatrix} ,
\end{equation}
where $|\Psi\ra=\frac{1}{\surd{d}}\sum_i|i\ra$ and $\omega=\exp(2\pi i/d)$.

\subsection{Decoherence in discrete-time walks}
Decoherence in any quantum system is a process which destroys quantum superpositions. Any quantum system will have decoherence (quantum noise). Thus any practical implementation scheme of quantum walks must deal with decoherence. In general, it can be thought of as any completely positive and trace preserving (CPTP) map. An arbitrary CPTP map can be represented using the operator sum (OSR) or Kraus representation. 
\be
\mathcal{D}(\rho)=\sum_i \hat{A}_i \rho \hat{A}_i^\dag ,
\ee
where the operators $\hat{A}_i$ are the Kraus operators and satisfy the relation: $\sum_i \hat{A}_i^\dag \hat{A}_i=\hat{I}$. Thus in a discrete-time quantum walk with decoherence, we can assume that a step of the evolution is of the form
\be
\mathcal{E}(\rho)=\mathcal{D}\circ\mathcal{U}(\rho),
\ee
where $\mathcal{U}(\rho)=\hat{U}\rho\hat{U}^\dag$ as before.

\subsection{Continuous-time walks}

Continuous time quantum walks were defined by Farhi and Gutmann in \cite{FG98}. For an undirected graph $G(V,E)$, the unitary evolution operator is defined as $\hat{U}=\exp(i\hat{H}t)$, where $\hat{H}$ is obtained from the adjacency matrix of the graph. Here again, the vertices of the graph form a basis for the Hilbert space on which $\hat{U}$ is defined. This gives rise to the following Schr\"{o}dinger equation:
\begin{equation}
i\frac{d}{dt} \la v|\psi (t)\ra=\la v|\hat{H}|\psi (t) \ra.
\end{equation}
This walk has a structure very similar to that of continuous time Markov chains. $\hat{H}$ is defined as 
\begin{eqnarray}
 \label{Continuous_Ham} \hat{H}_{i,j}=\left\{
\begin{array}{cl}
   -\gamma &  \ i \neq j \mbox{ if nodes }i\mbox{ and }j\mbox{ connected} \\
         0 &  \ i \neq j \mbox{ if nodes }i\mbox{ and }j\mbox{ not connected} \\
d_i \gamma &  \ i=j
\end{array} \right.
\end{eqnarray}
where $\gamma$ is the jumping rate from a vertex to its neighbor i.e., the transitions between connected vertices happen with a probability $\gamma$ per unit time. 

But for a regular graph we can take $\hat{H}$ to be the adjacency matrix because $d_i=d$, where $d$ is the degree of the graph. This means that the Hamiltonian can be written as $\hat{H}=\gamma (D-A)$, where $D=dI$ and $A$ is the adjacency matrix of the graph. The matrix $D$ would lead to a trivial phase factor and can be dropped. $\hat{H}$ is a symmetric matrix (and hence $\hat{U}$ unitary) if the graph is undirected. Therefore, for a regular and undirected graph, the adjacency matrix $\hat{H}$, which acts as the Hamiltonian, is of the form:
\begin{equation}
H_{i,j}=\biggl\{ \begin{array}{cc} 1 & {\rm if}\ i\ {\rm and}\ j\ {\rm share\ an\ edge,} \\
0 & {\rm otherwise.} \end{array}
\end{equation}
As can be seen, this walk has no coin and so the Hilbert space on which $\hat{U}$ acts is only the vertex space $\mathcal{H}^p$.

\subsection{Decoherence in continuous-time walks}
Continuous-time quantum walks evolve by an application of a Hamiltonian as defined in Eq. (\ref{Continuous_Ham}). Decoherence in this scenario would give rise to an evolution which can be described using the Lindblad semigroup master equation as
\be
\dot{\rho}(t)=[\hat{H}, \rho(t)] + \sum_i (\hat{L}_i\rho(t)\hat{L}^\dag_i + \hat{L}^\dag_i\hat{L}_i\rho(t) + \rho(t)\hat{L}^\dag_i\hat{L}_i) ,
\ee
where $\hat{L}_i$ are the Lindblad operators. The Lindblad operators represent the environmental interactions which lead to decoherence.

\section{Cayley graphs}
\begin{figure}[tbh]
\begin{center}
\includegraphics[scale=0.6]{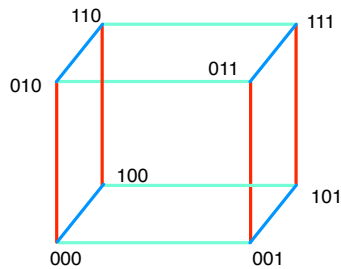}
\end{center}
\caption{The hypercube in three dimensions as an example of a Cayley graph.} \label{hypercube}
\end{figure}
Cayley graphs are defined in terms of a group $G$ and a set $S$ consisting of elements from $G$ such that the identity element $e\notin S$. Given $G$ and $S$, the resulting (right)-Cayley graph $\Gamma(G,S)$ is one whose vertices are labeled by the group elements, i.e., there is one vertex for every group element, and two vertices $g$ and $h$ are connected by a directed edge from $g$ to $h$ if $g^{-1}h\in S$, (see \cite{GroTuc87}). Another way to look at this definition is that from any vertex $g$ of a Cayley graph, there are $|S|$ outgoing edges, one to each of the vertices $gs$, $\forall s\in S$. A Cayley graph will be connected if and only if the set $S$ is a generating set for $G$, and it will be undirected if $s^{-1}\in S$, $\forall s\in S$. The degree of such a graph is $|S|$, the cardinality of the generating set. Finally, a $d$-regular Cayley graph can be $d$-colored if $s^2=1$, $\forall s\in S$, i.e., $s^{-1}=s$. Examples of Cayley graphs on which quantum walks have been studied include the line $\Gamma(\mathcal{Z},\{1,-1\})$; the cycle $\Gamma(\mathcal{Z}_n,\{1,-1\})$; the hypercube $\Gamma(\mathcal{Z}_2^n,X)$ where the set $X$ is the set of canonical generators $\{(1,0,0,\cdots,0),(0,1,0,\cdots,0),\dots,(0,0,0,\cdots,1)\}$; and the graph on the symmetric group $\Gamma (S_n,Y)$, where $Y$ is a generating set for $S_n$.  Let us look at the hypercube as an example of a Cayley graph where quantum walks have been extensively studied (see Fig.~(\ref{hypercube})). The hypercube has $|\mathcal{Z}_2^n|=2^n$ vertices each with a degree of $|X|=n$. The vertices can be labeled by an $n$-bit string from $(0,0,\cdots,0)$ through $(1,1,\cdots,1)$. Two vertices are adjacent if they differ only by a single bit. Vertex $\vec{v}$ is connected to $n$ vertices given by $\vec{v}\oplus\vec{s}$, $\forall \vec{s}\in X$, where $\oplus$ stands for the bit-wise XOR of the bit strings $\vec{v}$ and $\vec{s}$. One important property of the hypercube is that it can be $n$-colored, since $\vec{s}\oplus\vec{s} = \vec{e}$, $\forall \vec{s}\in X$ where $\vec{e} = (0,0,\cdots,0)$ is the identity element. The unitary evolution operator for a discrete walk on the hypercube becomes $\hat{U}=\hat{S}(\hat{I}\otimes \hat{C})$, where $\hat{S}$ has the form
\[
\hat{S}=\sum_{\vec{s}}\sum_{\vec{v}}|\vec{v}\oplus\vec{s}\ra\la \vec{v}|\otimes |\vec{s}\ra\la\vec{s}|.
\] Since the vertices of the hypercube are bit strings, and adjacent vertices are those that differ by one bit, the shift matrix of the discrete walk on the hypercube has a natural form given by
\begin{eqnarray}
\hat{S} &=& \hat{X}\otimes\hat{I}\otimes\dots\otimes\hat{I}\otimes |\vec{s}_1\ra\la\vec{s}_1|
+ \hat{I}\otimes\hat{X}\otimes\dots\otimes\hat{I}\otimes\hat{|}\vec{s}_2\ra\la \vec{s}_2| \nonumber\\
&& + \ldots + \hat{I}\otimes\hat{I}\otimes\dots\otimes\hat{X}\otimes |\vec{s}_n\ra\la \vec{s}_n|,
\label{S_Pauli}
\end{eqnarray}
where $\hat{X}$ stands for the Pauli $\sigma_x$ operator. This structure of $\hat{S}$ reflects the property of the hypercube that moving along an edge from $\vec{v}$ corresponds to flipping one bit of $\vec{v}$. This structure is also useful in determining its group of symmetries as we shall see later. 

\chapter{Hitting times}
\section{Classical hitting time}
Given a regular undirected graph and a particle which starts at
some vertex, the classical random walk is defined as before.  At
each vertex, the particle moves along any edge incident on the
vertex with some predefined probability.  This procedure is then
repeated at the new vertex.  The walk continues until the particle
arrives at (``hits'') a certain vertex (called the ``final vertex'') for
the first time.  The {\it hitting time} is defined as the average
time until the particle hits the final vertex:
\begin{equation}
\tau(v) = \sum_{t=0}^\infty t p_v(t) ,
\end{equation}
where $\tau(v)$ is the hitting time given that the walk starts at
vertex $v$ and $p_v(t)$ is the probability that the particle hits
the final vertex for the first time at time step $t$ (first crossing
probability) given that it was at $v$ at $t=0$.

Let us now specialize to the case of the hypercube, where the
the final vertex is assumed to be $11\cdots1$.  We would like to
find the hitting time starting from $00\cdots0$.
For the classical walk on the hypercube, one can
arrive at a recursive relation involving the hitting time. First,
from the symmetry of the hypercube one can conclude that the
hitting time depends only on the {\it Hamming weight} of the starting vertex
rather than the vertex itself.  The Hamming weight is the number of
1's in the string of bits.  At Hamming weight $x$, there are
$C^n_x=n!/x!(n-x)!$ vertices.  The probability to walk
to a vertex with weight $x+1$ is $(n-x)/n$, and the probability
to walk to a vertex with weight $x-1$ is $x/n$.
So, if $\tau(x)$ denotes the hitting time
starting at any vertex with Hamming weight $x$, then
\begin{equation}
\tau(x)=\frac{n-x}{n}\tau(x+1)+\frac{x}{n}\tau(x-1)+1,
\end{equation}
with the boundary condition $\tau(n)=0$.  This simplifies to
\begin{equation}
\Delta(x)=\frac{n-x-1}{x+1}\Delta(x+1)-\frac{n}{x+1},
\end{equation}
where $\Delta(x)=\tau(x)-\tau(x+1)$. Using this recursive formula,
we obtain
\begin{equation}
\tau(0)=\sum_{x=0}^{n-1}\Delta(x)=\sum_{x=0}^{n-1}\frac{\sum_{j=0}^{x-1}C_{x-j}^n+1}{C_x^{n-1}},
\end{equation}
This sum can readily be evaluated for reasonable sizes of $n$ and in fact this sum scales as $\approx 2^n$. We use
this expression to compare the classical hitting time to the
quantum hitting time. We define the hitting time of a quantum walk next.

\section{Hitting time for quantum walks}
The hitting time $\tau_h$ of a classical random walk is defined as the average time for the walk to hit a designated `final' vertex $v_f$ given that the walk began with some initial distribution $p_i$:
\begin{equation}\label{ht1}
\tau_h = \sum_{t=0}^\infty t p(t),
\end{equation}
where $p(t)$ is the probability of being in the final vertex for the first time at time step $t$. In order to carry this notion of hitting time over to the quantum case, we need to make the meaning of $p(t)$ more precise. In particular, we need to define clearly what ``for the first time'' means for a quantum walk.  As described in \cite{KB05}, we do this by performing a projective measurement of the particle at every step of the walk to see if the particle has reached the final vertex or not. The measurement $M$ which is used has projectors $\hat{P}_f$ and $\hat{Q}_f=\hat{I}-\hat{P}_f$ representing the particle being found or not found at the final vertex, respectively.  The projector is defined $\hat{P}_f=|x_f\ra\la x_f|\otimes\hat{I}_c$, where $|x_f\ra$ is the final vertex state and $\hat{I}_c$ is the identity operator on the coin space. Using this definition, each step of the measured walk consists of an application of the unitary evolution operator $\hat{U}$ followed by the measurement $M$.

Now we can use the same expression (\ref{ht1}) for the hitting time, where the probability $p(t)$ becomes
\begin{equation}\label{prob.eqn}
p(t)=\tr\{\hat{P}_f\hat{U}[\hat{Q}_f\hat{U}]^{t-1}
\rho_0[\hat{U^{\dag}}\hat{Q}_f]^{t-1}\hat{U^{\dag}}\hat{P}_f\} .
\end{equation}
To explicitly sum the series in Eq. (\ref{ht1}) using the expression for $p(t)$ in Eq. (\ref{prob.eqn}), we rewrite the expression in terms of {\it superoperators} (linear transformations on operators) $\mathcal{N}$ and $\mathcal{Y}$, defined by
\begin{eqnarray}\label{superops}
\mathcal{N}\rho = \hat{Q}_f\hat{U}\rho\hat{U^{\dag}}\hat{Q}_f \nonumber\\
\mathcal{Y}\rho=\hat{P}_f\hat{U}\rho\hat{U^{\dag}}\hat{P}_f.
\end{eqnarray}
In terms of $\mathcal{N}$ and $\mathcal{Y}$,
$p(t)=\tr\{\mathcal{Y}\mathcal{N}^{t-1}\rho_0\}$.
We introduce a new superoperator $\mathcal{O}(l)$ which depends on a real parameter $l$:
\begin{equation}
\mathcal{O}(l) = l\sum_{t=1}^\infty (l\mathcal{N})^{t-1} ,
\label{superop_sum}
\end{equation}
which is a function of the parameter $l$.  The hitting time now becomes
\begin{equation}
\tau_h = \frac{d}{dl} \tr\{ \mathcal{Y} \mathcal{O}(l) \rho_0 \} \biggr|_{l=1}.
\label{derivative_form}
\end{equation}

If the superoperator $\mathcal{I}-l\mathcal{N}$ is invertible, then we can replace the sum Eq. (\ref{superop_sum}) with the closed form
\begin{equation}
\mathcal{O}(l) = l(\mathcal{I}-l\mathcal{N})^{-1}.
\end{equation}
(The case when $\mathcal{I}-l\mathcal{N}$ is not invertible is discussed in detail later.) The derivative in Eq. (\ref{derivative_form}) is
\begin{equation}
\frac{d\mathcal{O}}{dt}(1)
= (\mathcal{I}-\mathcal{N})^{-1}+\mathcal{N}(\mathcal{I}-\mathcal{N})^{-2}
= (\mathcal{I}-\mathcal{N})^{-2}.
\end{equation}
This gives us the following expression for the hitting time:
\begin{equation}
\tau_h = \tr\{\mathcal{Y}(\mathcal{I}-\mathcal{N})^{-2}\rho_0\}.
\label{closed_form_tau}
\end{equation}

To evaluate Eq. (\ref{closed_form_tau}), we write these superoperators as matrices using Roth's lemma \cite{Roth34}. As shown in \cite{KB05}, we can then {\it vectorize} the density operators and operators on states, and write the action of superoperators as simple matrix multiplication.  Any matrix can be vectorized by turning its rows into columns and
stacking them up one by one, so that a $D\times D$ matrix becomes
a column vector of size $D^2$.  For example:
\begin{equation}
\begin{pmatrix}
a_{11} & a_{12} & a_{13} \\
a_{21} & a_{22} & a_{23} \\
a_{31} & a_{32} & a_{33}
\end{pmatrix}
\rightarrow
\begin{pmatrix}
a_{11} \\
a_{12} \\
a_{13} \\
a_{21} \\
a_{22} \\
a_{23} \\
a_{31} \\
a_{32} \\
a_{33}
\end{pmatrix} . \nonumber
\end{equation}
Consequently the superoperators become matrices
of size $D^2\times D^2$. This method of vectorization takes operators on one Hilbert space $\mathcal{H}$ to vectors in another Hilbert space $\mathcal{H}'=\mathcal{H}\otimes\mathcal{H}^\ast$ and so superoperators in $\mathcal{H}$ are operators on $\mathcal{H}'$. Note that a basis $\{|u_{i j}\ra\}$ for $\mathcal{H}'$ can be obtained from a basis $\{|v_i\ra\}$ for $\mathcal{H}$ by defining
\begin{equation}
|u_{i j}\ra=|v_i\ra\otimes |v_j\ra^\ast .
\end{equation}
For our superoperators $\mathcal{N}$ and $\mathcal{Y}$ we then get
\begin{eqnarray}
(\mathcal{N}\rho)^v &=& \left[ (\hat{Q}_f\hat{U})\otimes(\hat{Q}_f\hat{U})^\ast \right] \rho^v ,
\nonumber \\
(\mathcal{Y}\rho)^v &=& \left[ (\hat{P}_f\hat{U})\otimes(\hat{P}_f\hat{U})^\ast \right] \rho^v .
\end{eqnarray}
Let $\mathbf{N}=(\hat{Q}_f\hat{U})\otimes(\hat{Q}_f\hat{U})^\ast$ and $\mathbf{Y}=(\hat{P}_f\hat{U})\otimes(\hat{P}_f\hat{U})^\ast$. The hitting time becomes
\begin{equation}\label{hittime}
\tau_h = I^v \cdot \left( \mathbf{Y}(\mathbf{I}-\mathbf{N})^{-2}\rho^v \right) .
\end{equation}
Using this vectorization transformation, we treat the superoperators as operators on a larger Hilbert space and thus can find their inverses (when they exist). But, the expression in Eq. (\ref{hittime}) is not always well defined because the matrix $\mathbf{I}-\mathbf{N}$ may not be invertible. We will show that when it is not invertible, it means that the hitting time becomes infinite for some initial states, and vice versa. This property of quantum walks having infinite hitting times does not have a classical analogue.

\section{Other definitions of hitting time}

Two definitions of the quantum hitting time were given in
\cite{Kem03b}:  {\it one-shot hitting time} and {\it concurrent hitting
time}. The one-shot hitting time is defined for an unmeasured walk.
It is the time at which the probability of being in the final
state is greater than some given value. More precisely, given some
probability $p$, the one-shot hitting time is defined as the lowest time
$\tau_{\rm sh}(p)$ such that
\begin{equation}
|\la x_f|\hat{U}^{\tau_{sh}(p)}|x_0\ra|^2\geq p
\end{equation}
where $x_f$ and $x_0$ are the final and initial states, and
$\hat{U}$ is the evolution operator (as defined above). Essentially this same definition of hitting time was
used in the analysis of the continuous-time walk on the hypercube in \cite{Alagic05}.
This definition is useful if it is known that at some time the probability to be
in the final state will be higher than some reasonable value; but for a general graph,
this is not guaranteed.

The concurrent hitting time, by contrast, is defined for a measured quantum
walk. Given a probability $p$, it is the time $\tau_c(p)$ such that
the measured walk has a probability greater than $p$ of stopping
at a time less than $\tau_c(p)$. It has been proved that the concurrent
hitting time is $\tau_c(p)=\frac{\pi}{2}$ for $p=\Omega(\frac{1}{nlog^2n})$ for a
hypercube of dimension $n$ (for the symmetric initial condition used in this paper).
Since we consider only the measured quantum walk in our analysis, we
compare our numerical results to the numerical simulation of the
concurrent hitting time and the bound on it derived in
\cite{Kem03b}. In the next section, this is redefined in terms
of the residual probability $1-p$ and plotted against the hitting
time defined in the previous section.

If we think of quantum walks as a possible route to new algorithms,
then the concurrent hitting time corresponds to the time needed to
find a solution with probability greater than p.  The definition of hitting
time used here corresponds more to a typical running time
for the algorithm.  Both definitions could prove useful for particular purposes.
In particular, both definitions give different characterizations of the
distribution $p(t)$---the probability of first being in the final vertex at time
$t$---in different ways.  A complete understanding of the time needed to find
a solution would require knowledge of the entire distribution $p(t)$ (or
$\tau_c(p)$ for all $p$), which is unlikely to be achievable, in practice;
but $\tau_h$ and $\tau_c(p)$ (for fixed $p$) give different windows on this function.

\section{Results for the Grover coin}
\begin{figure}[tbh]
\begin{center}\includegraphics[width=4in]{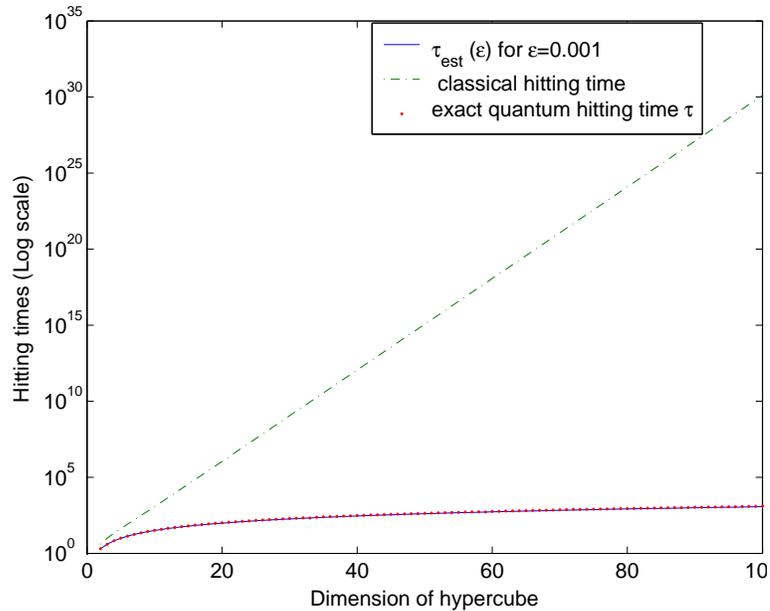}\end{center}
\caption{Hitting times of Classical and Quantum walks (semi-log scale)} \label{htQandC}
\end{figure}

We calculated the average hitting time by evaluating the above expression
(\ref{hittime}) in Matlab.  Because of the multiple tensor products,
the size of the matrices $\mathcal{N}$ and $\mathcal{Y}$
is $(2^n n)^2\times(2^n n)^2$. As $n$ increases, the size of these matrices increases exponentially, and the matrix inversion that Eq.~(\ref{hittime}) demands is not easy to compute. We can always compute an {\it estimate} of the quantum hitting time (strictly, a lower bound), obtained by iterating the quantum walk for a large number of steps.  To get this lower bound, we define $\tau_{\rm est}(\epsilon)$ using Eq.~(\ref{ht1}), and summing the series up to a finite number of terms:
\begin{equation}
\tau_{\rm est}(\epsilon) = \sum_{t=1}^{\tau_c(1-\epsilon)} t p(t),
\end{equation}
where the concurrent hitting time $\tau_c(1-\epsilon)$, as defined in \cite{Kem03a}, is the shortest time $T$ for which
\[
\sum_t^T p(t) \ge 1-\epsilon .
\]
This comparison of the exact and approximate values of the average hitting time is useful, because the matrices in Eq.~(\ref{hittime}) grow faster than those used in summing Eq.~(\ref{ht1}), which means
that for some graphs it may be impractical to calculate the exact value of $\tau_h$, but still possible to calculate the lower bound $\tau_{\rm est}(\epsilon)$.  However, for the quantum walk on the hypercube with the Grover coin, when the initial state of the walk is a particular symmetric state, it is possible to reduce the matrix sizes considerably since  the walk remains in a lower dimensional subspace of the original space, as explained below. For such an initial state we can compute the exact average hitting time (Eq.~(\ref{hittime})) up to a large number of dimensions.  Figure \ref{htQandC} shows both the classical and quantum walks on the hypercube for dimensions up to 100.  The exact average hitting time $\tau_h$ is plotted as a dotted line, and the lower bound $\tau_{\rm est}(\epsilon)$ for $\epsilon=0.001$ is plotted as a solid line. These two lines almost coincide in the graph, and we can see that when $\epsilon$ is small enough summing the series gives a very good estimate of $\tau_h$.  We conjecture that this will remain true for more general graphs as well.  In comparing classical and quantum results, the average hitting time for the quantum walk is a low order polynomial, whereas the classical walk grows exponentially with dimension, so there is a very dramatic speed-up in the quantum case.

The simplification that makes this computation tractable is in the case where
the coin-flip unitary is the {\it Grover coin} $\hat{G}$
defined in (\ref{Grover_matrix}), and the specific starting
state is $\rho=|0\ra\la0| \otimes |\phi\ra\la\phi|$,
where $|\phi\ra=\frac{1}{\surd{n}}\sum_{i=1}^n|i\ra$.  It was first observed in \cite{MooRus02} that for the walk on the hypercube
with this initial state and the Grover coin, the state remains always
in a $2n$-dimensional subspace, where the
walk is on a line with $n+1$ points.  This is rather similar to the
simplification made in the classical case, when we kept track
only of the Hamming weight of the current vertex.
With this simplification, the operators in Eq. (\ref{closed_form_tau})
reduce to $(2n)^2\times(2n)^2$
matrices obtained in \cite{SKW03}, which makes explicit calculations possible even for
high dimensional hypercubes.

We write a set of basis states for this subspace as
$|R,0\ra,|L,1\ra,|R,1\ra,\dots,|R,n-1\ra,|L,n\ra$, where the first label
says whether the state is ``right-going'' or ``left-going'', and the
second label gives the Hamming weight of the state.
The initial state is $|R,0\ra$, and the final state is $|L,n\ra$.
(Note that there are no states $|L,0\ra$ or $|R,n\ra$.)  Restricted
to this subspace, the $\hat{S}$ and $\hat{C}$ matrices become
\begin{equation}\label{S}
S=\sum_{x=0}^n|R,x\ra\la L,x+1|+|L,x+1\ra\la R,x|
\end{equation}
and
\begin{eqnarray}\label{C}
C|L,x\ra&=&-\cos\omega_x|L,x\ra+\sin\omega_x|R,x\ra \nonumber,
\\
C|R,x\ra&=&\sin\omega_x|L,x\ra+\cos\omega_x|R,x\ra,
\end{eqnarray}
where $\cos\omega_x = 1-2x/n$.  We see that for the walk in this
subspace, the coin flip is no longer independent of the position; this
is quite analogous to the reduction of the classical walk to the Hamming weight,
in which the probabilities favor walking toward $x=n/2$.

\begin{figure}[tbh]
\begin{center}
\includegraphics[width=4in]{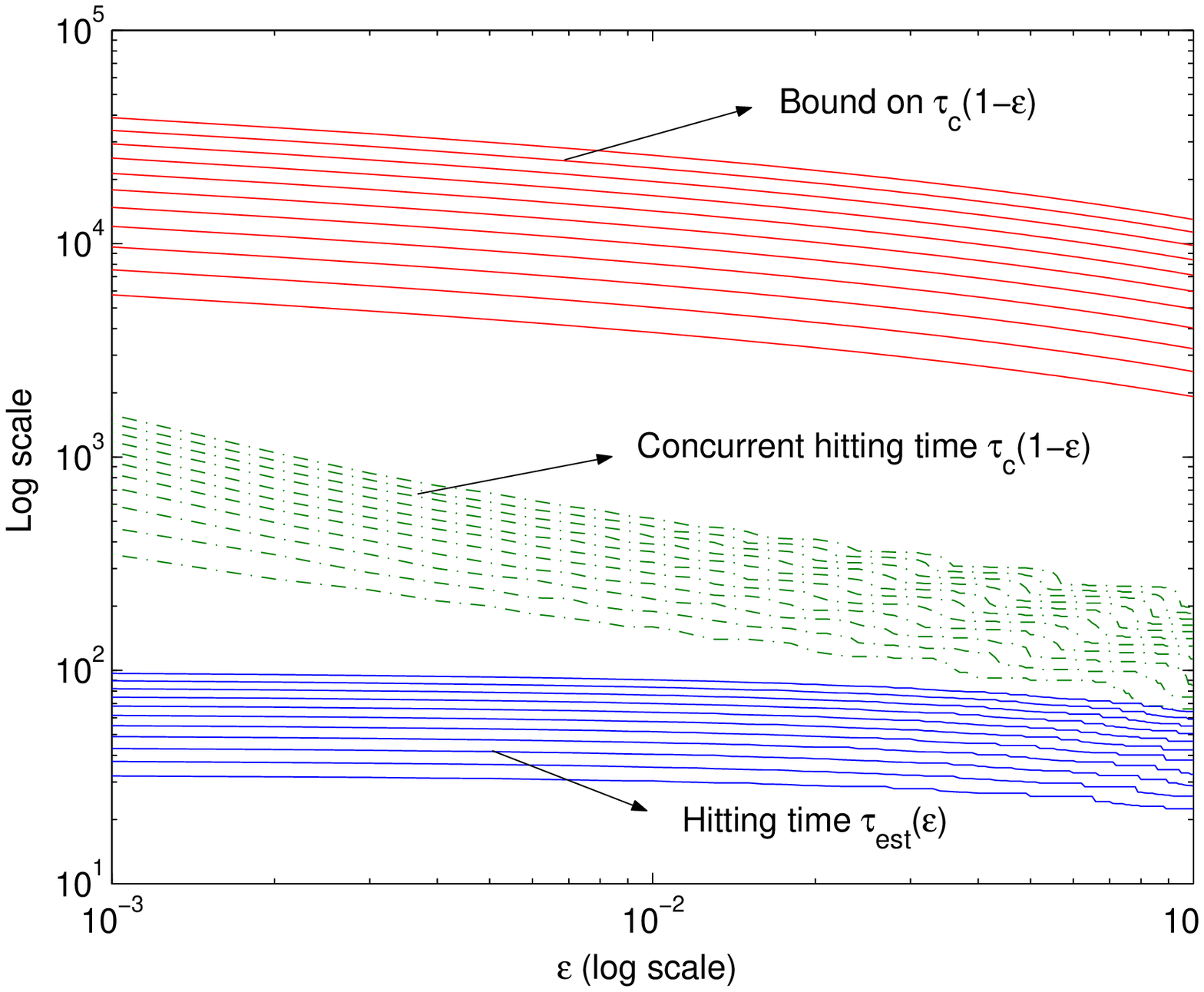}
\end{center}
\caption{Comparison of Quantum hitting time, concurrent hitting time and
the bound on the concurrent hitting time for dimensions 10 to 20(log scale)}
\label{epsilon10_20}
\end{figure}

\begin{figure}[tbh]
\begin{center}
\includegraphics[width=4in]{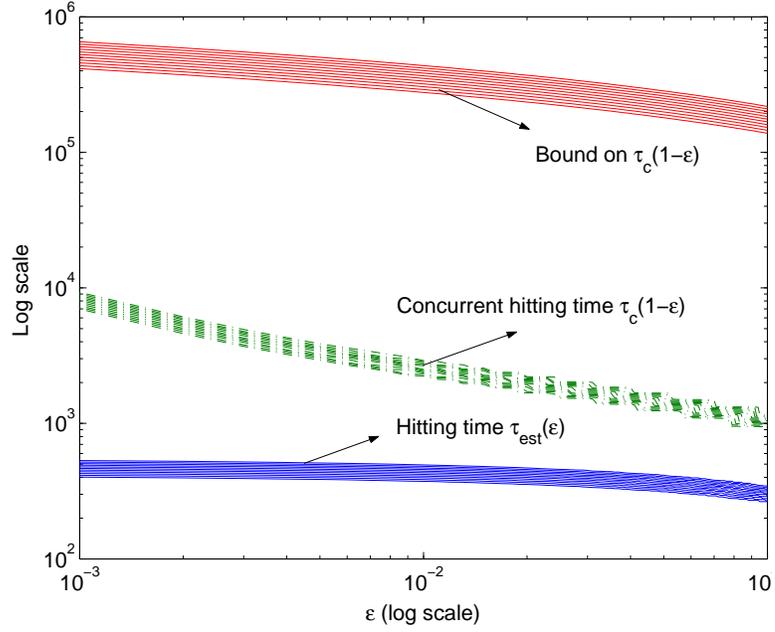}
\end{center}
\caption{Comparison of Quantum hitting time, concurrent hitting time and
the bound on the concurrent hitting time for dimensions 50 to 60(log scale)}
\label{epsilon50_60}
\end{figure}

How does the average hitting time compare to the concurrent hitting time for the walk on the hypercube?
Figure \ref{epsilon10_20} plots the estimate of the average hitting time $\tau_{\rm est}(\epsilon)$, the
concurrent hitting time $\tau_c(1-\epsilon)$, and the bound on $\tau_c(1-\epsilon)$ (obtained in \cite{Kem03b}) as a function of $\epsilon$ for dimensions from 10 to 20. (Both the axes are in log scale.)  Figure \ref{epsilon50_60} plots the same for dimensions from 50 to 60. We can see from these two figures that both $\tau_c(1-\epsilon)$ and the bound on it become less tight for higher dimensions, and both are much longer than the bound on the average hitting time $\tau_{\rm est}(\epsilon)$.  This, together with the comparison of $\tau_{\rm est}(\epsilon)$ to the exact value $\tau_h$, strongly indicates that the average walk ends much faster than these bounds might suggest, and also that the estimate $\tau_{\rm est}(\epsilon)$ is quite insensitive to the choice of $\epsilon$, at least for the hypercube.

\subsection{Hitting time in the presence of decoherence}
This decoherence can be thought of as acting on the system at every step of the measured walk. Thus the evolution of the walk can be decomposed into three parts--the unitary evolution operator followed by the decoherence  (CPTP) map and then the partial measurement on the final vertex all taking place in sequence. The combined effect can be seen as the application of a superoperator at every step and it can be written as
\ber
\rho(t+1)&=&\mathcal{E}(\rho(t)) \nonumber \\
&=&\mathcal{M}\circ\mathcal{D}\circ\mathcal{U}(\rho(t))\nonumber \\
&=&\hat{M}_j\left (\sum_i \hat{A}_i (\hat{U}\rho(t) \hat{U}^\dag) \hat{A}_i^\dag \right)\hat{M}_j^\dag,
\eer
where $\hat{M}_j$ is the measurement operator with the outcome $j\in\{0,1\}$ with $\hat{M}_0=\hat{P}_f$ and $\hat{M}_1=\hat{Q}_f$.
The state of the system after $n$ steps can be written as 
\be
\rho(t)=\mathcal{E}^t(\rho(0)). 
\ee

The formula for hitting time in Eq. (\ref{hittime}) assumes a purely unitary evolution and partial measurement. We now modify this formula to include the effect of decoherence on the walk. In the presence of decoherence, the evolution without measurement is described by
\be
\rho(t+1)=\mathcal{D}\circ\mathcal{U}(\rho(t)) .
\ee
Therefore, we only need to replace the superoperator $\mathcal{U}$ to $\mathcal{D}\circ\mathcal{U}$. This means that the vectorized quantities $\mathbf{Y}$ and $\mathbf{N}$ are modified accordingly. They become
\ber
\mathbf{Y}_D&=&(\hat{P}_f\otimes\hat{P}^\ast_f)(\sum_i \hat{A}_i\otimes\hat{A}_i^\ast)(\hat{U}\otimes\hat{U}^\ast) \nonumber \\
\mathbf{N}_D&=&(\hat{Q}_f\otimes\hat{Q}^\ast_f)(\sum_i \hat{A}_i\otimes\hat{A}_i^\ast)(\hat{U}\otimes\hat{U}^\ast) ,
\eer
and we can write $\mathbf{D}=\sum_i \hat{A}_i\otimes\hat{A}_i^\ast$ as the decoherence superoperator. The structure of the formula for hitting time given by Eq. (\ref{hittime}) remains the same i.e.
\be
\tau_h = I^v \cdot \left( \mathbf{Y}_D(\mathbf{I}-\mathbf{N}_D)^{-2}\rho_0^v \right) .
\ee
 
\subsection{Dephasing in the position and coin space}
\begin{figure}[tbh]
\begin{center}
\includegraphics[width=5in]{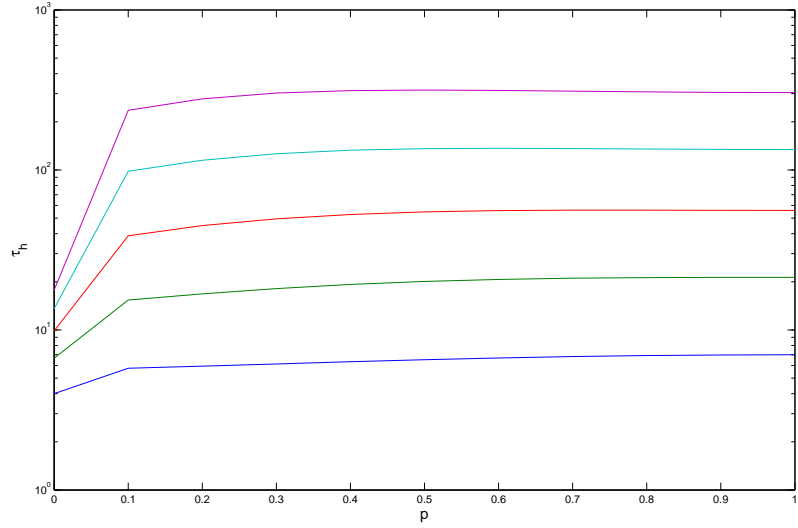}
\end{center}
\caption{Hitting time (log scale) vs $p$ for dimensions $n=3$ to $n=7$ with dephasing in the position and coin space.}\label{ht_vc}
\end{figure}

\begin{figure}[tbh]
\begin{center}
\includegraphics[width=5in]{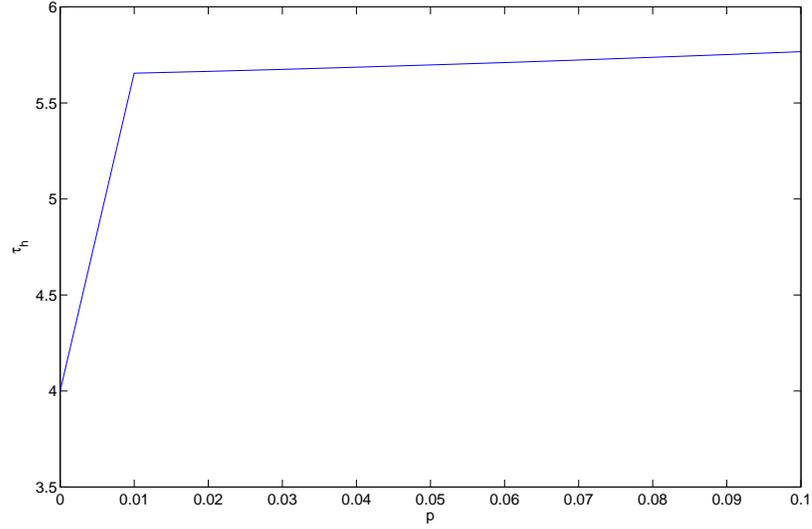}
\end{center}
\caption{Hitting time (log scale) vs $p$ for dimensions $n=3$ for small values of decoherence.}\label{ht_vc_001}
\end{figure}

\begin{figure}[tbh]
\begin{center}
\includegraphics[width=5in]{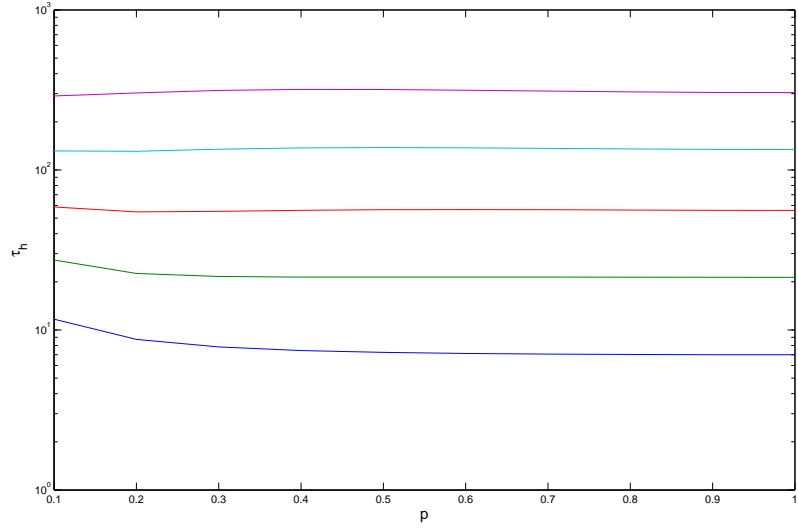}
\end{center}
\caption{Hitting time (log scale) vs $p$ for dimensions $n=3$ for small values of decoherence for an initial state which has an infinite hitting time.}\label{ht_vc_init2}
\end{figure}

Complete dephasing has been considered as a kind of decoherence for the discrete-time quantum walk on the line \cite{KT03,BCA03a,BCA03b} and the hypercube in \cite{KT03}. Here we analyze the effect of this kind of dephasing on the hitting time. Consider the discrete-time quantum walk on the hypercube. The evolution operator is given by $\hat{U}=\hat{S}(\hat{I}\otimes\hat{C})$ where $\hat{S}$ is given by Eq. (\ref{S_Pauli}) and $\hat{C}$ is the Grover coin given by Eq. (\ref{Grover_matrix}). The decoherence we consider is the dephasing map given by
\be
\mathcal{D}(\rho)=(1-p)\rho + \sum_i p \hat{\mathbb{P}}_i\rho\hat{\mathbb{P}}_i,
\ee
where $\hat{\mathbb{P}}_i$ is a projector onto a basis element of the form $|v\ra\otimes |c\ra$, $v$ corresponds to a vertex and $c$ to a direction in the coin space. The initial condition chosen is $|00\dots 0\ra\otimes (1/\sqrt{n})\sum_i|i\ra$, ie., the $\vec{0}$ vertex and the equal superposition of all directions. This initial condition lies in the subspace that gives fast hitting times for the non-decohering walk. We analyze it in the presence of decoherence. Fig.~(\ref{ht_vc}) plots the hitting time against the decoherence parameter $p$ for dimensions $n=3$ to $n=7$. Observe that the hitting time has an initial jump from $p=0$ to $p=0.1$ after which it increases smoothly. This jump increases from $n=3$ to $n=7$ suggesting that for higher dimensions it gets worse. Fig. (\ref{ht_vc_001}) plots the hitting time for $p$ ranging from $0.01$ to $0.1$ and we can see that even for a little decoherence i.e. $p=0.01$, the hitting time jumps and then increases uniformly. Fig.~(\ref{ht_vc_init2}) plots the hitting time against the decoherence parameter for the initial state given by $|00\dots 0\ra\otimes |1\ra$. We shall show later that this initial state has an infinite hitting time for the quantum walk with the Grover coin. As can be seen it becomes finite in the presence of decoherence. This shows how this kind of decoherence can turn purely quantum behavior into something classical.

\subsection{Dephasing in the coin space}
\begin{figure}[tbh]
\begin{center}
\includegraphics[width=5in]{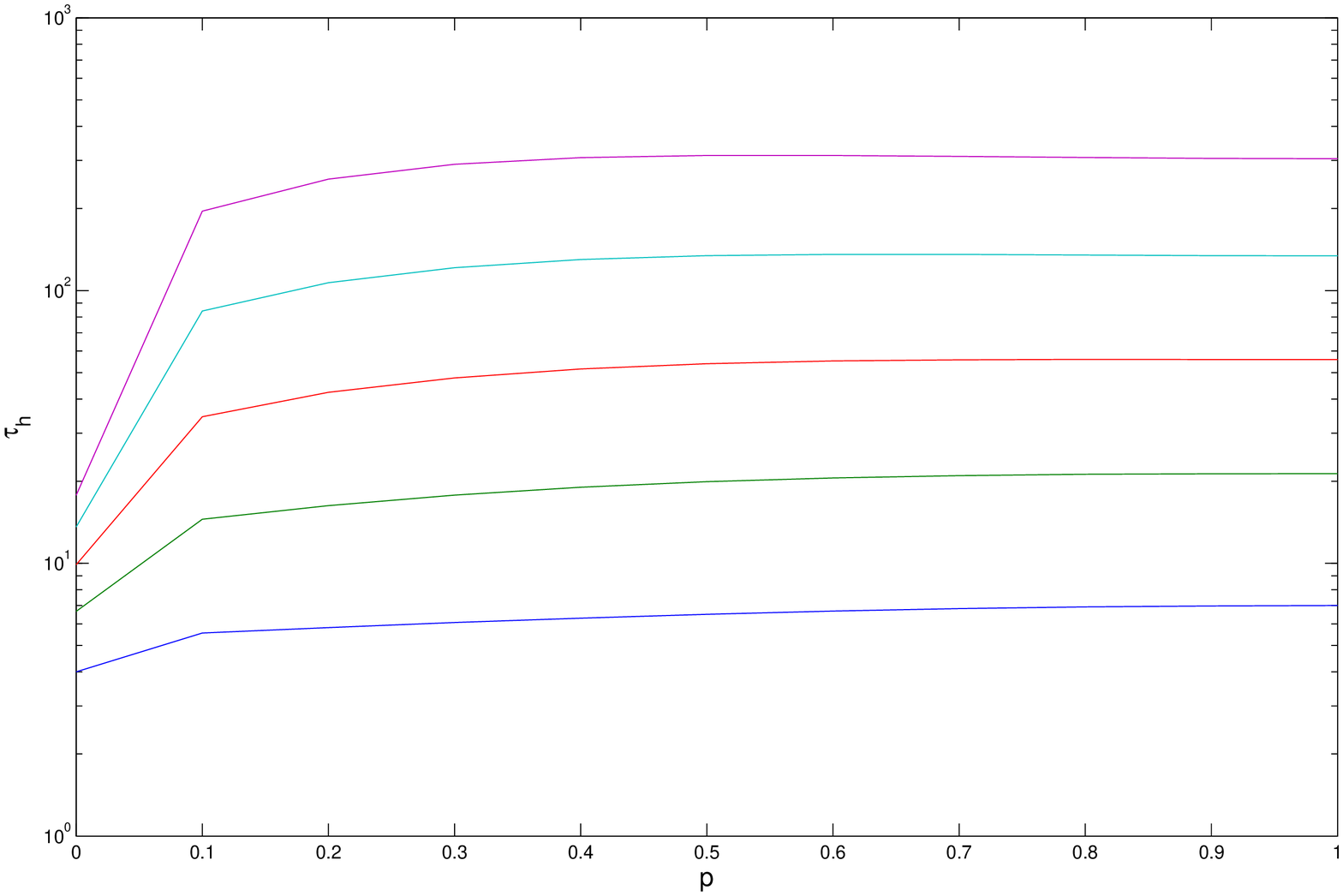}
\end{center}
\caption{Hitting time (log scale) vs $p$ for dimensions $n=3$ to $n=7$ with dephasing in the coin space.}\label{ht_c}
\end{figure}

Now we consider a decoherence process that acts only in the coin space. In particular, we consider a dephasing map given by
\be
\mathcal{D}(\rho)=(1-p)\rho + \sum_i p (\hat{I}\otimes\hat{\mathbb{P}}_i) \rho (\hat{I}\otimes\hat{\mathbb{P}}_i),
\ee
where $\hat{\mathbb{P}}_i$ are projectors only onto the coin basis states. Fig. (\ref{ht_c}) plots the hitting time against the parameter $p$. The behavior for this type of decoherence is qualitatively the same as before. The hitting time is a little less than in the previous case except at $p=0$ and $p=1$. At these two points the hitting time is equal to the one in the previous kind of dephasing precess.

\subsection{Dephasing in the position space}
\begin{figure}[tbh]
\begin{center}
\includegraphics[width=5in]{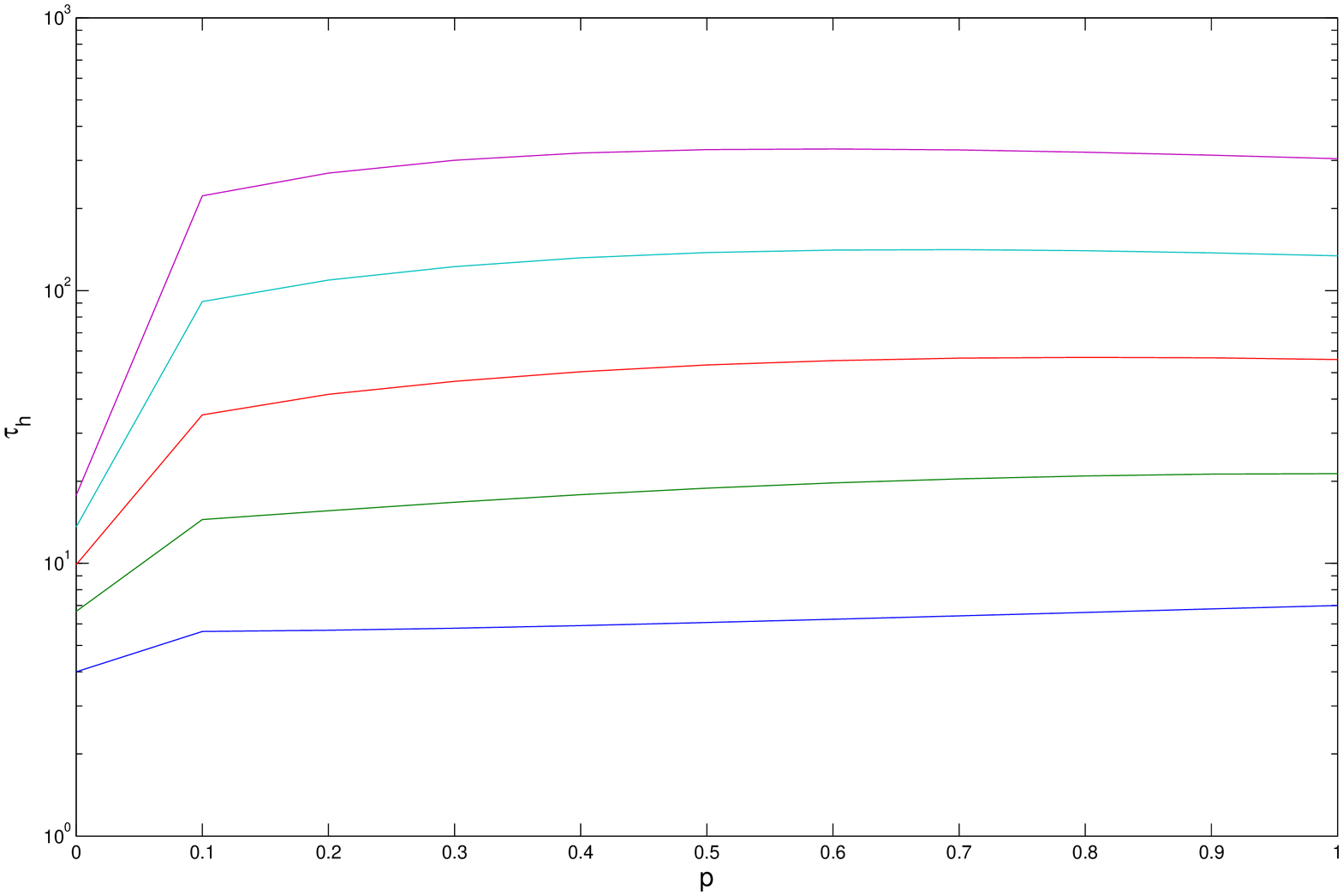}
\end{center}
\caption{Hitting time (log scale) vs $p$ for dimensions $n=3$ to $n=7$ with dephasing in the position space.}\label{ht_v}
\end{figure}

Finally, we consider a decoherence process that acts only in the position space. The dephasing map in this case is given by
\be
\mathcal{D}(\rho)=(1-p)\rho + \sum_i p (\hat{\mathbb{P}}_i\otimes\hat{I}) \rho (\hat{\mathbb{P}}_i\otimes\hat{I}),
\ee
where $\hat{\mathbb{P}}_i$ are projectors only onto the position basis states. Fig. (\ref{ht_v}) shows the variation of hitting time with the decoherence parameter. Note that when there is no dephasing and pure dephasing, i.e. $p=0$ and $p=1$, then the hitting time is the same whether the dephasing is in the position space, coin space or both spaces.

\subsection{Analysis of the slope of hitting time}
\begin{figure}[tbh]
\begin{center}
\includegraphics[width=5in]{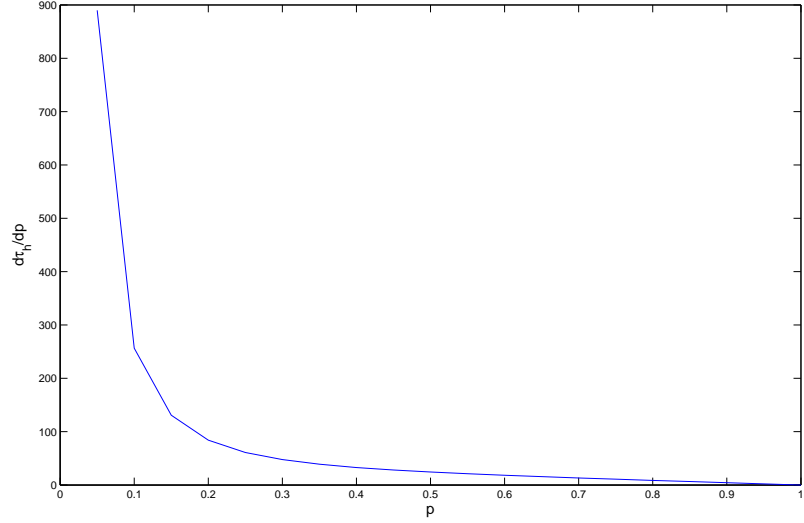}
\end{center}
\caption{Slope of the hitting time vs $p$ for the dimension $n=3$ with dephasing in the position and coin space.}\label{sl_ht_vc}
\end{figure}

Here we analyze the slope of the expression for hitting time with respect to the decoherence parameter $p$ in order to determine the behavior of the jump for small values of $p$. The hitting time from Eq. (\ref{hittime}) is
\be
\tau_h = I^v \cdot \left( \mathbf{Y}(\mathbf{I}-\mathbf{N})^{-2}\rho_0^v \right),
\ee
where the matrices $\mathbf{Y}$ and $\mathbf{N}$ are defined as
\ber
\mathbf{Y}&=&(\hat{P}_f\otimes\hat{P}^\ast_f)((1-p)\hat{I}\otimes\hat{I}+p \sum_i \hat{P}_i\otimes\hat{P}_i^\ast)(\hat{U}\otimes\hat{U}^\ast) \nonumber \\
\mathbf{N}&=&(\hat{Q}_f\otimes\hat{Q}^\ast_f)((1-p)\hat{I}\otimes\hat{I}+p \sum_i \hat{P}_i\otimes\hat{P}_i^\ast)(\hat{U}\otimes\hat{U}^\ast) \nonumber.
\eer
Here we consider complete dephasing i.e., $\hat{P}_i$ is the projector from dephasing in the position and coin spaces. Differentiating this expression w.r.t $p$, we get
\be
\frac{d\tau_h}{dp}=I^v \cdot \left( \frac{d\mathbf{Y}}{dp}(\mathbf{I}-\mathbf{N})^{-2}\rho_0^v \right) + 2 I^v \cdot \left( \mathbf{Y}(\mathbf{I}-\mathbf{N})^{-3}\frac{d\mathbf{N}}{dp}\rho_0^v \right) \nonumber .
\ee
The derivative of $\mathbf{Y}$ is given by
\be
\frac{d\mathbf{Y}}{dp}=(\hat{P}_f\otimes\hat{P}^\ast_f)(\frac{d\mathbf{D}}{dp})(\hat{U}\otimes\hat{U}^\ast) ,
\ee 
where
\be
\frac{d\mathbf{D}}{dp}=-\hat{I}\otimes\hat{I} + \sum_i \hat{P}_i\otimes\hat{P}_i^\ast .
\ee
We can define a similar expression for the derivative of $\mathbf{N}$. In Fig. (\ref{sl_ht_vc}) we plot this slope for various values of the parameter $p$. Note that for $p=0$ this expression has the same problem i.e., of non-invertible matrices as the expression for hitting time.

\section{Prior work}
Kendon \cite{Ken06} has written a good review of the work done so far in analyzing decoherence in quantum walks on many graphs such as the line, cycle and the hypercube. Here we review some of the prior work on the hypercube. Decoherence in discrete-time quantum walks on the hypercube has been studied by Kendon and Tregenna \cite{KT03}, where they determine its effect on one-shot and concurrent hitting times defined in \cite{Kem03b}. It has been proved that the concurrent hitting time is $\tau_c(p_0)=\frac{\pi}{2}$ for $p_0=\Omega(\frac{1}{n\log^2n})$ for a hypercube of dimension $n$ (for the symmetric initial condition used in this paper). Kendon and Tregenna observe that for the range $0\leq p\leq 1/n$, where $n$ is the dimension of the hypercube, the quantum speed-up is preserved. However, we notice in our simulations that for the definition of hitting time considered here (defined in \cite{KB05}), the situation is different. The quantum speed-up is not preserved even for small values of $p$, which seems to show that this definition of the hitting time is more sensitive to decoherence as it captures the time-averaged dynamics.

Decoherence in continuous-time quantum walks on the hypercube has been analyzed in \cite{Alagic05}. The kind of decoherence considered is dephasing in the position basis (recall that there is no coin space in this kind of walk). The evolution of this walk can be described by the Hamiltonian
\be
\hat{H}= \hat{X}\otimes\hat{I}\otimes\dots\otimes\hat{I}+ \hat{I}\otimes\hat{X}\otimes\dots\otimes\hat{I}  \otimes + \ldots + \hat{I}\otimes\hat{I}\otimes\dots\otimes\hat{X},
\ee 
where $\hat{X}$ is the Pauli $\sigma_x$ operator. The hopping rate is chosen as $\gamma=k/n$, where $n$ is the dimension of the hypercube and $k$ is the total energy of the system. In order to incorporate decoherence, a superoperator picture was used. In this picture, the non-decohering walk has the superoperator $\hat{U}_t\otimes\hat{U}_t^\dag$, where $\hat{U}_t=\exp(-it\hat{H})$. In the presence of decoherence this superoperator becomes
\be
S_t=\prod_{j=1}^n [\identity\otimes\identity]\otimes\dots\otimes e^{\hat{A}}\otimes\dots\otimes [\identity\otimes\identity] ,
\ee
where
\be
\hat{A}=\gamma t[(\identity\otimes i\hat{X})-(i\hat{X}\otimes\identity)-p(\identity\otimes\identity)+ p(\Pi_1\otimes\Pi_1)+p(\Pi_0\otimes\Pi_0)] ,
\ee
$\Pi_0=|0\ra\la 0|$ and $\Pi_1=|1\ra\la 1|$. It has been observed in \cite{Alagic05} that for $p<4k$ the walk has quantum behavior and for $p>4k$ it has classical behavior.

\section{Results for the DFT coin}

The hitting time for a discrete-time quantum walk on the hypercube using the DFT coin, by contrast
to the Grover coin case considered above, can actually be infinite. For n=4 we
will demonstrate that for the same initial condition as with the Grover coin, the
hitting time for a quantum walk using the DFT coin is infinity.
This is because there exist eigenvalues of the evolution
operator $\mathcal{\hat{U}}$ whose eigenvectors have an overlap
with the initial state, but have no overlap with the final vertex
for any state of the coin. In this subsection we specialize to the DFT coin, but a more detailed and general explanation of infinite hitting times based on the symmetry of the walk is presented in the next chapter.

Suppose there are states $|\phi\ra$ which have no overlap with the
final vertex,  $\la\phi|\hat{P}_f|\phi\ra = 0$, and which are eigenstates
of the evolution operator $\hat{U}$:   $\hat{U}|\phi\ra = \exp(i\theta)|\phi\ra$.
If the system is in the state $|\phi\ra$, clearly there is no probability to
ever detect the particle in the final vertex.  
Let $\mathcal{\hat{P}}$ be a projector onto all such states $|\phi\ra$.
Then $\mathcal{\hat{P}}\hat{P}_f = \hat{P}_f\mathcal{\hat{P}} = 0$, and
$[\mathcal{\hat{P}}, \hat{U}] = 0$, where the commutator is defined as $[A,B]=AB-BA$.  One can write the initial state as a superposition of vectors in this subspace and its orthogonal complement:
\begin{equation}
|\Psi\ra=\mathcal{\hat{P}}|\Psi\ra+(\mathcal{\hat{I}}-\mathcal{\hat{P}})|\Psi\ra .
\end{equation}
Any state that begins in the subspace selected by $\mathcal{\hat{P}}$
will remain there for all time, and any
state in the orthogonal complement will stay there; this follows from the fact
that the projectors commute with both the unitary transformation $\hat{U}$
and the measurement operator $\hat{P}_f$.  As one starts the walk, the
probability that the particle never reaches the final state is
$\la\Psi|\hat{U^{\dag^t}}\mathcal{\hat{P}}\hat{U}^t|\Psi\ra$,
which is $\la\Psi|\mathcal{\hat{P}}|\Psi\ra$.

In order for this probability to be nonzero, there must be eigenstates of
the unitary evolution operator $\hat{U}$ which have no amplitude for
the final vertex.  We can readily demonstrate this for the hypercube with
the DFT coin.  Consider the 4-dimensional hypercube.  Numerically diagonalizing
the evolution operator $\hat{U}$, we find it has $i,-i,1$ and
$-1$ among its eigenvalues, each with a degeneracy of 8.
Since the subspace corresponding to the final vertex is 4-dimensional,
it is clearly possible to construct a superposition of eigenvectors of any of these
eigenvalues so that it has no overlap with the final vertex in any
coin state.  For each of the four degenerate eigenvalues we can construct
a 4-dimensional subspace of eigenvectors with no overlap with the final
vertex, giving a 16-dimensional space for all such eigenvectors.
By numerically constructing an orthonormal basis for this space, we can
find an expression for the projector $\mathcal{\hat{P}}$ and measure
its overlap with the initial state.

We considered in particular the
initial state where the particle was located at the $|00\dots0\ra$ vertex
and the coin is in an equal superposition of basis states $|0\ra,\ldots,|n-1\ra$.
For the hypercube with $n=4$ and the given initial state,
the probability $\la\Psi|\mathcal{\hat{P}}|\Psi\ra$ is
$0.4286$, which exactly matches the total probability to never
hit the final node after a large number of iterations in our
numerical simulations.  Thus, the probability is close to half that the
particle never reaches the final state and the hitting time
becomes infinity.

This demonstrates a property of quantum walks
not seen in their classical counterparts:  for certain initial
conditions, there is a nonzero probability that the particle never
reaches the final state, even though the initial and final states
of the graph are connected.  For a quantum walk with substantial
degeneracy, this phenomenon is likely to be generic.  It might
be possible to make the hitting time finite by choosing an appropriate
initial condition---clearly this happens for the Grover coin---but for
some coins this may require an initial condition which is not
localized on one vertex.  From our simulations, it it seems that for
higher dimensions the DFT coin behaves similarly to $n=4$.
For example, for $n=5$, our simulations show that the
probability to hit the final node increases slowly but does not
reach 1 even after many time steps. This could be due to the
fact that the final vertex has no overlap with some eigenvectors
of the evolution operator (as for $n=4$), and additionally that
it overlaps very little for some other eigenvector.  This would make
the probability increase slowly but never reach 1.

\section{Results for a quantum walk on a distorted hypercube}

If, as seems likely, the dramatic speed-ups (and slow downs) of
quantum walks over their classical counterparts depend on the
symmetry of the graph, it should be instructive to see the effect
of deviations from that symmetry.  In this section, we look at
results for the measured walk using the Grover coin on a
distorted hypercube. The distorted hypercube is defined by
constructing the usual hypercube, and then switching two
of the connections.  Pick 4 vertices which form a face--for example,
$(0\dots00),(0\dots01),(0\dots10),(0\dots11)$. Calling these vertices
$A,B,C,D$ for short, we distort the hypercube by connecting $A$ to $D$
and $B$ to $C$, and removing the edges between $A$ and $B$ and
between $C$ and $D$. This is still a regular graph, and the same
quantum walk can be used without having to redefine the evolution
operator.  Unlike the usual hypercube, it is no longer a bipartite graph,
and the walk can no longer be reduced to a walk in Hamming weight.

\begin{figure}[tbh]
\begin{center}
\includegraphics[width=4in]{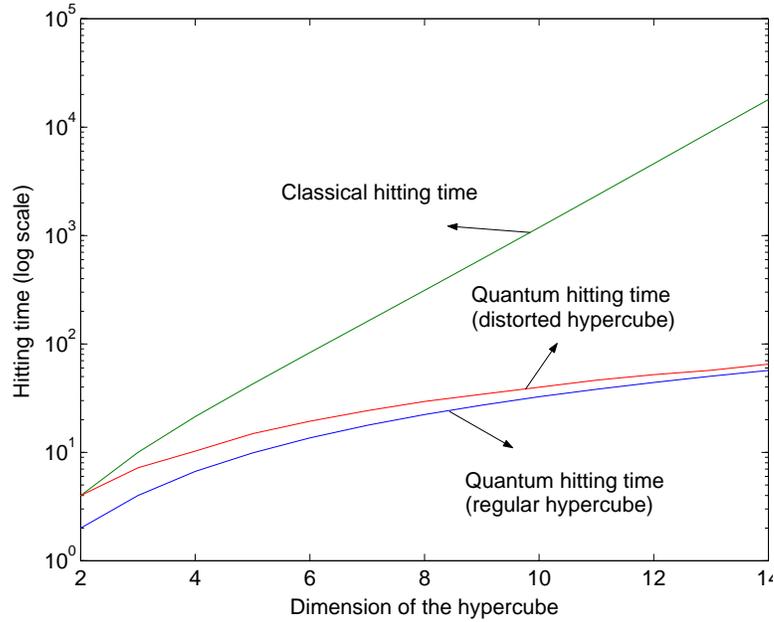}
\end{center}
\caption{Comparison of hitting times on the regular and
distorted hypercubes} \label{skewed}
\end{figure}

Figure \ref{skewed} plots the hitting time of a quantum walk on a
distorted hypercube together with that of a classical walk and a
quantum walk on regular hypercubes for comparison. The hitting
time for a quantum walk on a distorted hypercube is more than that
of a quantum walk on a regular hypercube, but still is much
smaller than the hitting time of a classical walk. In fact, as the
dimension increases one can see that the hitting times of the
quantum walk on the distorted and regular hypercubes converge
towards each other.  This presumably reflects the fact that for
higher dimensions the symmetry is mostly unchanged.

\section{Infinite hitting times}

We will now show that it is possible for hitting times to be {\it infinite}, and derive a sufficient condition for the unitary evolution operator to allow infinite hitting times. As before, we begin by forming the projector $\hat{P}$ onto the subspace spanned by all eigenstates of $\hat{U}$ which have no overlap with the final vertex.  This projector is orthogonal to the projector onto the final vertex, $\hat{P}\hat{P}_f = \hat{P}_f\hat{P} = 0$, and commutes with $\hat{U}$, $[\hat{U},\hat{P}]=0$.  We assume (for the moment) that this projector is nonzero; later, we will find a sufficient condition for this to be true, and exhibit quantum walks which satisfy this condition.  We can write any initial state as a superposition of a state in the subspace projected onto by $\hat{P}$ and a state orthogonal to it, giving the decomposition
\begin{equation}
|\Psi\ra=\hat{P} |\Psi\ra + (\hat{I}-\hat{P}) |\Psi\ra .
\end{equation}
It is easy to see that if $|\Psi\ra$ lies entirely inside $\hat{P}$, i.e., $\hat{P}|\Psi\ra=|\Psi\ra$, then under the evolution the subsequent states will never have any component in the final state and the probability defined in eq. (\ref{prob.eqn}) will be zero. Indeed, since $[\hat{P},\hat{U}]=0$ and $[\hat{P},\hat{Q}_f]=0$,
\begin{eqnarray}
p(t) &=& \tr\{\hat{P}_f\hat{U}[\hat{Q}_f\hat{U}]^{t-1}
\rho_0[\hat{U^{\dag}}\hat{Q}_f]^{t-1}\hat{U^{\dag}}\hat{P}_f\}\\
&=& \tr\{\hat{P}_f\hat{U}[\hat{Q}_f\hat{U}]^{t-1}
\hat{P}\rho_0\hat{P}[\hat{U^{\dag}}\hat{Q}_f]^{t-1}\hat{U^{\dag}}\hat{P}_f\}\\
&=& \tr\{\hat{P}_f\hat{P}\hat{U}[\hat{Q}_f\hat{U}]^{t-1}
\rho_0[\hat{U^{\dag}}\hat{Q}_f]^{t-1}\hat{U^{\dag}}\hat{P}^{\dag}\hat{P}_f\}\\
&=& 0 ,
\end{eqnarray}
where $\rho_0=|\Psi\ra\la\Psi|$. Therefore, the hitting time for this initial state is infinite. More generally, if $|\Psi\ra$ has nonzero overlap with $\hat{P}$, $\hat{P}|\Psi\ra \ne 0$, then that component of $|\Psi\ra$ can never reach the final vertex. The probability of ever hitting the final vertex if one starts with this initial state is
\begin{equation}
p=|\la\Psi|(\hat{I}-\hat{P})|\Psi\ra|^2 < 1,
\end{equation}
and the hitting time is again infinite.

To construct this projector, we look at the spectral decomposition of $\hat{U}$. If $\hat{U}$ has at least one sufficiently degenerate eigenspace, then we can construct a subspace of this eigenspace which has a zero overlap with the final state. For instance, consider one such degenerate eigenspace which has a degeneracy of $k$. Since the vector space at the final vertex is $d$ dimensional (i.e., it has $d$ coin degrees of freedom), we would be solving the following $d\times k$ system of homogeneous equations:
\begin{equation}
a_1\begin{pmatrix}
v_{N-d+1}^1\\
\vdots\\
v_N^1\end{pmatrix}+
a_2\begin{pmatrix}
v_{N-d+1}^2\\
\vdots\\
v_N^2\end{pmatrix}+
\cdots+
a_k\begin{pmatrix}
v_{N-d+1}^k\\
\vdots\\
v_N^k\end{pmatrix}=0 .
\end{equation}
Here, we use a labeling where the final vertex in some coin state occupies the last $d$ entries of the eigenvectors. The subscript refers to the component of the eigenvector, and the superscript distinguishes the eigenvectors in the degenerate eigenspace. This system is under-determined if $k>d$, and it will always have a nontrivial solution---in fact, it will have a space of solutions of dimension $k-d$. Therefore, it is sufficient that there exist at least one eigenspace of $\hat{U}$ with dimension greater than the dimension of the coin, in order to have a nonzero projector $\hat{P}$.  If there is more than one degenerate eigenvalue with multiplicity greater than $d$, the subspace projected onto by $\hat{P}$ will include {\it all} the eigenvectors of $\hat{U}$ which have no overlap with the final vertex.

The condition derived above is closely related to the question of invertibility of $\mathbf{I}-\mathbf{N}$ in Eq.~(\ref{hittime}) of the previous section. Here we show that $\mathbf{I}-\mathbf{N}$ is not invertible if and only if the projector $\hat{P}$ is nonzero. Furthermore, in the case when $\mathbf{I}-\mathbf{N}$ is not invertible, the hitting time for a state or density matrix whose support has no overlap with $\hat{P}$ is calculated by replacing the inverse of $\mathbf{I}-\mathbf{N}$ with its pseudo-inverse in Eq.~(\ref{hittime}).

Assume that the projector $\hat{P}$ is nonzero. Then there is at least one eigenvector $|v\ra$ of $\hat{U}$ such that $\hat{P}_f |v\ra=0$. Therefore,
\begin{equation}
(\hat{Q}_f \hat{U}\otimes \hat{Q}_f^\ast \hat{U}^\ast) (|v\ra\otimes |v\ra^\ast) = (\hat{Q}_f \otimes \hat{Q}_f^\ast)  (|v\ra\otimes |v\ra^\ast) = |v\ra\otimes |v\ra^\ast ,
\end{equation}
since $\hat{U} |v\ra=\exp(i\theta) |v\ra$ and $\hat{U}^\ast |v\ra^\ast = \exp(-i\theta) |v\ra^\ast$, and $\hat{P}_f |v\ra=0$ implies that $\hat{Q}_f |v\ra = |v\ra$. Therefore, $(\mathbf{I}-\mathbf{N})|v\ra\otimes |v\ra^\ast=0$, $\mathbf{I}-\mathbf{N}$ has a nonzero nullspace, and hence is not invertible.  This proves the ``if'' direction.

To prove the ``only if'' direction, assume that $\mathbf{I}-\mathbf{N}$ is not invertible. This implies that there exists a normalized vector $|u\ra \in {\cal H}\otimes{\cal H}^\ast$ such that
\begin{equation}
(\hat{I}-\hat{Q}_f \hat{U}\otimes \hat{Q}_f^\ast \hat{U}^\ast) |u\ra = 0 \Longrightarrow
\hat{Q}_f \hat{U}\otimes \hat{Q}_f^\ast \hat{U}^\ast |u\ra=|u\ra .
\label{N_nullspace}
\end{equation}
The vector $|u\ra$ is an eigenvector of $\hat{Q}_f \hat{U}\otimes \hat{Q}_f^\ast \hat{U}^\ast$ with eigenvalue 1.  Since $\hat{Q}_f\otimes\hat{Q}_f^\ast$ is a projector, the vector $|u\ra$ must therefore lie in the eigenspace of eigenvalue $1$ of both $\hat{Q}_f\otimes\hat{Q}_f^\ast$ and $\hat{U}\otimes\hat{U}^\ast$.  This can only be true if $|u\ra$ is of the form:
\begin{equation}
|u\ra=\sum_{i,j}a_{i j}|v_i\ra\otimes |v_j\ra^\ast ,
\end{equation}
where the $\{|v_i\ra\}$ are eigenvectors of $\hat{U}$, and $a_{ij}$ is only nonzero if $|v_i\ra$ and $|v_j\ra$ lie in the same eigenspace of $\hat{U}$ and $\hat{P}_f|v_i\ra = \hat{P}_f|v_j\ra = 0$. Note that the vector $|u\ra$ need not correspond to a physical state in the Hilbert space of the walk. But the existence of such a $|u\ra$ means that the projector $\hat{P}$ is nonzero, since there must exist at least one $|v_i\ra$ which has a zero overlap with $\hat{P}_f$.  This proves the ``only if'' direction.

We have shown that if $\mathbf{I}-\mathbf{N}$ is not invertible then the projector $\hat{P}$ is nonzero and vice-versa. Now we will see that if a density operator $\rho$ is orthogonal to $\hat{P}$, $\rho\hat{P} = \hat{P}\rho = 0$, its corresponding vector $\rho^v$ lies outside the null space of $\mathbf{I}-\mathbf{N}$. If $\rho$ is orthogonal to $\hat{P}$, then when written in the eigenbasis $\{|v_i\ra\}$ of $\hat{U}$, the diagonal components $\la v_i|\rho|v_k\ra$ are nonzero if and only if $\hat{P}_f|v_i\ra\neq 0$. This implies that for the corresponding vectorized quantity $\rho^v$, we have 
\begin{equation}
I^v.(\hat{P}_f\otimes\hat{P}_f^\ast)\rho^v\neq 0 . 
\end{equation}
For any $\rho^v$,
\begin{equation}
I^v.(\hat{P}_f\otimes\hat{P}_f^\ast) (\hat{Q}_f\hat{U}\otimes\hat{Q}_f^\ast\hat{U}^\ast) \rho^v = 0 .
\end{equation}
Therefore, $\mathbf{N}\rho^v \ne \rho^v$ i.e., $\rho^v$ does not lie in the null space of $\mathbf{I}-\mathbf{N}$. Moreover, if $\rho$ is orthogonal to $\hat{P}$, then so are $\mathcal{N}\rho$ and $\mathcal{Y}\mathcal{N}^t\rho$. This is easy to see, since from Eq.~(\ref{superops}) we get
\begin{equation}
\hat{P}\mathcal{N}\rho = \hat{P}(\hat{Q}_f\hat{U}\rho\hat{U}^{\dag}\hat{Q}_f) = \hat{Q}_f\hat{U}\hat{P}\rho\hat{U}^{\dag}\hat{Q}_f = 0 ,
\end{equation}
since $\hat{P}\hat{Q}_f=\hat{P}$ and $[\hat{U},\hat{P}]=0$. Similarly, we obtain $\hat{P}\mathcal{Y}\rho = 0$, since
\begin{equation}
\hat{P}\mathcal{Y}\rho=\hat{P}(\hat{P}_f\hat{U}\rho\hat{U}^{\dag}\hat{P}_f)=0 .
\end{equation}
Therefore, if $\rho$ is orthogonal to $\hat{P}$, then all the terms of the type $\mathcal{Y}\mathcal{N}^{t-1}\rho$ for all $t$ and hence all the terms inside the trace of Eq. (\ref{derivative_form}) are orthogonal to $\hat{P}$, which means that the vectorized versions of all terms inside the trace in Eq.~(\ref{derivative_form}) lie outside the null space of $\mathbf{I}-\mathbf{N}$. Thus, for states that do not overlap with $\hat{P}$, the hitting time is finite as would be expected, and is given by the same formula Eq.~(\ref{hittime}) with the inverse replaced by a pseudo-inverse. 

We focus on the discrete time quantum walk for the remainder of this section, considering the walk on the hypercube in particular. It was observed in numerical simulations mentioned earlier that for a walk on the hypercube with the DFT coin, an initial state given by
\begin{equation}\label{init.state}
|\Psi\ra=|00\cdots 0\ra\otimes \frac{1}{\surd{d}}\sum_i|i\ra
\end{equation}
has an infinite hitting time. The phenomenon of infinite hitting times is not restricted to the walk with the DFT coin, however.  Numerical simulations, followed by analytical calculations, have shown that it also occurs with the Grover coin, but for different initial states. In fact, it turns out that for the Grover coin, the symmetric initial state (\ref{init.state}) is the {\it only} initial state localized at the vertex $|00\cdots0\ra$ that has a {\it finite} hitting time. Any other superposition of the coin states for that vertex will give an {\it infinite} hitting time, because all such states have a nonzero overlap with $\hat{P}$.

Given any vertex $v$ on a graph, it is natural to ask if there exists any superposition of its coin states which overlaps with $\hat{P}$, and for which coin state the overlap is maximum and for which it is minimum (or zero). We can write the projector $\hat{P}$ in the form
\begin{equation}
\hat{P}=\sum_{i,j,k,l} A_{ijkl}|x_i\ra\la x_j|\otimes |k\ra\la l|,
\end{equation}
where $\{|x_i\ra\}$ are the vertices and $\{|k\ra\}$ are the directions. Suppose the initial state is
\begin{equation}\label{alpha}
|\Psi\ra=|v\ra\otimes\sum_i\alpha_i|i\ra \equiv |v\ra \otimes |\alpha\ra.
\end{equation}
Its overlap with the projector $\hat{P}$ is given by
\begin{equation}
\la\Psi|\hat{P}|\Psi\ra=\sum_{k,l} A_{vvkl} \alpha_k^*\alpha_l .
\end{equation}
To find the superposition of coin states such that the overall initial state has the least (or greatest) overlap with $\hat{P}$, define the matrix
\begin{equation}\label{C_mtx}
\hat{C}_v = \tr_{\rm vertices}\left\{\hat{P}(|v\ra\la v|\otimes\hat{I}_{\rm coin})\right\} ,\ \ (\hat{C}_v)_{kl} = A_{vvkl} .
\end{equation}
The overlap of the initial state with $\hat{P}$ can be written in terms of this matrix as
\begin{equation}
\la\Psi|\hat{P}|\Psi\ra=\la\alpha|\hat{C}_v|\alpha\ra .
\end{equation}
The matrix $\hat{C}_v$ is Hermitian and positive, and hence has a spectral decomposition into a complete orthonormal basis of eigenstates with non-negative eigenvalues.  Assuming that $\{\lambda_i,|e_i\ra\}$ is the spectral decomposition of $\hat{C}_v$, we can rewrite the overlap as
\begin{equation}
\la\Psi|\hat{P}|\Psi\ra=\sum_i \lambda_i |\la\alpha|e_i\ra|^2 .
\end{equation}
From the above expression, we see that the overlap is maximum (or minimum) if $|\alpha\ra$ is in the direction of the eigenvector with the largest (or smallest) eigenvalue, and zero if $|\alpha\ra$ is an eigenvector with zero eigenvalue. Therefore, if $\hat{C}_v$ does not have a zero eigenvalue (i.e., is positive definite), then for that vertex every superposition of coin states will overlap with $\hat{P}$.  In other words, the hitting time will be infinity if one starts at that vertex no matter what coin state one chooses. Numerical calculations for the Grover and the DFT coins on the hypercube show that for the vertex $|00\cdots 0\ra$, an equal superposition of coin states is the only superposition that has a zero overlap with $\hat{P}$ for the Grover coin, and no superposition of coin states has a zero overlap for the DFT coin. Moreover, $\tr\{\hat{P}\}$ for the Grover coin on the hypercube for $n=4$ is $32$, which is fully half the dimension of the total space (dim=$2^4\cdot4=64$). These examples suggest that infinite hitting times may be a generic phenomenon on graphs with symmetry.

\section{Discussion}

In this chapter, we have examined the definition of hitting time for measured quantum walks and analyzed some of its properties. We used this definition to
obtain an expression for hitting time which is valid on any general
graph as long as the unitary evolution operator $\hat{U}$ of the walk
is defined. We simulated this hitting time for
a measured quantum walk using the Grover coin and compared it to
the classical hitting time and to the bounds obtained on it; the quantum
hitting time is exponentially smaller than the classical
hitting time. We also showed that the bounds on the hitting time
obtained in \cite{Kem03b} become less tight as the
dimension increases. We then investigated the effect of decoherence on the hitting time of a discrete quantum walk. We have observed that it affects the hitting time more dramatically than it does for the definitions proposed in earlier work. Classical behavior occurs for small values of the decoherence parameter and the hitting time is worse than it is for a corresponding classical walk for higher values of this parameter. We observed that in the case when the quantum walk has infinite hitting times (a phenomenon which is consequence of the quantum behavior of the walk), decoherence makes the hitting time finite and thus make the walk classical. We analyzed the effect of different kinds of decoherence and noticed that for dephasing (whether in the position or coin space or both), the hitting time deteriorates in more or less the same way. 

These results on fast and infinite hitting times show that simply making a walk quantum does not guarantee a speed-up
over the classical case.  We demonstrated that the hitting time for quantum
walks can depend sensitively on the initial condition, unlike classical walks.
For certain initial states, the DFT walk can have {\it infinite} hitting
time, a phenomenon not possible in classical random walks. This
dependence on the initial state varies with the coin used, since
for the same initial state the Grover walk has a polynomial
hitting time. This infinite hitting time is directly related to the
degeneracy of the eigenvalues of the evolution operator.
If the evolution operator is highly degenerate, then it is very likely
that there exist initial states which give infinite hitting times.

The cause of the speed-up in quantum hitting time may not be completely clear at the moment, but it will be shown in the next two chapters that the symmetry of the graph plays a major role in both the speed-up and slow down of the quantum walk. For now, we can see this as an interference effect. In the faster quantum walk, the different paths leading to the final vertex interfere constructively, enhancing the probability of arrival; paths which lead to ``wrong'' vertices interfere destructively, reducing the probability of meandering around in the graph for long times. Unlike a classical random walk, the quantum walk is sensitive to the presence of a global symmetry which is not apparent at a purely local level.  This phenomenon leads to the speed-up of the continuous-time quantum walk on the glued-trees graph as well \cite{CCDFGS03}.

However, this same reason is undoubtedly the culprit in the {\it slow-down}
observed for the DFT walk.  The existence of states which never arrive at
the final vertex is made possible by the degeneracy of the evolution operator
$\hat{U}$---a degeneracy which arises due to the symmetry of the graph.
The existence of states which never arrive at the final vertex can also be seen
as an interference effect, only in this case the interference of paths which lead
to the final vertex is {\it destructive}:  all amplitude to make a transition to
the final vertex cancels out.

The connection to symmetry is supported by the quantum walk on the distorted
hypercube. We observe that the hitting time is worse than that of
the usual hypercube, but still much smaller than that of a
classical walk. The curve of the hitting time on the distorted
hypercube seems to converge slowly to that of a quantum walk on
the regular hypercube. This is probably because the distortion used in our simulations is very mild. As the dimension grows, and
with it the number of edges and vertices, this distortion has less
effect on the overall symmetry.

We hasten to add that symmetry of the graph is not the sole reason
for speed-ups in quantum walks.  A polynomial speed-up has been
demonstrated in the quantum walk versions of the search of an 
unstructured database \cite{SKW03} and the element distinctness
problem \cite{Ambainis03}.  However, the dramatic exponential speed-ups
have all been demonstrated in highly symmetric graphs. In the next two chapters, we explore this idea in more detail and show how symmetry can have a strong influence on the quantum walk.

\chapter{Symmetry in quantum walks}
We saw in the previous chapter that degeneracy of the evolution operator leads to infinite hitting times. One of the main sources of degeneracy in quantum mechanics is {\it symmetry}. Since the $\hat{S}$ matrix which makes up part of the evolution operator encodes the connections of the graph, it is natural to expect that symmetries of the graph will produce symmetries of the evolution operator.  Here we analyze this idea.

Is it sufficient to consider only the symmetries of the graph?  Apart from the symmetries induced from the graph, the evolution operator may have additional symmetries of its own which lead to additional degeneracy. But such symmetries are difficult to analyze in generality, and depend on the details of how one defines the quantum walk. First, there is a choice between the discrete and the continuous walk. Second, for the discrete walk, if the structure $\hat{U}=\hat{S}(\hat{I}\otimes \hat{C})$ is used, there is still the freedom to use any unitary matrix as the coin.  In order to generalize our discussion of symmetries to any walk, we restrict attention to the symmetries induced by the graph alone. This naturally leads us to the question:  ``Are there graphs with sufficient symmetry such that, for any walk that is defined on the graph, the resultant evolution operator will have enough degeneracy to give rise to infinite hitting times?''  It turns out that such graphs do exist, and we give an example of such a class of graphs.  We will comment briefly on the effect of additional symmetries of the coin for the hypercube later in this chapter. We now look at the automorphism groups of Cayley graphs which provide the examples of graphs with infinite hitting times.

\section{Automorphism groups of Cayley graphs}
An {\it automorphism} of a graph is a permutation of its vertices such that it leaves the graph unchanged. The set of all such permutations is the {\it automorphism group} of the graph. When the edge labels or colors in the graph are important, as in the case of a discrete quantum walk, we restrict ourselves to those permutations which preserve the edge labels. In other words, an edge connecting two vertices has the same label before and after the permutation. Such automorphisms are called {\it direction-preserving}. In general, we could consider automorphisms where we permute the direction labels along with the vertices to obtain the same graph with the same coloring. This would form a larger group $G$ of which the direction-preserving automorphisms are a subgroup $H$.

Since the vertex Hilbert space $\mathcal{H}^v$ has its basis elements in one-to-one correspondence with the vertices of the graph, and the coin Hilbert space has a basis in correspondence with the direction labels, the automorphisms (which are just permutations of vertices and directions) are permutation matrices. In fact, these are all the permutation matrices on $\mathcal{H}^v\otimes\mathcal{H}^c$ that leave $\hat{S}$ unchanged, i.e., $\{$all $\hat{P}\, |\, \hat{P}\hat{S}\hat{P}^{\dag}=\hat{S}$, where $\hat{P}$ is a permutation matrix$\}$. In this representation, any direction-preserving automorphism has the structure $\hat{P}_v\otimes \hat{I}_c$, where $\hat{P}_v$ acts solely on $\mathcal{H}^v$ and $\hat{I}_c$ on $\mathcal{H}^c$. Such automorphisms become important if we wish to consider the symmetries of $\hat{U} \equiv  \hat{S}(\hat{I}\otimes\hat{C})$. Clearly, any automorphism of this type is a symmetry of $\hat{U}$, since
\begin{equation}
(\hat{P}_v\otimes \hat{I}_c) \left[ \hat{S}(\hat{I}\otimes\hat{C}) \right] (\hat{P}_v\otimes \hat{I}_c)^{\dag} = \left[ (\hat{P}_v\otimes \hat{I}_c)\hat{S}(\hat{P}_v\otimes \hat{I}_c)^{\dag} \right] (\hat{I}\otimes\hat{C}) = \hat{S}(\hat{I}\otimes\hat{C}) .
\end{equation}
Elements of $G$ in general do not act trivially on the coin space. Because of this, they need not be symmetries of $\hat{U}$ unless the coin flip operator $\hat{C}$ respects these symmetries. 

\begin{figure}[tbh]
\begin{center}
\includegraphics[width=4in]{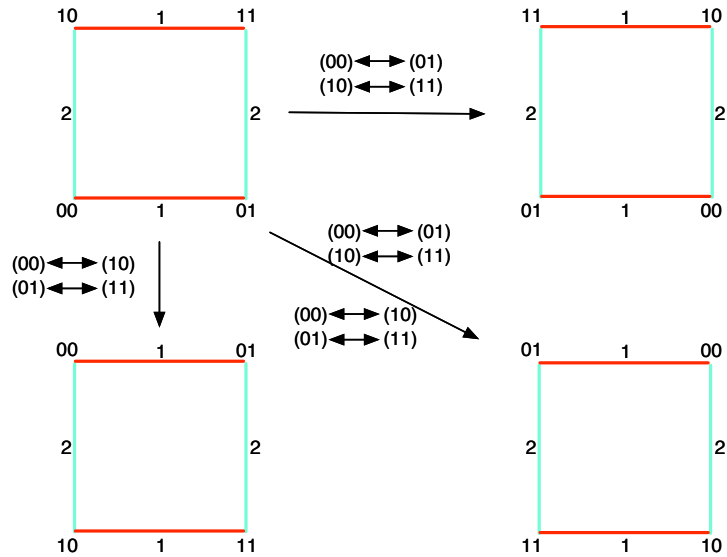}
\end{center}
\caption{The direction-preserving automorphism group of the n=2 hypercube.} \label{fig2}
\end{figure}

To illustrate all this, consider the example of a hypercube in 2 dimensions (i.e., a square). The vertex labels are $\{(00),(01),(10),(11)\}$ (which also form a basis for $\mathcal{H}^v$); the edges connecting $(00)$ to $(01)$ and $(10)$ to $(11)$ are both labeled $1$, and the edges connecting $(00)$ to $(10)$ and $(01)$ to $(11)$ are both labeled $2$. Thus, the transformation $(00)\leftrightarrow (01)$ and $(10) \leftrightarrow (11)$, or the transformation $(00)\leftrightarrow (10)$ and $(01) \leftrightarrow (11)$, or both together, are automorphisms of this graph which need no permutation of the directions. Together with the identity automorphism (which permutes nothing), these permutations form the direction-preserving subgroup $H$. In a matrix representation on the Hilbert space $\mathcal{H}^v\otimes\mathcal{H}^c$, they are,
\begin{equation}\nonumber
\begin{pmatrix}
0 & 1 & 0 & 0\\
1 & 0 & 0 & 0\\
0 & 0 & 0 & 1\\
0 & 0 & 1 & 0
\end{pmatrix}\otimes \hat{I}_c,
\begin{pmatrix}
0 & 0 & 1 & 0\\
0 & 0 & 0 & 1\\
1 & 0 & 0 & 0\\
0 & 1 & 0 & 0
\end{pmatrix}\otimes \hat{I}_c,
\begin{pmatrix}
0 & 0 & 0 & 1\\
0 & 0 & 1 & 0\\
0 & 1 & 0 & 0\\
1 & 0 & 0 & 0
\end{pmatrix}\otimes \hat{I}_c,
\begin{pmatrix}
1 & 0 & 0 & 0\\
0 & 1 & 0 & 0\\
0 & 0 & 1 & 0\\
0 & 0 & 0 & 1
\end{pmatrix}\otimes \hat{I}_c
\end{equation}
where $\hat{I}_c$ is the $2\times 2$ identity matrix acting on the coin space. These permutations can be easily seen to be $H=\{\hat{I}\otimes\hat{X}\otimes\hat{I}, \hat{X}\otimes\hat{I}\otimes\hat{I}, \hat{X}\otimes\hat{X}\otimes\hat{I}, \hat{I}\otimes\hat{I}\otimes\hat{I} \}$.  Just as in the representation of $\hat{S}$ matrix in terms of the Pauli $\hat{X}$ operators given by Eq.~(\ref{S_Pauli}), this group denotes a bit flip in the first, second or both bits of each vertex, together with the identity, which gives no flip.  (See Fig.~\ref{fig2}.)

\begin{figure}[h]
\begin{center}
\includegraphics[width=4in]{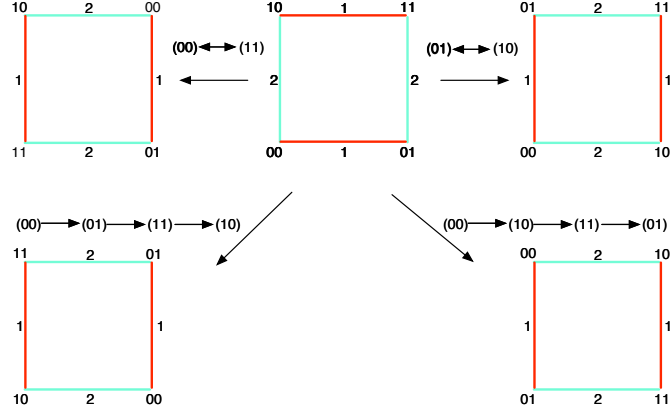}
\end{center}
\caption{Automorphisms which interchange directions for the n=2 hypercube.} \label{fig3}
\end{figure}

The permutation $(10)\leftrightarrow (01)$, reflecting along the diagonal while keeping $(00)$ and $(11)$ fixed, will be an automorphism only if we interchange the directions $1\leftrightarrow 2$. Similarly, the permutations $(00)\leftrightarrow (11)$, $(00)\rightarrow (01)\rightarrow (11)\rightarrow (10)$ and $(00)\rightarrow (10)\rightarrow (11)\rightarrow (01)$ are automorphisms when we interchange the two directions. If we view these permutations along with those obtained above, we obtain a new group $G$ for which $H$ is a subgroup. In a matrix representation, the new automorphisms are,
\begin{equation}\nonumber
\begin{pmatrix}
0 & 1 & 0 & 0\\
0 & 0 & 0 & 1\\
1 & 0 & 0 & 0\\
0 & 0 & 1 & 0
\end{pmatrix}\otimes \hat{X}_c,
\begin{pmatrix}
1 & 0 & 0 & 0\\
0 & 0 & 1 & 0\\
0 & 1 & 0 & 0\\
0 & 0 & 0 & 1
\end{pmatrix}\otimes \hat{X}_c,
\begin{pmatrix}
0 & 0 & 1 & 0\\
1 & 0 & 0 & 0\\
0 & 0 & 0 & 1\\
0 & 1 & 0 & 0
\end{pmatrix}\otimes \hat{X}_c,
\begin{pmatrix}
0 & 0 & 0 & 1\\
0 & 1 & 0 & 0\\
0 & 0 & 1 & 0\\
1 & 0 & 0 & 0
\end{pmatrix}\otimes \hat{X}_c,
\end{equation}
where $\hat{X}_c$ acts on the coin space and corresponds to an interchange of the two directions.  (See Fig.~\ref{fig3}.)  These four elements of $G$ need not be symmetries of $\hat{U}$, since the coin need not be symmetric under conjugation with $\hat{X}_c$. However, for the hypercube, if we use the Grover diffusion matrix as the coin, then the automorphism group $G$ is indeed its group of symmetries, since the Grover coin is symmetric under any permutation of its basis elements. The symmetry group of the evolution operator would be $H$ if the DFT coin is used, since the DFT does not have permutation symmetry. It is important to note that the symmetry group defined above is not the only thing that influences the degeneracy of the evolution operator. Degeneracy of the coin flip operator $\hat{C}$ can also induce degeneracy in the evolution operator, and the coin may be degenerate even if it does not have permutation symmetry (like the DFT coin). We discuss this in more detail later after describing the relationship between symmetry and degeneracy.

It can be shown that the direction-preserving automorphism group $H$ for any Cayley graph is isomorphic to the group on which the graph is defined. This is because any direction-preserving automorphism of a Cayley graph is a left translation by a  group element, and conversely all left translations are direction-preserving automorphisms. The first part of the statement is easy to see. Consider any left translation $L_a:G\rightarrow G$ which has the action $L_a(g)=a g$, for all $g\in G$. Now, given vertices $g$ and $h$ in $G$, they are connected by an edge from $g$ to $h$ if $g^{-1}h=s$, where $s\in S$. Clearly, after the transformation we still have $(a g)^{-1}(ah)=g^{-1}h=s$ and hence this automorphism preserves the direction labels. Since the group elements are basis states of the vertex Hilbert space, a left translation by a group element corresponds to a permutation matrix on this Hilbert space. The fact that every direction-preserving automorphism of a Cayley graph is a left translation by a group element becomes important in the discussion of regular and irreducible representations later in this chapter.

Finally, let us explicitly construct the representation of the automorphism group for the hypercube, which has $H\cong\mathcal{Z}_2^n$ and $G\cong H\cdot S_n$. In terms of the Pauli operators the representation of $H$ is $\{\hat{I}\hat{I}\hat{I}\cdots \hat{I}\otimes\hat{I}_c, \hat{X}\hat{I}\hat{I}\cdots \hat{I}\otimes\hat{I}_c, \hat{X}\hat{X}\hat{I}\cdots \hat{I}\otimes\hat{I}_c,\dots,\hat{X}\hat{X}\hat{X}\cdots \hat{X}\otimes\hat{I}_c\}$, where the tensor product symbol has been dropped in the vertex space, and $\hat{I}_c$ is the identity operator in the coin space. In fact, the representation of $H$ for any Cayley graph will be of the form $\hat{P}\otimes \hat{I}$, where $\hat{P}$ is a permutation matrix on the vertex space and $\hat{I}$ is the identity on the coin space. The group $G$ for the hypercube will become $H\cdot S_n=\{h \cdot \pi | h\in H, \pi\in S_n\}$, where $S_n$ is the permutation group on $n$ elements which is assumed to act on the direction labels. 

We now briefly review linear representations of finite groups, and describe how the symmetry group of $\hat{U}$ affects its degeneracy by determining the dimensions of the irreducible representations of the group.  (For further details, see for example \cite{S94} or \cite{Ser77}.)

\section{Representations of finite groups}

A linear representation of a finite group $G$ on a finite-dimensional vector space $V$ is a map $\sigma :G \rightarrow GL(V)$ , such that $\sigma(g h)=\sigma(g)\sigma(h)$.  $GL(V)$ is the space of invertible linear maps of $V$ onto itself.  If $V$ is $d$-dimensional, and we choose a basis of $d$ vectors in $V$, then $\sigma (g)$ for $g\in G$ becomes a $d\times d$ invertible matrix. The trace of this matrix is called the {\it character} of the representation. Therefore, characters are maps $\chi: G\rightarrow \mathbb{C}$ with $\chi(g)=\tr(\sigma(g))$.  Two representations $\sigma_1$ and $\sigma_2$ of the group $G$ on vector spaces $V_1$ and $V_2$, respectively, are considered equivalent if there exists an invertible linear map $\tau :V_1\rightarrow V_2$ such that $\tau\circ\sigma_1(g)=\sigma_2(g)\circ\tau$, for all $g\in G$. The characters of equivalent representations are equal, a fact which follows from the cyclic property of the trace operator.

Assume that the vector space $V$ has an inner product defined on it. Since the group $G$ is assumed finite, it can be shown that any representation $\sigma$ is equivalent to a {\it unitary} representation---there exists a basis for $V$ in which $\sigma(g)$ is a unitary matrix for all $g\in G$ (see \cite{Ser77}). A vector space $W$ is said to be {\it invariant} or {\it stable} under the action of $G$ if $x\in W\Rightarrow \sigma(g)x\in W$ for all $g\in G$. If the vector space $V$ has a subspace $W$ which is invariant under the action of $G$, then it can be shown that its orthogonal complement $W^\perp$ is also invariant under $G$, and so $V$ can be decomposed
\begin{equation}
V=W\oplus W^\perp .
\end{equation}
This means that the representation $\sigma$ as a matrix on $V$ can be written in a block diagonal form consisting of two blocks as
\begin{equation}
\sigma=\begin{pmatrix}
\sigma |_W & 0\\
0 & \sigma |_{W^\perp}
\end{pmatrix} ,
\end{equation}
where $\sigma |_W$ and $\sigma |_{W^\perp}$ are the restrictions of $\sigma$ to the subspaces $W$ and $W^\perp$. 

The linear map $\sigma: G\rightarrow GL(V)$ is called an {\it irreducible} representation (irrep), if it is a representation and no non-trivial subspace of $V$ is stable under the action of $G$.  Equivalently, it is an irreducible representation if it is not a direct sum of two representations. {\it Any} representation on $V$ can, by an appropriate choice of basis, be written as a representation in block diagonal form, where each block corresponds to an irrep.  So $V$ can be decomposed in the following way:
\begin{equation}\label{V_decomp}
V=W_1\oplus W_2\oplus\cdots\oplus W_k ,
\end{equation}
where each of the $W_i$ is stable under the action of $\sigma(G)$.  This decomposition is unique, up to reordering of the spaces and an overall equivalence transformation.  By an abuse of notation, we will use the same labels ($V$ or $W_i$) to refer both to the vector space and to the group representation on that space. Each component $W_i$ of the decomposition in (\ref{V_decomp}) is isomorphic to an irrep, and the number of such $W_i$ isomorphic to a given irrep does not depend on the details of the decomposition.

If we define an inner product for characters,
\begin{equation}
(\phi|\chi)=(1/|G|)\sum_{g\in G} \phi(g)\chi(g)^\ast ,
\end{equation}
it can be shown that the number of times an irrep with character $\chi_i$ occurs in a representation with a character $\chi$ is given by $(\chi|\chi_i)$ \cite{Ser77}.

We now define a {\it regular} representation of a group $G$. Suppose $|G|=n$, and let $V$ be an $n$-dimensional vector space. Let $\{e_t\}$ define a basis for V, which is labeled by the group elements of $t\in G$. A regular representation of $G$ is a map $\sigma :G\rightarrow V$ such that $\sigma(g) e_t = e_{gt}$. It can easily be seen that the action of $G$ on the basis vectors is a left translation, and in matrix form the representations of the group elements will be permutation matrices. An important property of the regular representation is that its decomposition into irreps contains {\it all} the irreps of the group in it, and each irrep has a multiplicity equal to its dimension \cite{S94}. Therefore, we can write
\begin{equation}
V=n_1W_1\oplus n_2W_2\oplus\cdots\oplus n_kW_k ,
\end{equation}
where we use the notation $nW$ to mean $n$ copies of the space $W$, $nW = W\oplus W\oplus\cdots W$.  The $\{W_i\}$ are all the inequivalent irreps of the group, and $n_i$ is the dimension of $W_i$.

It will be useful to note that all Abelian groups have one-dimensional irreps. The converse of this statement is also true:  all groups which have only one-dimensional irreps are Abelian. Finally, suppose $\sigma_1 :G\rightarrow V_1$ and $\sigma_2 :G\rightarrow V_2$ are two linear representations of $G$ on $V_1$ and $V_2$, then $\sigma_1\otimes\sigma_2 :G\rightarrow V_1\otimes V_2$ is a representation of $G$ on $V_1\otimes V_2$.  (The tensor product $W^1_i\otimes W^2_j$ of two irreps $W^1_i$ and $W^2_j$ of $V_1$ and $V_2$, respectively, need not be an irrep of $V_1\otimes V_2$, however.)

Now consider the unitary operator $\hat{U}$ on a finite dimensional vector space $V$ which has a {\it group of symmetries} $G$. This means that the matrices $\sigma(g)$ representating the elements of $g\in G$ on $V$ all commute with $\hat{U}$:  $[\sigma(g),\hat{U}]=0$.  Since $\hat{U}$ is unitary, we can decompose $V$ into a direct sum of eigenspaces of $\hat{U}$:
\begin{equation}\label{U_eig}
V=U_1\oplus U_2\oplus\cdots\oplus U_m .
\end{equation}
We can also decompose $V$ into a direct sum of irreps of $G$:
\begin{equation}
V=W_1\oplus W_2\oplus\cdots\oplus W_k .
\end{equation}
It can be shown using Schur's lemma (see \cite{S94}) that since $G$ is the group of symmetries of $\hat{U}$, each irrep of $G$ must lie entirely inside some eigenspace of $\hat{U}$. Therefore, for some $i$ and $j$, if $W_i\subset U_j$ then the degeneracy of $U_j$ is at least equal to the dimension of $W_i$. Using this fact, we show now that if a graph has sufficient symmetry (in a particular sense), then it will lead to quantum walks with infinite hitting times.

\section{Discrete-time walks on Cayley graphs}

It was observed earlier that the direction-preserving automorphism group of a Cayley graph is the group of left translations of the group elements. Since every group element corresponds to a vertex of a Cayley graph and every vertex corresponds to a basis element on the vertex Hilbert space ($\mathcal{H}^v$) of the walk, the automorphism group of a Cayley graph is a regular representation of $G$ on $\mathcal{H}^v$. Every direction-preserving automorphism of a Cayley graph will induce a representation on this Hilbert space which looks like $\sigma(g)\otimes \hat{I}_c$, where $\sigma(g)$ is the regular representation of $G$ on $\mathcal{H}^v$ and $\hat{I}_c$ is the identity on the coin space. (Note that $\sigma(g) \otimes \hat{I}_c$ is {\it not} a regular representation of $G$.) In order to prove that this walk has an infinite hitting time for certain initial states, we need to show that $\hat{U}$ has at least one degenerate eigenspace whose dimension is greater than the dimension of the coin. Since every irrep lies completely inside an eigenspace, if one of the irreps occurring in $\sigma\otimes\hat{I}$ has a dimension greater than the dimension of the coin, then we can say that the eigenspace containing that irrep has a degeneracy greater than the dimension of the coin. We now show that every irrep of $G$ occurs in $\sigma\otimes\hat{I}$. We have,
\begin{equation}
\tr(\sigma\otimes\hat{I}) = d\chi ,
\end{equation}
where $\chi=\tr(\sigma)$ and $d$ is the dimension of the coin. If $\chi_i$ is any irreducible character of $G$, then
\begin{equation}
(\tr(\sigma\otimes\hat{I})|\chi_i) = d(\chi|\chi_i) .
\end{equation}
Since $\chi$ is the character of the regular representation of $G$, $(\chi|\chi_i)\neq 0$ for any irreducible character $\chi_i$. Therefore, all irreps of $G$ occur in $\sigma\otimes\hat{I}$, and if any irrep of $G$ has a dimension greater than the dimension of the coin then there is an eigenspace of $\hat{U}$ whose dimension is greater than the dimension of the coin.  So for Cayley graphs, a sufficient condition for any discrete-time quantum walk to have infinite hitting times for some initial conditions is that the group used to define the graph have an irrep with dimension greater than the degree of the graph.

For a regular graph which is not a Cayley graph, this can be modified as follows. A discrete time walk defined on a graph will have an infinite hitting time for certain initial states if at least one irrep occurring in the induced representation (on the Hilbert space of the walk) of the direction-preserving automorphism group of the graph has a dimension greater than the degree of the final vertex.  This is a somewhat more difficult to evaluate, since unlike a Cayley graph, the induced representation of the symmetry group for a general graph is not guaranteed to include every irrep.  But in principle it is not difficult to check.

For an example of a graph with infinite hitting times, consider the Cayley graph on the symmetric group $S_n$:  $\Gamma (S_n,X)$, where $X$ is a generating set for $S_n$. In order to use the form $\hat{U}=\hat{S}(\hat{I}\otimes\hat{C})$, the graph needs to be $|X|$-colored and so we chose a generating set whose elements $x$ are such that $x^2=e$, where $e$ is the identity element. For $S_n$, such a generating set is any set of $n-1$  transpositions, e.g., $\{(1,2),(2,3),\dots,(n-1,n)$.  These form the basis for the coin space, and so the dimension of the coin is the cardinality of the generating set:  $|X|=n-1$. Therefore, a symmetric group $S_n$ which has an irrep with a dimension greater than $n-1$ will have infinite hitting times for some initial conditions for any coin matrix. It turns out that for $n\geq 5$ any symmetric group $S_n$ possesses this property \cite{JK}, and so the corresponding Cayley graph $\Gamma (S_n,X)$ will have infinite hitting times.

All this indicates that it is not so much the {\it size} of the symmetry group that matters for infinite hitting times, but rather the {\it kind} of group, or more precisely the size and number of irreps of the group occurring in the induced representation. Consider the hypercube, which has the group $H\cong \mathcal{Z}_2^n$ as its symmetry group (if the coin has no permutation symmetry like the DFT coin). This group is Abelian and hence has only one-dimensional irreps. So one would expect the unitary evolution operator having this as its symmetry group to have very little or no degeneracy. But when the DFT coin is used the evolution operator has sufficient degeneracy to have infinite hitting times. This is because it is the degeneracy of the DFT coin, rather than the symmetry group, that makes the evolution operator degenerate. This is supported by numerical evidence which shows that when a randomly generated non-degenerate or slightly degenerate unitary coin is used instead of the DFT, the unitary evolution operator has a very small or no degeneracy. The degeneracy of the evolution operator with an Abelian symmetry group seems to come only from the coin.

\section{Continuous-time walks on Cayley graphs}

In the prior discussion of infinite hitting times we have used the definition (\ref{ht1}) for the hitting time, which is only well-defined for discrete time quantum walks.  We have not described a suitable measurement process to define the hitting time for a continuous time walk; nor is it obvious how to do so in the quantum case, where the presence or absence of measurements has a profound effect on the dynamics.  Any notion of hitting time for the continuous case, however, must include a measurement performed on the final vertex at some time which will verify if the particle has arrived there or not.  This leaves an ambiguity in the definition of hitting time for {\it finite} hitting times, but the notion of an {\it infinite} hitting time still has an intuitive definition:  a continuous time quantum walk has infinite hitting time if, for any set of measurements on the final vertex at any sequence of times, there is always a bounded, nonzero probability that the particle will {\it never} be found at the final vertex.

Continuous time quantum walks do not have a coin matrix, and their evolution operator for undirected graphs is $\hat{U}(t) = \exp(i\hat{H}t)$ where $\hat{H}$ is the adjacency matrix of the graph. Since there is no coin, the degree of freedom at any given vertex is one dimensional. Any eigenspace of $\hat{U}$ with a degeneracy greater than one can therefore contribute to the projector $\hat{P}$ having zero overlap with the final vertex and commuting with $\hat{U}$.   Whenever a measurement on the final vertex is performed, the measurement operators will commute with $\hat{P}$ since $\hat{P}|x_f\ra\la x_f|=0$, where $|x_f\ra$ is the final vertex state.  A nonzero $\hat{P}$ necessarily means an infinite hitting time for initial states that overlap with it. Therefore, only if $\hat{U}$ is completely non-degenerate, and none of its eigenvectors have a zero overlap with the final state, will there be finite hitting times for all initial states.  This is a lesser degree of degeneracy than is needed in the discrete time case, and we therefore expect infinite hitting times to be even more common in continuous time walks than discrete time walks.

Making the connection to symmetry again, for a continuous walk to have infinite hitting times, a sufficient condition is that at least one irrep with dimension greater than one occurs in the induced representation of the automorphism group of the graph. Consider once more the example of Cayley graphs.  As discussed in the case of discrete time quantum walks above,  the induced representation of the automorphism group $G$ on the Hilbert space $\mathcal{H}^v$ is the regular representation. Since there is no coin, $\mathcal{H}^v$ is the Hilbert space of the walk. All the irreps of the group appear in this representation because it is regular. Therefore, if any of the irreps of $G$ has a dimension greater than one, then the walk will have infinite hitting times for certain starting states. Since only Abelian groups have all their irreps of dimension one, any Cayley graph defined on a non-Abelian group will have infinite hitting times for the continuous walk.
%

\section{Discussion}
In this chapter, we have examined the role of symmetry in shaping the properties of the quantum walk. The role of symmetry is not restricted to producing infinite hitting times. Symmetry can also be related to the exponentially {\it fast} hitting times observed in \cite{Kem03b,KB05}. For one of the cases examined in \cite{KB05}---the discrete walk on the hypercube with the Grover coin---we observe an exponentially smaller hitting time than the classical walk on the same graph. But this happens only for the symmetric initial state $|\Psi\ra=|00\cdots 0\ra\otimes\frac{1}{\sqrt{d}}\sum_i|i\ra$. Other superpositions of coin states do not have this speed-up, but rather lead to infinite hitting times, because their overlap with $\hat{P}$ is nonzero.

We noted that the group of symmetries of the $n$-dimensional hypercube is $G=H\cdot S_n$ when one takes into account the direction labels, where $H$ is the normal subgroup of direction-preserving automorphisms, and $S_n$ is the permutation group on $n$ elements (in this case, the different graph directions), and is also a subgroup of $G$. It was observed in \cite{SKW03} that $|\Psi\ra$ is the simultaneous eigenstate of eigenvalue $1$ of the subgroup $S_n$ (more precisely, the simultaneous eigenstate of the representation operators of the subgroup). Since every element of $S_n$ commutes with $\hat{U}$ for the walk with the Grover coin, a state that begins in an eigenspace of the permutation group will remain in the same eigenspace at all times.  That is,
\begin{equation}
U^t|\Psi\ra = U^t \sigma(g) |\Psi\ra = \sigma(g) U^t |\Psi\ra ,
\end{equation}
where $\sigma$ is the representation of $G$ on the Hilbert space of $\hat{U}$ and $g\in S_n$. Thus, $\hat{U}^t|\Psi\ra$ is an eigenstate of $\sigma(g)$ with eigenvalue $1$.  This eigenspace has dimension $2n$.  It turns out that the final vertex with an equal superposition of coin states $|11\cdots 1\ra\otimes\frac{1}{\sqrt{d}}\sum_i|i\ra$ also lies in this eigenspace. Since the walk never escapes this subspace to explore other parts of Hilbert space, it leads to an exponentially fast hitting time.  For a measured walk, if the symmetry subgroup commutes with the measurement operators (as is true in this case), then the same argument holds. This shows that those symmetries of the graph which are passed on to the evolution operator can create subspaces to which the walk may be confined. If the final vertex has no overlap with such a subspace for any coin state, then a walk starting in that subspace will have an infinite hitting time.  Otherwise, the hitting time will be finite, or even exponentially small depending on the dimension of this subspace relative to the full Hilbert space.

We should also point out that the conditions derived above for infinite hitting times are sufficient for a particular graph to have infinite hitting times; but they are not necessary conditions.  For example, the symmetry group of the hypercube is Abelian, and hence does not imply that the evolution operator $\hat{U}$ must be degenerate.  Nevertheless, infinite hitting times are observed for quantum walks on the hypercube, due to the fact that the choices of coin flip operator (the Grover coin or the DFT) both have their own symmetries, which increases the total degeneracy of the evolution operator.  Infinite hitting times are therefore likely to be even more common than the conditions derived here would suggest.

One can make very plausible intuitive arguments that both infinite hitting times and exponentially fast hitting times are related to symmetry.  This makes it seem likely that the ideal problem to be solved by a quantum walk would be a problem with {\it global} symmetry, but in which this symmetry is not apparent at the local scale.  In the next chapter a framework which explains fast hitting times observed in quantum walks, is presented. Using symmetry of the graph again, we show that the quantum walk that possesses these symmetries can be confined to a smaller graph called a quotient graph. If this new graph is exponentially smaller than the original graph, then it can lead to fast hitting times.

\chapter{Quotient graphs}
The idea of restricting a search to an invariant subspace of the full search space has proved very fruitful in Grover's search algorithm. In both the quantum walk-based algorithm on the hypercube in \cite{SKW03} and the ``glued-trees" graph in \cite{CCDFGS03}, the quantum algorithm works very fast by searching a smaller space, where it is known that the solution lies in this space. In this chapter, we explore this concept for quantum walks on more general graphs. Using symmetry arguments, we show that it is possible to find invariant subspaces of the total Hilbert space on which the walk is defined. The automorphism group of the graph produces a group of symmetries of the evolution operator for the walk.  This group of symmetries in turn determines the invariant subspace of the walk. If the initial state is in this subspace, the quantum walk effectively evolves on a different graph---a {\it quotient} graph, which can in some cases be much smaller than the original graph. Here, we give a general construction of quotient graphs, given the original graph and a subgroup of its automorphism group. We determine the structure of the quantum walk on the quotient graph. We then apply the analysis of hitting times developed in \cite{KB05} and \cite{KB06} to quotient graphs, and investigate the possibility of both infinite hitting times and reduced hitting times on quotient graphs.

\section{Action of an automorphism group}
Consider any undirected graph $\Gamma$ and let $V(\Gamma)$ and $E(\Gamma)$ denote its vertex and edge sets. Let the graph be colored, not necessarily consistently i.e., the edge between vertices $v_i$ and $v_j$ may be colored with a color $c_k$ in the direction from $v_i\rightarrow v_j$ and a color $c_l$ from $v_j\rightarrow v_i$. This creates the Hilbert space of positions and colors (or directions) and the total space $\mathcal{H}$ is spanned by basis vectors $|v_1,c_1\ra,\dots |v_n,c_m\ra$. Let this set of basis vectors be $X$. The set of colors at each vertex is not the same for all the vertices since the graph may be irregular. We assume that vertices having the same degree have the same set of colors. Denote by $C_v$ the set of colors used to color edges going from the vertex $v$. Thus, the shift matrix for this graph is,
\begin{equation}
S=\sum_{v_i\in V(\Gamma),k\in C_{v_i}} |v_j,c_l\ra\la v_i,c_k| .
\end{equation}
This matrix encodes the structure of the graph $\Gamma$ which includes edge colors. An automorphism of the graph $\Gamma$, as defined above is a permutation matrix which preserves $S$ under conjugation i.e., a matrix $P$ such that $PSP^\dag=S$. The set of automorphisms form a group which we denote by $G$.

Now consider a subgroup (not necessarily proper) $H$ of this automorphism group. We would like to know what kind of {\it action} this subgroup has on the graph and hence on the Hilbert space. First, we define what is meant by the term {\it action} \cite{Rot95}. 
\begin{definition}
If $X$ is a set and $G$ is a group, then $X$ is a $G${\it -set} if there is a function $\alpha : G\times X\rightarrow X$ (called a left action), denoted by $\alpha:(g,x)\rightarrow gx$, such that :
\begin{itemize}
\item $1x=x$, for all $x\in X$; and
\item $g(hx)=(gh)x$, for all $g,h \in G$ and $x\in X$ .
\end{itemize}
\end{definition}
\begin{definition}
If $X$ is a $G$-set and $x\in X$, then the $G$-{\it orbit} (or just {\it orbit}) of $x$ is
\begin{equation}
\mathcal{O}(x)=\{gx:g\in G\}\subset X .
\end{equation}
\end{definition}
The set of orbits of a $G$-set $X$ form a partition and the orbits correspond to the equivalence classes under the equivalence relation $x\equiv y$ defined by $y=gx$ for some $g\in G$. We can define the action of the subgroup $H$ of the permutation group on the set of basis elements $X$ of the Hilbert space $\mathcal{H}$ as the multiplication of its {\it matrix representation} $\sigma (H)$ (in the basis given by the vectors $X$) with a basis vector. This is a well-defined action since $\sigma (1)|x\ra=|x\ra$ and $\sigma (g) (\sigma (h) |x\ra) = (\sigma (g) \sigma (h))|x\ra = \sigma(gh) |x\ra$. Therefore, the set $X$ is partitioned into orbits under the action of $H$.

Since $H$ is a subgroup of the automorphism group, these orbits can be related to the graph $\Gamma$ through the following results.
\begin{theorem}
If $|v,c_i\ra$ and $|v,c_j\ra$ are in different orbits, then the set of all the vertices in the orbits of $|v,c_i\ra$ and $|v,c_j\ra$ are the same.
\end{theorem}
\begin{proof}
If the graph $\Gamma$ is irregular (regular graphs are just a special case), then clearly any automorphism takes a given vertex to another vertex of the same degree. Thus, automorphisms permute vertices of a certain degree among themselves. Therefore, on the Hilbert space $\mathcal{H}$, the matrix representation of any automorphism can be written as
\begin{equation}
P=\bigoplus_{d\in D} P_d ,
\end{equation}
where the set $D$ contains all the different degrees in the graph. Consider the subspace of vertices of a given degree $d$ which can be written as $\mathcal{H}_d^V\otimes \mathcal{H}_d^C$. Now, if any given permutation takes $|v_1,c_i\ra$ to $|v_2,c_j\ra$, then it takes all the basis vectors associated with $v_1$ to those of $v_2$. Thus, the set of all vertices that lie in the orbit of $|v_1,c_i\ra$ must be the same as the set of vertices that lie in the orbit of $|v_1,c_j\ra$ (if $|v_1,c_i\ra$ and $|v_1,c_j\ra$ lie in the same orbit, then this is trivially true). Since $v_1$ is arbitrary, the set of vertices in the two orbits must be the same.
\end{proof}
By an abuse of language, say that a vector $|v_1,c_1\ra$ is ``connected" to $|v_2,c_2\ra$ if the edge colored $c_1$ from vertex $v_1$ on the graph is connected to vertex $v_2$ along the color $c_2$ (i.e., the term $|v_2,c_2\ra\la v_1,c_1|$ occurs in $S$).
\begin{theorem}
If $|v_1,c_1\ra$ and $|v_2,c_2\ra$ are ``connected," $|v_1,c_1\ra$ lies in the orbit $\mathcal{O}_1$ and $|v_2,c_2\ra$ lies in orbit $\mathcal{O}_2$ (not necessarily distinct from $\mathcal{O}_1$), then each of the remaining vectors of $\mathcal{O}_1$ are ``connected" to some vector  of $\mathcal{O}_2$.
\end{theorem}
\begin{proof}
If $|v_1,c_1\ra$ and $|v_2,c_2\ra$ are connected, then there is a term of the type $|v_2,c_2\ra\la v_1,c_1|$ in $S$. When we conjugate by some automorphism $h\in H$ i.e., perform $\sigma(h)S\sigma(h)^T$, then this term transforms to $\sigma(h)|v_2,c_2\ra\la v_1,c_1|\sigma(h)^T$ and this must be a term in $S$ because $\sigma(h)S\sigma(h)^T=S$. This means that the vector that $|v_1,c_1\ra$ gets taken to, is connected to the vector that $|v_2,c_2\ra$ gets taken to by $\sigma (h)$. Since this is true for all $h\in H$, all the vectors in the orbit of $|v_1,c_1\ra$ are ``connected" to some term in the orbit of $|v_2,c_2\ra$.
\end{proof}
The above result applies equally well to any vector in $\mathcal{O}_2$. Therefore, one can think of the orbits $\mathcal{O}_1$ and $\mathcal{O}_2$ as being ``connected".

\section{Quotient graphs and quantum walks}
Based on the action on a graph $\Gamma$ of the subgroup $H$ of its automorphism group, consider the following construction of a graph---a {\it quotient} graph. The set of vertices occuring in an orbit $\mathcal{O}$ is a single vertex $v_\mathcal{O}$ on the new graph and the number of orbits that have the same set of vertices is the degree of this new vertex. Thus, for a given vertex $v_\mathcal{O}$, the set of directions are the various orbits which correspond to the same vertex set. If an orbit $\mathcal{O}_1$ is ``connected" to $\mathcal{O}_2$, then the vertices $v_{\mathcal{O}_1}$ and $v_{\mathcal{O}_2}$ are connected in the quotient graph. If $\mathcal{O}_1$ and $\mathcal{O}_2$ are identical, this corresponds to a self loop for $v_{\mathcal{O}_1}$. This means that there can be self loops in the quotient graph even if there are none in the original graph. We denote the quotient graph obtained by the action of the subgroup $H$ on $\Gamma$ as $\Gamma /H$ or $\Gamma_H$.

Now consider a basis vector $|x\ra\equiv|v,c\ra$ and its $H$-orbit $\mathcal{O}_x=\{\sigma (h)|x\ra : h\in H\}$.  The vector $|\tilde{x}\ra\equiv\frac{1}{\sqrt{|\mathcal{O}_x|}}\sum_{h\in H} \sigma(h) |x\ra$ is an eigenvector of eigenvalue 1 of all the matrices $\sigma (h)$ for $h\in H$ since 
\begin{eqnarray}
\sigma(h)|\tilde{x}\ra &=& \frac{1}{\sqrt{|\mathcal{O}_x|}}\sum_{h'\in H} \sigma(h)\sigma(h')|x\ra = \frac{1}{\sqrt{|\mathcal{O}_x|}}\sum_{h'\in H} \sigma(hh')|x\ra \nonumber \\
&=& \frac{1}{\sqrt{|\mathcal{O}_x|}}\sum_{h''\in H} \sigma(h'')|x\ra = |\tilde{x}\ra .
\end{eqnarray}
Similarly, the vector $|\tilde{y}\ra$ formed from a vector $|y\ra$ of another orbit is also an eigenvector of eigenvalue 1. Each of these vectors $\{|\tilde{x}\ra\}$ are orthonormal, since they are formed from orbits and distinct orbits do not intersect and they span the simulataneous eigenspace of eigenvalue 1 of the matrices $\sigma(H)$. We denote the Hilbert space spanned by these vectors by $\mathcal{H} /H$ or $\mathcal{H}_H$. Note that the vectors $|x\ra$ are just representatives, and any vector in its orbit could be used to generate $|\tilde{x}\ra$. Since the $\{|\tilde{x}\ra\}$ are in one to one correspondence with the orbits, we let $|\mathcal{O}_x\ra$ denote a vector in $\mathcal{H}$ and $|\tilde{x}\ra$ denote the corresponding basis vector in $\mathcal{H}_H$. 

Each basis vector in this space corresponds to a vertex and direction on the quotient graph, just as the basis vectors of $\mathcal{H}$, namely $\{|v,c\ra\}$ represent a vertex and direction on $\Gamma$. Suppose that a vertex $\tilde{v}$ on $\Gamma_H$ comes from the set of vertices in the orbit $\mathcal{O}_1$ and that the orbits $\mathcal{O}_2,\dots ,\mathcal{O}_k$ are all the orbits with the same set of vertices. Since each of these orbits is ``connected" to some other orbit (either in this set or outside), the degree of $\tilde{v}$ is $k$. Therefore, all the basis vectors $|\mathcal{O}_1\ra,\dots ,|\mathcal{O}_k\ra$ can be associated with $\tilde{v}$ and the edges along which they are  ``connected" to other orbits, as the different directions. An alternate labelling of these vectors could be $|\tilde{v},c_1\ra,\dots ,|\tilde{v},c_k\ra$ and likewise for each vertex. Note that this does not produce any natural coloring scheme induced from $\Gamma$, on the edges of $\Gamma_H$. 

We now show that any discrete quantum walk on $\Gamma$ induces a discrete quantum walk on $\Gamma_H$ as long as $\hat{U}$ respects $H$ i.e., $\sigma(h)\hat{U}\sigma(h)^\dag=\hat{U}$ $\forall h\in H$. Let us define a discrete quantum walk as the application of any unitary $\hat{U}$ which takes a particle on a given vertex $v$ to some superposition of vertices that it is connected to and the directions of only those edges which connect them to $v$. On a basis state it acts as
\begin{equation}\label{Qwalk}
\hat{U}|v,c_i\ra=\sum_j a_j|v(c_j),c'_j\ra ,
\end{equation}
where $|v(c_j),c'_j\ra$ and $|v,c_j\ra$ are ``connected", $\sum_j |a_j|^2=1,$ $\forall j$ and the sum runs over all the colors of the edges on the side of $v$. Given this definition for a walk, we have the following results.
\begin{theorem}
Let $H$ be a subgroup of the automorphism group of $\Gamma$ and let $\hat{U}$ be a discrete quantum walk defined on $\Gamma$ such that $\hat{U}$ respects the symmetries of the subgroup i.e., $[\hat{U},\sigma(h)]=0$, $\forall h\in H$. If the initial state lies in the subspace spanned by all the orbits $\{|\mathcal{O}_i\ra\}$ under the action of $H$, then the walk is contained in the subspace.
\end{theorem}
\begin{proof}
We have
\begin{equation}
\hat{U}^t |\mathcal{O}_i\ra=\hat{U}^t \sigma(h) |\mathcal{O}_i\ra = \sigma(h)\hat{U}^t |\mathcal{O}_i\ra .
\end{equation}
This shows that since $|\mathcal{O}_i\ra$ lies in the eigenspace of eigenvalue 1, $\hat{U}^t |\mathcal{O}_i\ra$ also lies in the same space, which is spanned by $\{|\mathcal{O}_i\ra\}$.
\end{proof}
\begin{theorem}
Let $H$ be a subgroup of the automorphism group of $\Gamma$ and let $\hat{U}$ be a discrete quantum walk defined on $\Gamma$ such that $\hat{U}$ respects the symmetries of the subgroup i.e., $[\hat{U},\sigma(h)]=0$, $\forall h\in H$. If the initial state lies in the subspace spanned by all the $H$-orbits $\{|\mathcal{O}_i\ra\}$, then $\hat{U}$ induces a walk on $\Gamma_H$ in the Hilbert space $\mathcal{H}_H$.
\end{theorem}
\begin{proof}
In order to show that $\hat{U}$ induces a walk on $\Gamma_H$, we need to show that its action is similar to Eq.~(\ref{Qwalk}):
\begin{equation}
\hat{U}|\mathcal{O}\ra = \sum_j b_j |\mathcal{O}_j\ra ,
\end{equation}
where $|\mathcal{O}_j\ra$ are the orbits connected to $|\mathcal{O}\ra$. But this follows from the fact that  if the walk moves the particle from a vector to vectors ``connected" to it, then it does the same for superpostions of vectors or the orbit states $|\mathcal{O}\ra$.
\end{proof}

We can derive the structure of this induced walk from the original walk by making use of its action on the orbit states. The induced walk on the subspace $\mathcal{H}_H$ becomes $\hat{U}_H=\sum_{\tilde{x},\tilde{y}} \la \mathcal{O}_y |\hat{U}|\mathcal{O}_x\ra | \tilde{y}\ra\la \tilde{x}|$. This defines a unitary operator in $\mathcal{H}_H$ because,
\begin{eqnarray}
\hat{U}_H^\dag\hat{U}_H &=& \sum_{\tilde{x},\tilde{y},\tilde{y}'} \la \mathcal{O}_{y'} | \hat{U}^\dag | \mathcal{O}_x\ra\la \mathcal{O}_x| \hat{U} |\mathcal{O}_y\ra |\tilde{y}'\ra\la \tilde{y}| \nonumber \\
&=& \sum_{\tilde{x},\tilde{y},\tilde{y}'} \la \mathcal{O}_{y'} | \hat{U}^\dag P_H \hat{U}|\mathcal{O}_y\ra |\tilde{y}'\ra\la \tilde{y}| \nonumber \\
&= & I_H ,
\end{eqnarray}
since $\hat{U}$ commutes with $P_H$, where $P_H$ is the projector onto $\mathcal{H}_H$. Now consider the shift matrix of the walk. Its action on $\mathcal{H}_H$ is given by $\hat{S}_H=\sum_{\tilde{x},\tilde{y}} \la \mathcal{O}_y | \hat{S} | \mathcal{O}_x\ra |\tilde{y}\ra\la \tilde{x} |$. The expression $\la \mathcal{O}_y | \hat{S} | \mathcal{O}_x\ra$ is non-zero if and only if the two orbits are ``connected". If two orbits are connected then they must be a superposition of the same number of vectors i.e., $|\mathcal{O}_x|=|\mathcal{O}_y|$ and each vector in the superposition in $|\mathcal{O}_x\ra$ is connected to one vector in the superposition in $|\mathcal{O}_y\ra$. Therefore, 
\[
\la \mathcal{O}_y | \hat{S} | \mathcal{O}_x\ra= |\mathcal{O}_x|/ \sqrt{|\mathcal{O}_x||\mathcal{O}_y}| . 
\]
Thus, 
\begin{equation}
\hat{S}_H=\sum_{\tilde{x},\tilde{y}}|\tilde{y}\ra\la \tilde{x} | .
\end{equation}
This means that the action of $\hat{S}_H$ is very similar to the action of $\hat{S}$ in that it takes the walker from any vertex to the vertex it is connected to in the quotient graph. The action of the coin which was $\hat{I}\otimes\hat{C}$ on the original graph becomes $\hat{C}_H$ on the quotient graph so that $\hat{U}_H=\hat{S}_H\hat{C}_H$. Moreover, $\hat{C}_H$ can be decomposed as follows,
\be
\hat{C}_H=\hat{C}_1\oplus\hat{C}_2\oplus \dots\oplus \hat{C}_N ,
\ee
where $N$ is the total number of vertices of the quotient graph and each $\hat{C}_i$ acts only on the basis vectors associated with the vertex $v_i$ of the quotient graph and each $\hat{C}_i$ has a dimension $d_i$ which corresponds to the degree of the $v_i$. In the following examples, such a decomposition is provided along with a list of the basis vectors on the quotient graph such that $\hat{C}_1$ acts on the first $d_1$ basis vectors, $\hat{C}_2$ acts on the next $d_2$ vectors etc.

\section{Examples of quotient graphs}
In this section, we illustrate the above abstract formalism with some examples. In all of the examples we use the following notation to describe the subgroups used to find quotient graphs. The elements of the subgroups denote permutations of directions, but it is to be understood that this has to be done along with an appropriate permutation of vertices, which makes it an automorphism of the graph. Although such a permutation of vertices need not exist for every permutation of directions, they exist for the examples that we consider here. Moreover, this permutation of vertices can be specified simply:  permute the generators which are in one-to-one correspondence with the directions in the same way as the directions and this induces a permutation of vertices.

For example, let $(1,2)$ be a group element. This is the automorphism obtained by interchanging  directions $1$ and $2$ and interchanging generators $t_1$ and $t_2$ so that vertices such as $t_1t_2$ go to $t_2t_1$ etc. We do not consider direction preserving automorphisms in the following examples, since they tend to give rise to quotient graphs with self loops. Finally, we use cycle notation to denote permutations, i.e., $(1,2,3)$ means $1$ goes to $2$, $2$ goes to $3$ and $3$ goes to $1$.

\begin{figure}[tbh]
\begin{center}
\includegraphics[width=4in]{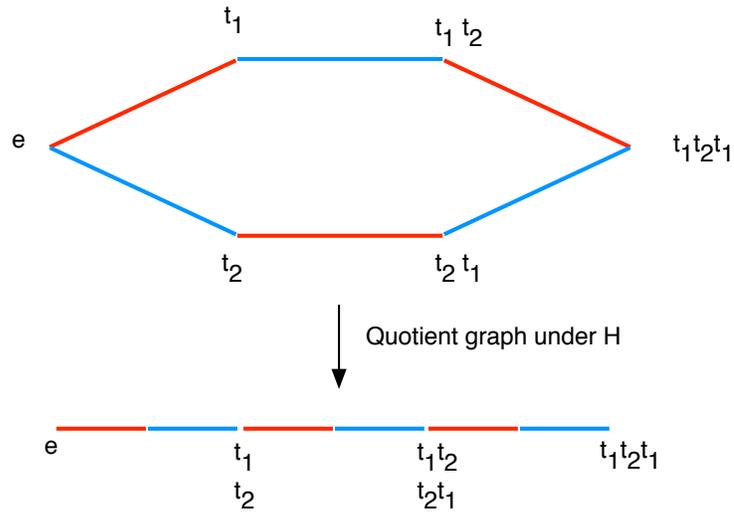}
\end{center}
\caption{The graph $\Gamma(S_3,\{(1,2),(2,3)\})$ and its quotient graph.} \label{S_3_2gen}
\end{figure}

\begin{example}[Example 1.]
As the first example, consider the Cayley graph $\Gamma(S_3,\{(1,2),(2,3)\})$, and let $\{t_1,t_2\}=\{(1,2),(2,3)\}$. The basis vectors of the Hilbert space of the walk are $\{|e,1\ra,|e,2\ra,|t_1,1\ra ,\dots ,|t_1t_2t_1,2\ra\}$. The automorphism group of this graph is $\text{Aut}(\Gamma(S_3,T))\simeq \text{R}(S_3)Z_2$. Consider the subgroup $H=Z_2$ which corresponds to interchanging the directions $1$ and $2$. The orbits under the action of this subgroup are
\begin{eqnarray*}
|\mathcal{O}_1\ra &=& (1/\sqrt{2}) (|e,1\ra + |e,2\ra),  \\
|\mathcal{O}_2\ra &=& (1/\sqrt{2}) (|t_1,1\ra + |t_2,2\ra),  \\
|\mathcal{O}_3\ra &=& (1/\sqrt{2}) (|t_1,2\ra + |t_2,1\ra),  \\
|\mathcal{O}_4\ra &=& (1/\sqrt{2}) (|t_1t_2,2\ra + |t_2t_1,1\ra),  \\
|\mathcal{O}_5\ra &=& (1/\sqrt{2}) (|t_1t_2,1\ra + |t_2t_1,2\ra),  \\
|\mathcal{O}_6\ra &=& (1/\sqrt{2}) (|t_1t_2t_1,1\ra + |t_1t_2t_1,2\ra). 
\end{eqnarray*}
The original and the quotient graph in this case are shown in Fig. ~\ref{S_3_2gen}. The unitary describing the quantum walk on $\Gamma$ is given by $\hat{U}=\hat{S}(\hat{I}\otimes \hat{C})$ where $\hat{S}=\hat{S}'+\hat{S}'^\dag$ and
\ber
\hat{S}'&=&|e,1\ra\la t_1,1| + |e,2\ra\la t_2,2| + |t_1,2\ra\la t_1t_2,2| +
|t_2,1\ra\la t_2t_1,1| + |t_1t_2,1\ra\la t_1t_2t_1,1| \nonumber \\
&+& |t_2t_1,2\ra\la t_1t_2t_1,2| . \nonumber
\eer
This becomes $\hat{S}_H$ on the quotient graph and is given by $\hat{S}_H=\hat{S}'_H+\hat{S}'^\dag_H$ and
\begin{equation}
\hat{S}'_H=|\mathcal{O}_1\ra\la \mathcal{O}_2| + |\mathcal{O}_3\ra\la \mathcal{O}_4| + |\mathcal{O}_5\ra\la \mathcal{O}_6| .
\end{equation}
This can also be written by giving new labels to the vertices and directions of the quotient graph:
\begin{equation}
\hat{S}'_H=|v_1,R\ra\la v_2,L| + |v_2,R\ra\la v_3,L| + |v_3,R\ra\la v_4,L| ,
\end{equation}
where we have relabeled $|\mathcal{O}_1\ra$ through $|\mathcal{O}_6\ra$ as $|v_1,R\ra$ through $|v_4,L\ra$. Note that there is no $|v_1,L\ra$ and $|v_4,R\ra$ which exactly corresponds to the way these vertices are connected in the quotient graph. Now, if we take the coin to be $C=\sigma_x$, the Pauli $X$ operator (which is also the Grover coin in two dimensions), then on the quotient graph the coin flip matrix $\hat{F}_H=(\hat{I}\otimes \hat{C})_H$ becomes
\begin{equation}
\hat{F}_H=|v_1,R\ra\la v_1,R| + |v_4,L\ra\la v_4,L| + \hat{F}'_H +\hat{F}'^\dag_H,
\end{equation}
where $\hat{F}'_H=|v_2,L\ra\la v_2,R| + |v_3,L\ra\la v_3,R|$. It can also be written as
\be
\hat{F}_H=\hat{1}\oplus\hat{C}'\oplus\hat{C}'\oplus\hat{1},
\ee
where $\hat{C}'=\hat{X}$, the Pauli $\sigma_x$ operator. Thus, the walk becomes $\hat{U}_H=\hat{S}_H\hat{F}_H$ i.e.,
\[
\hat{U}_H=\begin{pmatrix}
0 & 0 & 1 & 0 & 0 & 0\\
1 & 0 & 0 & 0 & 0 & 0\\
0 & 0 & 0 & 0 & 1 & 0\\
0 & 1 & 0 & 0 & 0 & 0\\
0 & 0 & 0 & 0 & 0 & 1\\
0 & 0 & 0 & 1 & 0 & 0
\end{pmatrix} .
\]
\end{example}

\begin{example}[Example 2.]
Now consider the Cayley graph $\Gamma (S_3,\{(1,2),(2,3),(1,3)\})$ where $\{(1,2),(2,3),(1,3)\}$ $=$ $\{t_1,t_2,t_3\}$. A subgroup of its automorphism group is $S_3$ which consists of all permutations of the three directions. Consider a subgroup of this consisting of $H_1=\{e,(1,2,3),(1,3,2)\}$. Under the action of this subgroup, the orbits are
\begin{eqnarray*}
|\mathcal{O}_1\ra &=& (1/\sqrt{3}) (|e,1\ra + |e,2\ra + |e,3\ra),  \\
|\mathcal{O}_2\ra &=& (1/\sqrt{3}) (|t_1,1\ra + |t_2,2\ra + |t_3,3\ra),  \\
|\mathcal{O}_3\ra &=& (1/\sqrt{3}) (|t_1,3\ra + |t_2,1\ra + |t_3,2\ra),  \\
|\mathcal{O}_4\ra &=& (1/\sqrt{3}) (|t_1,2\ra + |t_2,3\ra + |t_3,1\ra),  \\
|\mathcal{O}_5\ra &=& (1/\sqrt{3}) (|t_1t_2,1\ra + |t_1t_2,2\ra + |t_1t_2,3\ra),  \\
|\mathcal{O}_6\ra &=& (1/\sqrt{3}) (|t_2t_1,1\ra + |t_2t_1,2\ra + |t_2t_1,3\ra). 
\end{eqnarray*}
The shift matrix for this walk becomes $\hat{S}_{H_1}=\hat{S}'_{H_1}+\hat{S}'^\dag_{H_1}$, where
\begin{equation}
\hat{S}'_{H_1}=|\mathcal{O}_1\ra\la \mathcal{O}_2| + |\mathcal{O}_6\ra\la \mathcal{O}_3| + |\mathcal{O}_5\ra\la\mathcal{O}_4| .
\end{equation}
We can relabel the quotient graph as shown in Fig (\ref{S_3_3gen}). The matrix $\hat{S}'_{H_1}$ becomes
\begin{equation}
\hat{S}'_{H_1}=|v_1,1\ra\la v_2,1| + |v_3,1\ra\la v_2,2| + |v_4,1\ra\la v_2,3| .
\end{equation}
If we choose the Grover coin for the walk, the walk on the quotient graph becomes
\begin{equation}
\hat{U}_{H_1}=
\begin{pmatrix}
0 & -\frac{1}{3} & \frac{2}{3} & \frac{2}{3} & 0 & 0\\
1 & 0 & 0 & 0 & 0 & 0\\
0 & 0 & 0 & 0 & 0 & 1\\
0 & 0 & 0 & 0 & 1 & 0\\
0 & \frac{2}{3} & \frac{2}{3} & -\frac{1}{3} & 0 & 0\\
0 & \frac{2}{3} & -\frac{1}{3} & \frac{2}{3} & 0 & 0
\end{pmatrix}
\end{equation}
Fig.~\ref{S_3_3gen} also shows the quotient graphs for the above Cayley graph with subgroups $H_2\simeq S_3$ and $H_3\simeq \{e,(2,3)\}$. The basis states of the quotient Hilbert space $\mathcal{H}_{H_2}$ are
\begin{eqnarray*}
|\mathcal{O}_1\ra &=& (|e,1\ra+|e,2\ra+|e,3\ra)/\sqrt{3} ,\\
|\mathcal{O}_2\ra &=& (|t_1,1\ra+|t_2,2\ra+|t_3,3\ra)/\sqrt{3},\\
|\mathcal{O}_3\ra &=& (|t_1,2\ra+|t_1,3\ra+|t_2,1\ra+|t_2,3\ra \\
&+& |t_3,1\ra+|t_3,3\ra)/\sqrt{6},\\
|\mathcal{O}_4\ra &=& (|t_1t_2,1\ra+|t_1t_2,2\ra+|t_1t_2,3\ra + |t_2t_1,1\ra \\
&+& |t_2t_1,2\ra+|t_2t_1,3\ra)/\sqrt{6},
\end{eqnarray*}
and the basis states of $\mathcal{H}_{H_3}$ are
\begin{eqnarray*}
|\mathcal{O}_1\ra &=& |e,1\ra, \\
|\mathcal{O}_2\ra &=& (|e,2\ra+|e,3\ra)/\sqrt{2}, \\
|\mathcal{O}_3\ra &=& |t_1,1\ra, \\
|\mathcal{O}_4\ra &=& (|t_1,2\ra+|t_1,3\ra)/\sqrt{2}, \\
|\mathcal{O}_5\ra &=& (|t_2,2\ra+|t_3,3\ra)/\sqrt{2}, \\
|\mathcal{O}_6\ra &=& (|t_2,3\ra+|t_3,2\ra)/\sqrt{2}, \\
|\mathcal{O}_7\ra &=& (|t_2,1\ra+|t_3,1\ra)/\sqrt{2}, \\
|\mathcal{O}_8\ra &=& (|t_1t_2,3\ra+|t_2t_1,2\ra)/\sqrt{2}, \\
|\mathcal{O}_9\ra &=& (|t_1t_2,1\ra+|t_2t_1,1\ra)/\sqrt{2}, \\
|\mathcal{O}_{10}\ra &=& (|t_1t_2,2\ra+|t_2t_1,3\ra)/\sqrt{2}.
\end{eqnarray*}

\begin{figure}[tbh]
\begin{center}
\includegraphics[width=4in]{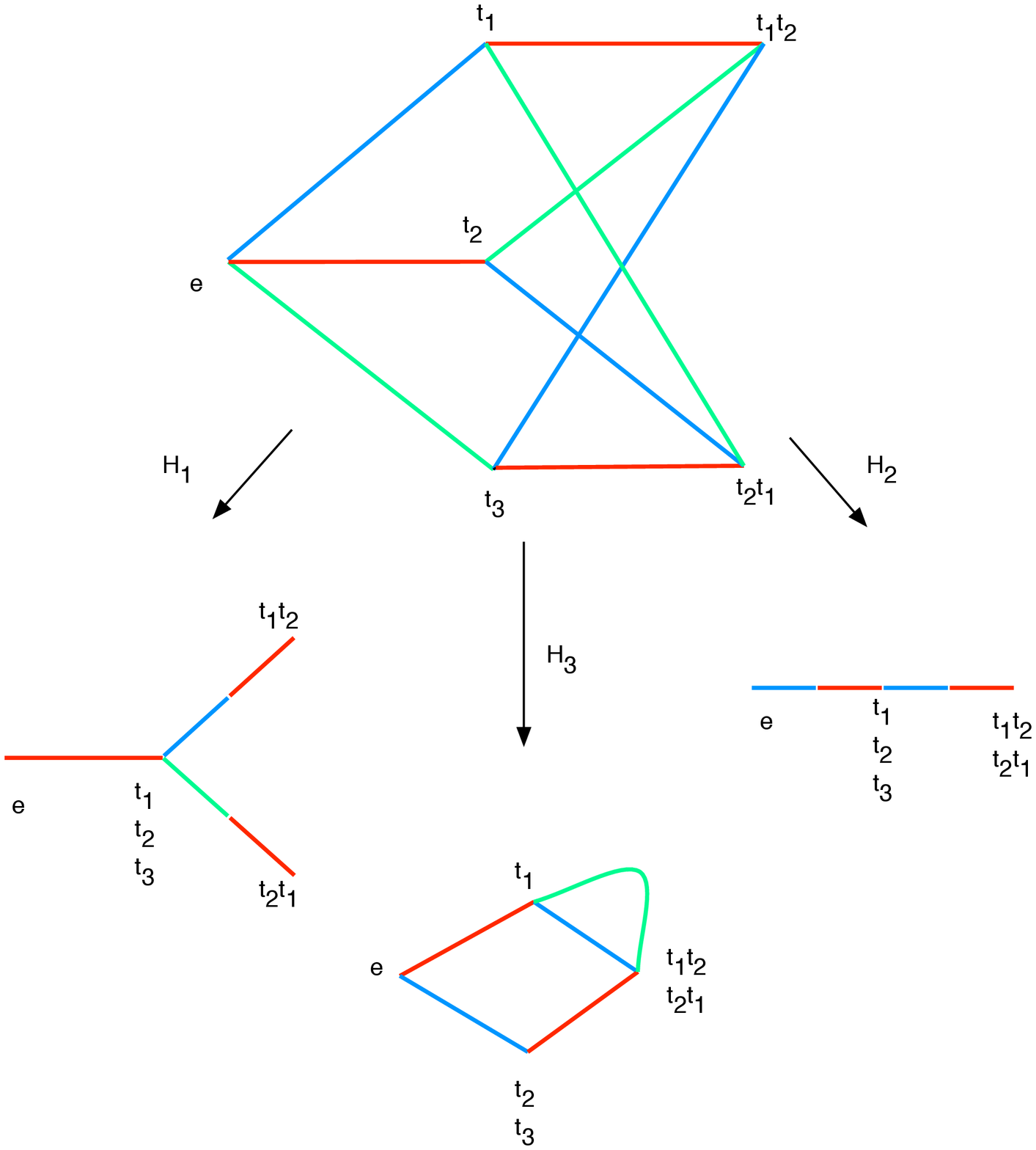}
\end{center}
\caption{The graph $\Gamma(S_3,\{(1,2),(1,3),(2,3)\})$ and its quotient graphs.} \label{S_3_3gen}
\end{figure}

The unitary corresponding to the walk on the quotient graph of $H_2$ is
\begin{equation}
\hat{U}_{H_2}=
\begin{pmatrix}
0 & -1/3 & 2\sqrt{2}/3 & 0\\
1 & 0 & 0 & 0\\
0 & 0 & 0 & 1\\
0 & -2\sqrt{2}/3 & 1/3 & 0
\end{pmatrix},
\end{equation}
and the one on the quotient graph of $H_3$ is
\be
\hat{U}_{H_3}=\hat{S}_{H_3}\cdot \hat{C}_{H_3} .
\ee
The matrices $\hat{S}_{H_3}$ and $\hat{C}_{H_3}$ are given by, $\hat{S}_{H_3}=\hat{S}'+\hat{S}'^\dag$ and $\hat{C}_{H_3}=\hat{C}'\oplus\hat{C}'\oplus \hat{C}''\oplus\hat{C}''$ where,
\begin{eqnarray*}
\hat{S}'&= & (|\mathcal{O}_1\ra\la\mathcal{O}_3| + |\mathcal{O}_2\ra\la\mathcal{O}_5| + |\mathcal{O}_4\ra\la\mathcal{O}_{10}| + |\mathcal{O}_6\ra\la\mathcal{O}_9| \\
&+& |\mathcal{O}_7\ra\la\mathcal{O}_8|,
\end{eqnarray*}
\[
\hat{C}'=\begin{pmatrix}-1/3 & 2\sqrt{2}/3\\ 2\sqrt{2}/3 & 1/3\end{pmatrix} 
\]
and
\[
\hat{C}''=\begin{pmatrix}-1/3 & 2/3 & 2/3 \\ 2/3 & -1/3 & 2/3 \\2/3 & 2/3 & -1/3\end{pmatrix} .
\]
\end{example}

\begin{example}[Example 3.]
In this example, we determine the quotient graph of $\Gamma(S_4,T)$ for $T=\{(1,2),(1,3),(1,4)\}=\{t_1,t_2,t_3\}$ under the subgroup $H\simeq S_3$ which corresponds to all possible permutations of the directions at each vertex. The original and the quotient graphs are shown in Fig (\ref{S_4_3gen}), where ``$t$" has been dropped in the vertex labels. There are 14 orbits under the action of this subgroup. These are
\begin{eqnarray*}
|\mathcal{O}_1\ra &=& (|e,1\ra + |e,2\ra+|e,3\ra)/\sqrt{3},\\
|\mathcal{O}_2\ra &=& (|t_1,1\ra + |t_2,2\ra+|t_3,3\ra)/\sqrt{3},\\
|\mathcal{O}_3\ra &=& (|t_1,2\ra + |t_2,1\ra+|t_3,2\ra \\
&+& |t_2,3\ra+ |t_1,3\ra+|t_3,1\ra)/\sqrt{6},\\
|\mathcal{O}_4\ra &=& (|t_1t_2,1\ra + |t_2t_1,2\ra+|t_3t_2,3\ra \\
&+ & |t_2t_3,2\ra+ |t_3t_1,3\ra+|t_1t_3,1\ra)/\sqrt{6},\\
|\mathcal{O}_5\ra &=& (|t_1t_2,2\ra + |t_2t_1,1\ra+|t_3t_2,2\ra \\
&+& |t_2t_3,3\ra+ |t_3t_1,1\ra+|t_1t_3,3\ra)/\sqrt{6},
\end{eqnarray*}
\begin{eqnarray*}
|\mathcal{O}_6\ra &=& (|t_1t_2,3\ra + |t_2t_1,3\ra+|t_3t_2,1\ra \\ 
&+& |t_2t_3,1\ra+ |t_3t_1,2\ra+|t_1t_3,2\ra)/\sqrt{6},\\
|\mathcal{O}_7\ra &=& (|t_1t_2t_1,1\ra + |t_1t_2t_1,2\ra+|t_3t_2t_3,3\ra \\ 
&+& |t_3t_2t_3,2\ra+ |t_1t_3t_1,3\ra+|t_1t_3t_1,1\ra)/\sqrt{6},\\
|\mathcal{O}_8\ra &=& (|t_1t_2t_1,3\ra + |t_3t_2t_3,1\ra+|t_1t_3t_1,2\ra)/\sqrt{3},\\
|\mathcal{O}_9\ra &=& (|t_1t_2t_3,1\ra + |t_2t_1t_3,2\ra+|t_3t_2t_1,3\ra \\
&+& |t_1t_3t_2,1\ra+ |t_2t_3t_1,2\ra+|t_3t_1t_2,3\ra)/\sqrt{6},\\
|\mathcal{O}_{10}\ra &=& (|t_1t_2t_3,2\ra + |t_1t_3t_2,3\ra+|t_2t_1t_3,1\ra \\
&+& |t_2t_3t_1,3\ra+ |t_3t_1t_2,1\ra+|t_3t_2t_1,2\ra)/\sqrt{6},
\end{eqnarray*}
\begin{eqnarray*}
|\mathcal{O}_{11}\ra &=& (|t_1t_2t_3,3\ra + |t_1t_3t_2,2\ra+|t_2t_1t_3,3\ra \\
&+& |t_2t_3t_1,1\ra+ |t_3t_1t_2,2\ra+|t_3t_2t_1,1\ra)/\sqrt{6},\\
|\mathcal{O}_{12}\ra &=& (|t_3t_1t_2t_1,1\ra + |t_2t_3t_2t_1,3\ra+|t_3t_1t_2t_1,2\ra \\
&+& |t_2t_3t_2t_1,2\ra+ |t_1t_3t_1t_2,1\ra+|t_1t_3t_1t_2,3\ra)/\sqrt{6},\\
|\mathcal{O}_{13}\ra &=& (|t_1,1\ra + |t_2,2\ra+|t_3,3\ra)/\sqrt{3},\\
|\mathcal{O}_{14}\ra &=& (|t_1t_3t_2t_1,1\ra + |t_1t_3t_2t_1,2\ra+|t_1t_3t_2t_1,3\ra \\
&+& |t_2t_3t_1t_2,1\ra+ |t_2t_3t_1t_2,2\ra+|t_2t_3t_1t_2,3\ra)/\sqrt{6} .
\end{eqnarray*}
The unitary walk on the quotient graph can be written as
\be
\hat{U}_H=\hat{S}_H\cdot \hat{C}_H .
\ee
The matrices $\hat{S}_H$ and $\hat{C}_H$ are given by, $\hat{S}_H=\hat{S}'+\hat{S}'^\dag$ and $\hat{C}_H=1\oplus\hat{C}'\oplus\hat{C}''\oplus\hat{C}'\oplus\hat{C}''\oplus\hat{C}'\oplus 1$ where,
\begin{eqnarray*}
\hat{S}'&= & (|\mathcal{O}_1\ra\la\mathcal{O}_2| + |\mathcal{O}_3\ra\la\mathcal{O}_4| + |\mathcal{O}_5\ra\la\mathcal{O}_7| + |\mathcal{O}_6\ra\la\mathcal{O}_9| \\
&+& |\mathcal{O}_8\ra\la\mathcal{O}_{12}| + |\mathcal{O}_{10}\ra\la\mathcal{O}_{13}| + |\mathcal{O}_{11}\ra\la\mathcal{O}_{14}|) ,
\end{eqnarray*}
\[
\hat{C}'=\begin{pmatrix}-1/3 & 2\sqrt{2}/3\\ 2\sqrt{2}/3 & 1/3\end{pmatrix} 
\]
and
\[
\hat{C}''=\begin{pmatrix}-1/3 & 2/3 & 2/3 \\ 2/3 & -1/3 & 2/3 \\2/3 & 2/3 & -1/3\end{pmatrix} .
\]

\begin{figure}[tbh]
\begin{center}
\includegraphics[width=4in]{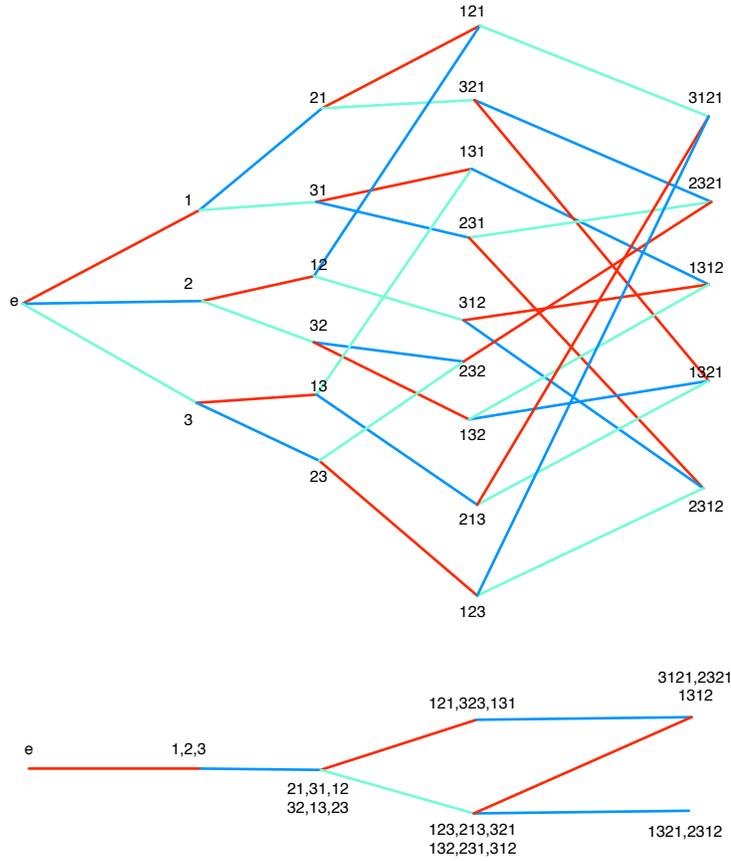}
\end{center}
\caption{The graph $\Gamma(S_4,\{(1,2),(1,3),(1,4)\})$ and it quotient graph under the subgroup $H$.} \label{S_4_3gen}
\end{figure}

\end{example}

\begin{example}[Example 4.]
Consider the hypercube. The automorphism group of the hypercube is $\text{Aut}(\Gamma(\mathcal{Z}_2^n,Y))\simeq \mathcal{Z}_2^nS_n$. We focus on the subgroup $H_1=S_n$ and look at the resulting quotient graph. We consider the case when $n=3$, but the procedure for a general $n$ is very similar. The subgroup $H_1$ consists of all possible permutations of $n$ directions. The orbits under the action of this subgroup are given by
\begin{eqnarray*}
|\mathcal{O}_1\ra &=& (|000,1\ra+|000,2\ra+|000,3\ra)/\sqrt{3}, \\
|\mathcal{O}_2\ra &=& (|001,1\ra+|010,2\ra+|100,3\ra)/\sqrt{3}, \\
|\mathcal{O}_3\ra &=& (|001\ra (|2\ra+|3\ra)+|010\ra(|1\ra+|3\ra)  \\
&+& |100\ra(|1\ra+|2\ra))/\sqrt{6},
\end{eqnarray*}
\begin{eqnarray*}
|\mathcal{O}_4\ra &=& (|011\ra(|1\ra+|2\ra)+|101\ra(|1\ra+|3\ra) \\
&+& |110\ra(|2\ra+|3\ra))/\sqrt{6}, \\
|\mathcal{O}_5\ra &=& (|011,3\ra+|101,2\ra+|110,1\ra)/\sqrt{3}, \\
|\mathcal{O}_6\ra &=& (|111,1\ra+|111,2\ra+|111,3\ra)/\sqrt{3} .
\end{eqnarray*}
The graph becomes a line as shown in Fig.~\ref{hypercube_line} and all the vertices of a certain Hamming weight collapse to a point. This fact hÊá first been observed in \cite{MooRus02}. In \cite{SKW03},  this idea was used to construct a search algorithm on the hypercube. As observed  in \cite{SKW03}, the states on the line can be relabeled $|0,R\ra, |1,L\ra, |1,R\ra, |2,L\ra, |2,R\ra, |3,L\ra$. For the general hypercube of dimension $n$, these states generalize to
\begin{eqnarray}
|x,R\ra &=& \sqrt{\frac{1}{(n-x){n\choose x}}}\sum_{|\vec{x}|=x}\sum_{x_d=0} |\vec{x},d\ra, \nonumber \\
|x,L\ra &=& \sqrt{\frac{1}{(x){n \choose x}}}\sum_{|\vec{x}|=x}\sum_{x_d=1} |\vec{x},d\ra ,
\end{eqnarray}
where $|\vec{x}|$ is the Hamming weight of $\vec{x}$.

\begin{figure}[tbh]
\begin{center}
\includegraphics[width=4in]{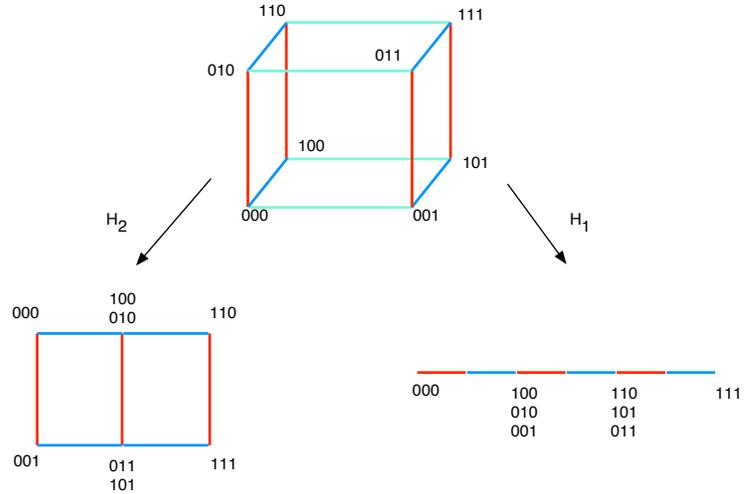}
\end{center}
\caption{The $n=3$ hypercube and its quotient graphs.} \label{hypercube_line}
\end{figure}

Under the action of a different subgroup $H_2=S_{n-1}$ consisting of permutations of $n-1$ directions and the corresponding permutations of vertices, the quotient graph is shown in Fig. (\ref{hypercube_line}). The basis states of $\mathcal{H}_{H_2}$ when $n=3$, are
\begin{eqnarray*}
|\mathcal{O}_1\ra &=& |000,1\ra, \\
|\mathcal{O}_2\ra & = & (|000,2\ra+|000,3\ra)/\sqrt{2}, \\
|\mathcal{O}_3\ra &=& |001,1\ra, \\
|\mathcal{O}_4\ra &=& (|001,2\ra+|001,3\ra)/\sqrt{2}, \\
|\mathcal{O}_5\ra &=& (|010, 2\ra + |100, 3\ra)/\sqrt{2},  \\
|\mathcal{O}_6\ra &=& (|010,1\ra + |100, 1\ra)/\sqrt{2}, \\
|\mathcal{O}_7\ra &=& (|010,3\ra+|100,2\ra)/\sqrt{2}, \\
|\mathcal{O}_8\ra &=& (|011, 2\ra + |101, 3\ra)/\sqrt{2},  \\
|\mathcal{O}_9\ra &=& (|011,1\ra + |101, 1\ra)/\sqrt{2}, \\
|\mathcal{O}_{10}\ra &=& (|011,3\ra+|101,2\ra)/\sqrt{2}, \\
|\mathcal{O}_{11}\ra & = & (|110,2\ra+|110,3\ra)/\sqrt{2}, \\
|\mathcal{O}_{12}\ra &=& |110,1\ra, \\
|\mathcal{O}_{13}\ra &=& (|111,2\ra+|111,3\ra)/\sqrt{2}, \\
|\mathcal{O}_{14}\ra &=& |111,1\ra .
\end{eqnarray*}

\begin{figure}[tbh]
\begin{center}
\includegraphics[scale=0.7]{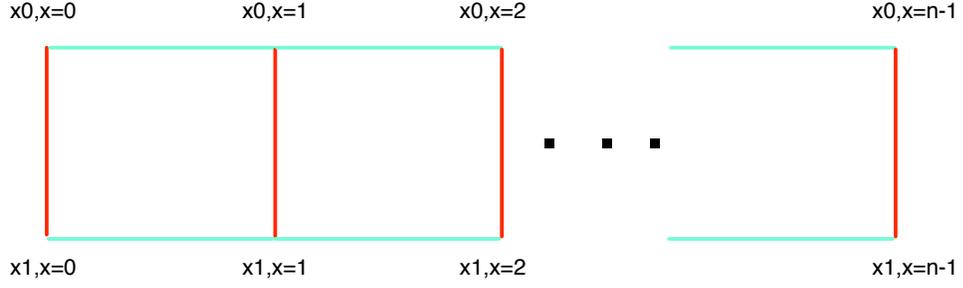}
\end{center}
\caption{The quotient graph of a general hypercube under the group $S_{n-1}$.} \label{hypercube_planar}
\end{figure}

For any general $n$, the graph is still planar as shown in Fig.~\ref{hypercube_planar} and there will be $6n-4$ basis states. They can be labeled as $|x0, L\ra,|x0, R\ra, |x0,D\ra, |x1,L\ra,|x1,R\ra,|x1,U\ra$, where $x$ is the Hamming weight of the last $n-1$ bits (which fall under the action of the subgroup $S_{n-1}$) and the bit next to it is the first bit. $L,R,U$ and $D$ stand for left, right, up and down respectively. They are given by
\begin{eqnarray}
|x0,R\ra &=& \sqrt{\frac{1}{(n-1-x){n-1\choose x}}}\sum_{|\vec{x}|=x}\sum_{x_d=0} |\vec{x}0,d\ra, \nonumber \\
|x0,L\ra &=& \sqrt{\frac{1}{(x){n-1 \choose x}}}\sum_{|\vec{x}|=x}\sum_{x_d=1} |\vec{x}0,d\ra \nonumber \\
|x0,D\ra &=& \sqrt{\frac{1}{{n-1 \choose x}}}\sum_{|\vec{x}|=x} |\vec{x}0,1\ra \nonumber \\
|x1,R\ra &=& \sqrt{\frac{1}{(n-1-x){n-1\choose x}}}\sum_{|\vec{x}|=x}\sum_{x_d=0} |\vec{x}1,d\ra, \nonumber \\
|x1,L\ra &=& \sqrt{\frac{1}{(x){n-1 \choose x}}}\sum_{|\vec{x}|=x}\sum_{x_d=1} |\vec{x}1,d\ra \nonumber \\
|x1,U\ra &=& \sqrt{\frac{1}{{n-1 \choose x}}}\sum_{|\vec{x}|=x} |\vec{x}1,1\ra.
\end{eqnarray}
Note that the states $|x0,U\ra$ and $|x1,D\ra$ do not exist. Moreover,  $|x0,L\ra$ and $|x1,L\ra$ do not exist when $x=0$ and $|x0,R\ra$ and $|x1,R\ra$ do not exist when $x=n-1$. The unitary matrices describing the walk on these graphs are
\be
\hat{U}_{H_1}=\begin{pmatrix}
0 & -1/3 & 2\sqrt{2}/3 & 0 & 0 & 0 \\
1 & 0 & 0 & 0 & 0 & 0 \\
0 & 0 & 0 & -1/3 & 2\sqrt{2}/3 & 0 \\
0 & 2\sqrt{2}/3 & 1/3 & 0 & 0 & 0 \\
0 & 0 & 0 & 0 & 0 & 1 \\
0 & 0 & 0 & 2\sqrt{2}/3 & 1/3 & 0
\end{pmatrix} ,
\ee
and
\be
\hat{U}_{H_2}=\hat{S}_{H_2}\cdot \hat{C}_{H_2} .
\ee
The matrices $\hat{S}_{H_2}$ and $\hat{C}_{H_2}$ are given by, $\hat{S}_{H_2}=\hat{S}'+\hat{S}'^\dag$ and $\hat{C}_{H_2}=\hat{C}'\oplus \hat{C}'\oplus\hat{C}''\oplus\hat{C}''\oplus \hat{C}'\oplus\hat{C}'$ where,
\begin{eqnarray*}
\hat{S}'&= & (|\mathcal{O}_1\ra\la\mathcal{O}_3| + |\mathcal{O}_2\ra\la\mathcal{O}_5| + |\mathcal{O}_4\ra\la\mathcal{O}_8| + |\mathcal{O}_6\ra\la\mathcal{O}_9| \\
&+& |\mathcal{O}_7\ra\la\mathcal{O}_{11}| + |\mathcal{O}_{10}\ra\la\mathcal{O}_{13}| + |\mathcal{O}_{12}\ra\la\mathcal{O}_{14}|) ,
\end{eqnarray*}
\[
\hat{C}'=\begin{pmatrix}-1/3 & 2\sqrt{2}/3\\ 2\sqrt{2}/3 & 1/3\end{pmatrix} 
\]
and
\[
\hat{C}''=\begin{pmatrix}-1/3 & 2/3 & 2/3 \\ 2/3 & -1/3 & 2/3 \\2/3 & 2/3 & -1/3\end{pmatrix} .
\]
\end{example}

\begin{example}[Example 5.]
While we have shown how to construct quotient graphs for discrete-time walks on Cayley graphs, the idea of a quotient graph is more general.  In this example, we consider the ``glued trees" graph shown in Fig.~\ref{Glued_trees}. This graph is not regular and hence not a Cayley graph. It is undirected, and so we can easily define a continuous walk on it. Because the continuous walk does not have a coin space, we need not consider permutations of directions in the automorphisms. Quantum walks on this graph were first analyzed in \cite{CFG02}, and it has been shown that quantum walks move exponentially faster on this graph from ``entrance" to ``exit" than classical walks. The main reason for this exponential speed up is that the quantum walk moves in a superposition of all the vertices in a given column. It can be seen that in any given column, the vertices which branch out from the same vertex in the previous column can be interchanged as long as the corresponding interchange on the other side of the central column takes place. Therefore, the automorphism group of this graph is $Z_2^k$, where $k$ is one half of the total number of vertices on one side of the central column. Under the action of these automorphisms, the vertices in each column form a single orbit, and hence collapse to a single point in the quotient graph. There are $2n+1$ orbits under the action of this subgroup, where the columns $j$ are such that $0\leq j\leq 2n$. The orbits can be written as
\be
|\mathcal{O}_j\ra=2^{-\min[j,2n-j]/2} \sum_{v \in\ {\rm column}\ {j}} |v\ra .
\ee
The Hamiltonian for the quantum walk on the quotient graph becomes \cite{KLMW06}
\begin{eqnarray}\label{Hamiltonian_line}
  \langle \tilde{j}|H|\tilde{j} \pm 1\rangle &=& -\sqrt{2}\gamma \nonumber\\
 \langle \tilde{j}|H|\tilde{j}\rangle &=& \left\{ \begin{array}{ll}
    2\gamma & {j}=0,n,2n \\
    3\gamma & {\rm otherwise,} \\
    \end{array} \right.
\end{eqnarray}
with all other matrix elements zero.  This is also shown in Fig.~\ref{Glued_trees} where the $\gamma$ has been dropped for brevity.

\begin{figure}[ht]
    \centering
    \setlength{\unitlength}{1.25cm}
    \begin{picture}(4,7.5)
        \put(-2,3.75){\line(1,2){1}}
        \put(-2,3.75){\line(1,-2){1}}
        \put(-1,1.75){\line(1,-1){1}}
        \put(-1,1.75){\line(1,1){1}}
        \put(-1,5.75){\line(1,1){1}}
        \put(-1,5.75){\line(1,-1){1}}
        \put(0,6.75){\line(2,1){1}}
        \put(0,6.75){\line(2,-1){1}}
        \put(0,4.75){\line(2,1){1}}
        \put(0,4.75){\line(2,-1){1}}
        \put(0,0.75){\line(2,1){1}}
        \put(0,0.75){\line(2,-1){1}}
        \put(0,2.75){\line(2,1){1}}
        \put(0,2.75){\line(2,-1){1}}
        \put(1,7.25){\line(4,1){1}}
        \put(1,7.25){\line(4,-1){1}}
        \put(1,6.25){\line(4,1){1}}
        \put(1,6.25){\line(4,-1){1}}
        \put(1,5.25){\line(4,1){1}}
        \put(1,5.25){\line(4,-1){1}}
        \put(1,4.25){\line(4,1){1}}
        \put(1,4.25){\line(4,-1){1}}
        \put(1,1.25){\line(4,1){1}}
        \put(1,1.25){\line(4,-1){1}}
        \put(1,0.25){\line(4,1){1}}
        \put(1,0.25){\line(4,-1){1}}
        \put(1,1.25){\line(4,1){1}}
        \put(1,1.25){\line(4,-1){1}}
        \put(1,2.25){\line(4,1){1}}
        \put(1,2.25){\line(4,-1){1}}
        \put(1,3.25){\line(4,1){1}}
        \put(1,3.25){\line(4,-1){1}}
        \put(6,3.75){\line(-1,2){1}}
        \put(6,3.75){\line(-1,-2){1}}
        \put(5,1.75){\line(-1,-1){1}}
        \put(5,5.75){\line(-1,1){1}}
        \put(5,1.75){\line(-1,1){1}}
        \put(5,5.75){\line(-1,-1){1}}
        \put(4,6.75){\line(-2,1){1}}
        \put(4,6.75){\line(-2,-1){1}}
        \put(4,4.75){\line(-2,1){1}}
        \put(4,4.75){\line(-2,-1){1}}
        \put(4,0.75){\line(-2,1){1}}
        \put(4,0.75){\line(-2,-1){1}}
        \put(4,2.75){\line(-2,1){1}}
        \put(4,2.75){\line(-2,-1){1}}
        \put(3,7.25){\line(-4,1){1}}
        \put(3,7.25){\line(-4,-1){1}}
        \put(3,6.25){\line(-4,1){1}}
        \put(3,6.25){\line(-4,-1){1}}
        \put(3,5.25){\line(-4,1){1}}
        \put(3,5.25){\line(-4,-1){1}}
        \put(3,4.25){\line(-4,1){1}}
        \put(3,4.25){\line(-4,-1){1}}
        \put(3,0.25){\line(-4,1){1}}
        \put(3,0.25){\line(-4,-1){1}}
        \put(3,1.25){\line(-4,1){1}}
        \put(3,1.25){\line(-4,-1){1}}
        \put(3,2.25){\line(-4,1){1}}
        \put(3,2.25){\line(-4,-1){1}}
        \put(3,3.25){\line(-4,1){1}}
        \put(3,3.25){\line(-4,-1){1}}
        \put(6,3.75){\circle*{0.2}}
        \put(-2,3.75){\circle*{0.2}}
        \put(5,5.75){\circle*{0.2}}
        \put(-1,5.75){\circle*{0.2}}
        \put(5,1.75){\circle*{0.2}}
        \put(-1,1.75){\circle*{0.2}}
        \put(4,6.75){\circle*{0.2}}
        \put(4,0.75){\circle*{0.2}}
        \put(0,6.75){\circle*{0.2}}
        \put(0,0.75){\circle*{0.2}}
        \put(4,4.75){\circle*{0.2}}
        \put(0,4.75){\circle*{0.2}}
        \put(4,2.75){\circle*{0.2}}
        \put(0,2.75){\circle*{0.2}}
        \put(3,7.25){\circle*{0.2}}
        \put(1,7.25){\circle*{0.2}}
        \put(3,0.25){\circle*{0.2}}
        \put(1,0.25){\circle*{0.2}}
        \put(3,6.25){\circle*{0.2}}
        \put(3,1.25){\circle*{0.2}}
        \put(1,6.25){\circle*{0.2}}
        \put(1,1.25){\circle*{0.2}}
        \put(3,5.25){\circle*{0.2}}
        \put(3,2.25){\circle*{0.2}}
        \put(1,5.25){\circle*{0.2}}
        \put(1,2.25){\circle*{0.2}}
        \put(3,4.25){\circle*{0.2}}
        \put(3,3.25){\circle*{0.2}}
        \put(1,4.25){\circle*{0.2}}
        \put(1,3.25){\circle*{0.2}}
        \put(2,7.5){\circle*{0.2}}
        \put(2,7){\circle*{0.2}}
        \put(2,6.5){\circle*{0.2}}
        \put(2,6){\circle*{0.2}}
        \put(2,5.5){\circle*{0.2}}
        \put(2,5){\circle*{0.2}}
        \put(2,4.5){\circle*{0.2}}
        \put(2,4){\circle*{0.2}}
        \put(2,3.5){\circle*{0.2}}
        \put(2,3){\circle*{0.2}}
        \put(2,2.5){\circle*{0.2}}
        \put(2,2){\circle*{0.2}}
        \put(2,1.5){\circle*{0.2}}
        \put(2,1){\circle*{0.2}}
        \put(2,0.5){\circle*{0.2}}
        \put(2,0){\circle*{0.2}}
    \end{picture}
\vspace{.2in}
    \setlength{\unitlength}{1.3cm}
    \begin{picture}(9,1)
        \put(0.55,0.5){\line(1,0){8}}
        \put(0.55,0.5){\circle*{0.2}}
        \put(1.55,0.5){\circle*{0.2}}
        \put(2.55,0.5){\circle*{0.2}}
        \put(3.55,0.5){\circle*{0.2}}
        \put(4.55,0.5){\circle*{0.2}}
        \put(5.55,0.5){\circle*{0.2}}
        \put(6.55,0.5){\circle*{0.2}}
        \put(7.55,0.5){\circle*{0.2}}
        \put(8.55,0.5){\circle*{0.2}}
        \put(0.6,0.75){{\scriptsize$-\sqrt{2}$}}
        \put(1.6,0.75){{\scriptsize$-\sqrt{2}$}}
        \put(2.6,0.75){{\scriptsize$-\sqrt{2}$}}
        \put(3.6,0.75){{\scriptsize$-\sqrt{2}$}}
        \put(4.6,0.75){{\scriptsize$-\sqrt{2}$}}
        \put(5.6,0.75){{\scriptsize$-\sqrt{2}$}}
        \put(6.6,0.75){{\scriptsize$-\sqrt{2}$}}
        \put(7.6,0.75){{\scriptsize$-\sqrt{2}$}}
        \put(0.475,0){{\scriptsize$2$}}
        \put(1.475,0){{\scriptsize$3$}}
        \put(2.475,0){{\scriptsize$3$}}
        \put(3.475,0){{\scriptsize$3$}}
        \put(4.475,0){{\scriptsize$2$}}
        \put(5.475,0){{\scriptsize$3$}}
        \put(6.475,0){{\scriptsize$3$}}
        \put(7.475,0){{\scriptsize$3$}}
        \put(8.475,0){{\scriptsize$2$}}
    \end{picture}
\caption{The glued trees graph and its quotient graph.}
\label{Glued_trees}
\end{figure}
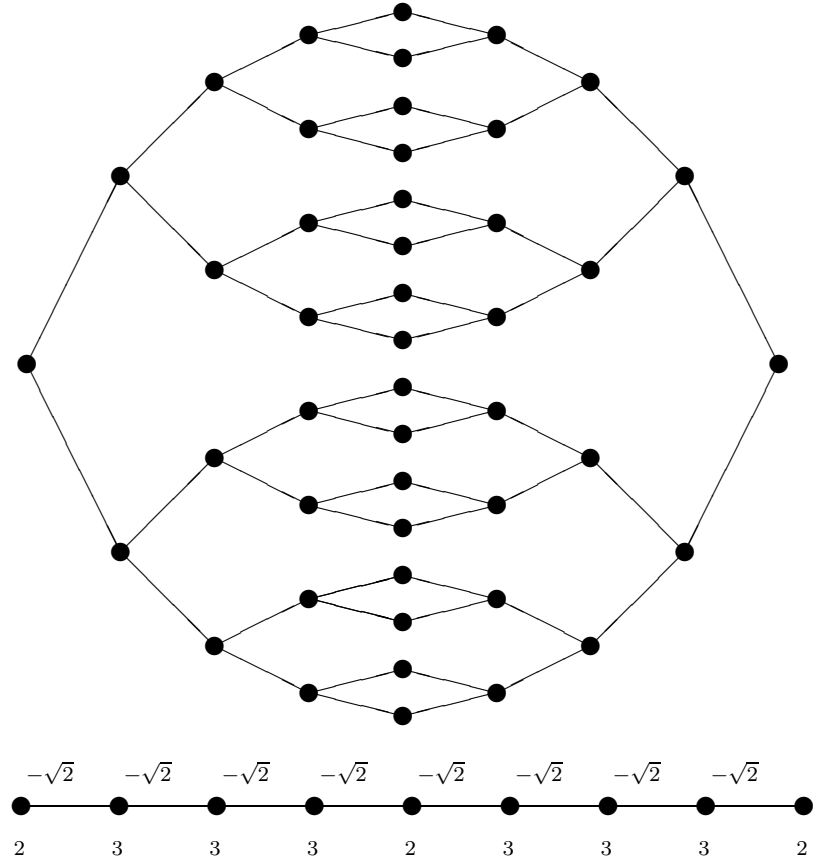
\end{example}

\section{Automorphism group of the quotient graph}
In this section we determine the automorphisms of the quotient graph which are induced from the automorphism group of the original graph and the subgroup used to obtain the quotient graph. In dealing with the automorphisms of $\Gamma$ we used permutations of vertices and edges, and this in turn corresponds to permutations of basis vectors which preserves the shift matrix. On the quotient graph, we define those permutations of {\it orbits} which preserve the $\hat{S}_H$ matrix as automorphisms, since there is no natural choice of edge colors. These permutations of orbits which preserve the new shift matrix also preserve the quotient graph.

Let $G_1$ be a set of automorphisms of $\Gamma$ which are of the following type. If they take a basis vector belonging to a $H$-orbit $\mathcal{O}_1$ to a basis vector belonging to $\mathcal{O}_2$, then they take every basis vector in $\mathcal{O}_1$ to some basis vector in $\mathcal{O}_2$. Clearly, all the automorphisms in $H$ are of this type, under the special case when $\mathcal{O}_1=\mathcal{O}_2$. It is also easy to verify that $G_1$ is a subgroup of $G$, and that $H$ is a subgroup of $G_1$. We now show that $H$ is a normal subgroup of $G_1$ i.e., $g h g^{-1} \in H$, $\forall g\in G_1$ and $\forall h\in H$.

\begin{theorem}
Given the group $G_1$ defined as above we have,
\begin{enumerate}
\item The subgroup $H$ is a normal subgroup of $G_1$.
\item $G_1$ is the largest subgroup of $G$ such that $H$ is a normal subgroup of $G_1$---that is, for any $g\in G$, if $g h g^{-1} \in H$ $\forall h\in H$ then $g\in G_1$.
\end{enumerate}
\end{theorem}
\begin{proof}
We show this by considering the action of all of these group elements on the set of basis vectors. 
\begin{enumerate}
\item Let $x$ be any basis element belonging to some $H$-orbit $\mathcal{O}_1$, and let $g$ take every element in $\mathcal{O}_2$ to some element in $\mathcal{O}_1$. Then $g h g^{-1} x = g h y$, where $y\in \mathcal{O}_2$. Now, $h y = z$ where $z\in \mathcal{O}_2$ since these orbits are formed under the action of $H$. Hence, $g z = x'$, where $x'\in \mathcal{O}_1$. But every $x'\in \mathcal{O}_1$ can be written as $h' x$ for some $h'\in H$. Thus, $g h g^{-1} \in H$. 
\item Consider some basis element $x \in \mathcal{O}_3$ and let $g x \in \mathcal{O}_4$. Since $g h x = h' g x$, $g (h x) = h' y = y'$, where $y,y' \in \mathcal{O}_4$. Therefore, $g$ takes $h x \in \mathcal{O}_3$ to $y'\in \mathcal{O}_4$, but since $h\in H$ is arbitrary, $g$ takes every element of $\mathcal{O}_3$ to some element of $\mathcal{O}_4$. It follows that $g\in G_1$.

\end{enumerate}
\end{proof}
Since $H$ is normal in $G_1$, the quotient set $G_1/H$, i.e., the set of all cosets $g H =\{g h | g\in H\}$, is a group. This group has a natural representation in the Hilbert space $\mathcal{H}_H$ as a permutation matrix in the basis where each orbit is a basis vector. 
\begin{theorem}
$G_1/H \subset \text{Aut}(\Gamma_H)$. 
\end{theorem}
\begin{proof}
Consider any automorphism $g\in G_1$ and let $\sigma(g)$ be its representation in $\mathcal{H}$. Then, we have $\sigma(g) S \sigma(g^{-1}) = S$. The projection of this into $\mathcal{H}_H$ is given by 
\begin{equation}
P_H \sigma(g) S \sigma(g^{-1}) P_H = P_H S P_H=S_H .
\end{equation}
The representation $\sigma(g)$ commutes with $P_H$, since it permutes all the vectors in an orbit to vectors in another orbit. Therefore,
\begin{equation}
P_H \sigma(g)P_H SP_H \sigma(g^{-1}) P_H =S_H .
\end{equation}
But as a representation, $P_H \sigma(g) P_H = \sigma(gH)$. This means that the representation of $gH \in G_1/H$ in $\mathcal{H}_H$ is a group of symmetries of $S_H$ and therefore $G_1/H \subset \text{Aut}(\Gamma)$.
\end{proof}
We see that the quotient graph is obtained from $\Gamma$ modulo the symmetries in $H$.

\section{Hitting time on quotient graphs}
In this subsection, we address the question of when quantum walks on quotient graphs have infinite hitting times. It is possible that for some subgroups, the walk on the quotient graph does not have infinite hitting times even if the walk on the original graph does. In order to carry over the discussion of hitting times to quotient graphs, we must keep in mind that the evolution operator is now followed by a measurement. To remain on the quotient graph (i.e., in the subspace given by $\hat{P}_H$), the measurement operators must commute with the symmetry operators $\sigma(h)$, $h \in H$.

If this condition is satisfied, then we can obtain a condition to check whether the quotient graph has initial states with infinite hitting times:  if the subspace of those initial states with infinite hitting times on the original graph whose projector is $\hat{P}$, has no nontrivial intersection with the subspace whose projector is $\hat{P}_H$ i.e., $\hat{P} \cap \hat{P}_H = \emptyset$, then the walk on the quotient graph does not have infinite hitting times.  If there is a nontrivial intersection, then it does.  (Here we have used the projectors onto the subspaces to denote the spaces themselves.) This condition can also be verified by obtaining the restriction of the evolution operator and the measurement operators onto the quotient graph. By diagonalizing the new unitary evolution operator and constructing the projector of states $\tilde{P}$ which have no overlap with the new final vertex state. The subspace of these states is exactly the intersection $\hat{P}\cap \hat{P}_H$. We will examine this condition for some of the examples considered above.

In the first example, we choose the final vertex to be $t_1t_2t_1$. The measurement operators are $\hat{P}_f=|t_1t_2t_1\ra\la t_1t_2t_1|\otimes \hat{I}$ and $\hat{I}-\hat{P}_f$. This measurement commutes with the subgroup chosen, and the quotient graph does not have infinite hitting times. This is because the original graph does not have infinite hitting times either i.e., $\hat{P}=\emptyset$.

For the second example, for the graph $\Gamma(S_3,\{(1,2),(2,3),(1,3)\})$, we used three different subgroups and form their quotient graphs. In order to determine whether the quotient graph has infinite hitting times for various subgroups, we must choose different final vertices for the different subgroups since the measurement must commute with the symmetries. Therefore, for $H_1$ we choose $t_1t_2$ as the final vertex, and the measurement operators are $|t_1t_2\ra\la t_1t_2|\otimes\hat{I}$ and its orthogonal complement. This measurement commutes with the subgroup $H_1$. For this final vertex and measurement, the original graph has infinite hitting times i.e., $\hat{P} \neq \emptyset$ and the quotient graph also has infinite hitting times i.e., $\hat{P}\cap \hat{P}_{H_1} \neq \emptyset$. In fact, using the $C$-matrix defined above in Eq.~(\ref{C_mtx}), we find that if the initial vertex is the identity $|e\ra$, then there is no superposition of coin states that has a finite hitting time, because $\hat{C}_v$ does not have a zero eigenvalue for $v=e$.

For the subgroup $H_2$, we choose the final vertices to be $t_1t_2$ and $t_2t_1$. Therefore, the measurement on the original graph must be a projective measurement with outcomes $\hat{P}_f=(|(t_1t_2\ra \la t_1t_2|+|t_2t_1\ra\la t_2t_1|)\otimes \hat{I}$ and its orthogonal complement. For this measurement and final vertices, the original graph has $\hat{P} = \emptyset$. Therefore, the quotient graph also does not have infinite hitting times. For the subgroup $H_3$, the measurement operators are $\hat{P}_f=(|(t_1t_2\ra \la t_1t_2|+|t_2t_1\ra\la t_2t_1|)\otimes \hat{I}$ and its orthogonal complement. For this measurement, neither the original graph nor the quotient graph have infinite hitting times.

In example 3, for the Cayley graph $\Gamma(S_4,\{(1,2),(1,3),(1,4)\})$, we choose the final vertices to be $|t_1t_3t_2t_1\ra$ and $|t_2t_3t_1t_2\ra$.  In this case, we find that while the original graph has infinite hitting times, the quotient graph does not. In fact, on the original graph, the equal superposition of all coin states at the vertex $|e\ra$ is the {\it only} superposition which does not have an infinite hitting time (i.e., the $\hat{C}_v$ matrix has only one zero eigenvalue with the equal superposition of coin states as its eigenvector).  It is precisely this vector which is included in the subspace of the quotient graph.  This is not a coincidence---in both cases, it is picked out by the symmetries of the graph.

In example 4, for the two different subgroups of the automorphism group considered for the hypercube, we find that the behavior of hitting times is very different. For the subgroup $H_1$, the quotient graph becomes a line with the vertex $\vec{0}=00\dots 0$ on one end and the vertex $\vec{1}=11\dots 1$ on the other. If one designates the final vertex to be $\vec{1}$ by choosing the measurement operators to be $\hat{P}_f=|\vec{1}\ra\la\vec{1}|\otimes\hat{I}$ and its orthogonal complement, then we find that this quotient graph does not have infinite hitting times for any initial state:  $\hat{P}\cap\hat{P}_{H_1}=\emptyset$. In fact, if the initial state is $|00\dots 0\ra\otimes\frac{1}{d}\sum_i |i\ra$, then the hitting time is polynomial in $d$, the dimension of the hypercube \cite{Kem03b,KB05}. Using the $C$-matrix for the original graph, we find that if the initial vertex is $00\dots 0$, then the equal superposition of all directions is the {\it only} zero eigenvector of $C_v$, which means that it is the only coin state that does {\it not} have an infinite hitting time.

On the other hand, choosing $11\dots 10$ as the final vertex and using the subgroup $H_2$, we find that the quotient graph shown in Fig.~\ref{hypercube_line} does have infinite hitting times for some initial states i.e., $\hat{P}\cap\hat{P}_{H_2}\neq\emptyset$. Using the $C$-matrix again, we find that if the initial vertex is $00\dots 0$, then the equal superposition of all directions once again is the only coin state that has no infinite hitting times. For every other superposition of coin states for that vertex (i.e., every other $|\alpha\ra$ in Eq.~(\ref{alpha})) $C_v$ has a nonzero eigenvalue.

\section{Discussion}

We have investigated the behavior of quantum walks on undirected graphs by making use of the automorphism group of the graph. Automorphisms of the graph may become symmetries of the discrete quantum walk, depending on the symmetries of the coin matrix. Quantum walks which respect the symmetries of some subgroup $H$ of this automorphism group have an invariant subspace in the total Hilbert space. We showed that the walk restricted to this subspace can be seen as a (different) quantum walk on a {\it quotient} graph, and that this graph can be constructed from the original graph given the subgroup $H$. The dynamics of the new walk can also be derived from the original walk and the subgroup. The quotient graph is obtained from the original graph by identifying vertices and edges which form an orbit under the action of $H$; this means that the quotient graph and the new quantum walk both have no symmetries coming from $H$. The new quantum walk only has the remaining automorphisms as its possible symmetries, and so it has, in a sense, ``used up'' the ones in $H$.

To discuss hitting times, we use the measured walk defined in \cite{Kem03b} and \cite{KB05}, which consists of the application of a unitary operator followed by a projective measurement at each time step. For the walk on the quotient graph to be preserved, the choice of measurement must commute with the symmetries in $H$.  This restriction is very important: even if the walk and initial state both have a larger group of symmetries, the walk will be on a quotient graph corresponding to a smaller subgroup $H$ if the measurement does not commute with the remaining elements of the larger group.

For instance, in Example 2, using the subgroup $H_1$, we obtained a walk on its quotient graph. Suppose the measurement is a projective measurement of the vertex $t_1t_2$. The initial state $|e\ra\otimes (|1\ra+|2\ra+|3\ra)/\sqrt(3)$, and the walk with the Grover coin $\hat{U}$ both have all the symmetries of  $H_2$.  But the measurement, which commutes with all the elements of $H_1$, does {\it not} commute with all those in $H_2$, and the effective walk will be on the quotient graph corresponding to $H_1$.

The remaining symmetries of the evolution operator can lead to degeneracy in its eigenspectrum \cite{KB06}, and may result in infinite hitting times on the quotient graph. In general, we found a condition to determine whether the walk on the quotient graph will have infinite hitting times:  given the original graph, the quantum walk and any subgroup $H$, one can determine the projector onto states with infinite hitting time $\hat{P}$, and the invariant subspace of the quotient graph $\hat{P}_H$.  If $\hat{P} \cap \hat{P}_H = \emptyset$ then the quotient graph does not have infinite hitting times.

Even when the hitting time is not infinite for an initial state on the quotient graph, it is possible that it could be extremely long.  It would be useful to have a criterion to pick out subgroups of the automorphism group whose quotient graphs have exponentially fast hitting times. For example, in the case of the hypercube, using the subgroup  $H_1$ (whose quotient graph is a line) turns out to give very fast hitting times. But on a general undirected graph it is not easy to determine whether there is a subgroup whose quotient graph gives fast hitting times.

To investigate this, we need to make the notion of ``fast" more precise. One way to define ``fast" for a parametrized class of graphs (such as the hypercube, where the parameter is the dimension) is to say that the hitting time must be $O(\log N)$ (exponentially smaller), where $N$ is the number of vertices of the graph. Using this notion, we can expect fast hitting times to exist in graphs which have quotient graphs with an exponentially smaller number of vertices. While this is not necessarily a sufficient condition for fast hitting times, it is interesting to observe that both the quantum walk search algorithm on the hypercube and the glued trees graph are examples of symmetric graphs where the quotient graph is exponentially smaller than the original graph. In the case of the hypercube, the hitting time for the effective walk on the quotient graph is exponentially smaller than the number of vertices in the original graph, and exponentially smaller than the classical hitting time; the same is true of the continuous-time walk on the glued-trees graph.  It is interesting to note that in both of these cases the quotient graph is a finite line. This seems to suggest that if we can identify graphs which have the line as a quotient graph, they may be fruitful ground to look for more examples of walks with fast hitting times.  This remains very much an open question, but it is our belief that graph symmetry is of vital importance in the understanding of hitting times for quantum walks, and that understanding the structure of quotient graphs is the key to further progress.

\chapter{Quantum walks in decoherence-free subspaces}
  We have seen in Chapter 3, that decoherence has a detrimental effect on hitting times of quantum walks, especially fast hitting times. This is because the examples of fast hitting times that have been found in the literature are due to the symmetry of the graph. This would not be the case in the presence of decoherence since it would break this symmetry and force the walk to leave this subspace. Infinite hitting times are also due to symmetry and its effect of confining the walk to a subspace. We have examined the effect of decoherence on infinite hitting times and we have seen that it makes them finite by breaking the symmetry. However, if the decoherence has symmetries of its own, then it can have invariant subspaces and the evolution in these subspaces can be purely unitary. In this chapter, we explore this possibility. We derive conditions on the decoherence such that it has enough symmetry to preserve the evolution in some subspace. This can be applied to preserve subspaces that lead to fast hitting times. The concept of decoherence-free subspaces (DFS) has been the subject of intense research in the last few years. Decoherence-free subspaces are a result of the symmetry associated with the decoherence operators and as such are subspaces in which the evolution is purely unitary. Refs. \cite{KBLW01,LW03} present overviews of the progress in this field. The conditions under which a given subspace becomes a DFS of the evolution is given for various different formulations of the evolution. Here, we first give the basic definitions and conditions of decoherence-free subspaces. Then we review the general theory of DFS using representation theory of complex algebras and apply it to quantum walks to arrive at a general condition for the subspace $\mathcal{H}_H$ to be a DFS (recall from Chapter 5, that the subspace $\mathcal{H}_H$ is the subspace of the quotient graph that gives fast hitting times for some graphs.) We also provide examples of decoherence which have the subspace $\mathcal{H}_H$ inside a decoherence-free subspace for both the discrete-time and continuous-time walks on the hypercube.

\section{Decoherence-free subspaces and quantum walks}
In order to apply the DFS formalism to the discrete-time quantum walk, consider an evolution given by the OSR
\be
\mathcal{D}(\rho)=\sum_i\hat{A}_i\rho\hat{A}_i^\dag,
\ee
where the Kraus operators satisfy the relation $\sum_i \hat{A}_i^\dag\hat{A}_i=\hat{I}$. 

\begin{definition}
A system with a Hilbert space $\mathcal{H}$ is said to have a decoherence free subspace $\mathcal{H}'$ if every pure state in this subspace is invariant under the OSR i.e., $\forall |j\ra\in \mathcal{H}'$ we have
\be
\sum_i \hat{A}_i |j\ra\la j|\hat{A}_i^\dag = |j\ra\la j|  .
\ee
\end{definition}
We now state the main theorem which gives the necessary and sufficient condition for a subspace to be a DFS.
\begin{theorem}
A subspace $\tilde{\mathcal{H}}$ is a decoherence-free subspace iff the Kraus operators act proportional to the identity on the subspace, i.e., $\forall$ $|j\ra\in\mathcal{H}'$
\be\label{DFS_degen}
\hat{A}_i |j\ra=c_i|j\ra .
\ee
\end{theorem}
The above theorem states that for $\mathcal{H}'$ to be a DFS of the given OSR, the entire subspace $\mathcal{H}'$ must lie inside a single degenerate eigenspace of each of the Kraus operators. This can be put in the language of the representation theory of the algebras, generated by the operators $\hat{A}_i$. We state the important results in this direction. First, it can be shown that the Kraus operators generate a complex, associative and $\dag$-closed algebra $\mathcal{A}$ ($\dag$-closed means that if $M\in\mathcal{A}$, then $M^\dag\in\mathcal{A}$.) We also assume that the identity lies in this algebra. We now have the following results.

\begin{theorem}
If $\mathcal{A}$ is a complex, associative, $\dag$-closed, algebra such that $I\in \mathcal{A}$, then $\mathcal{A}$ has the following decomposition:
\be\label{A_decomp}
\mathcal{A}\cong \bigoplus_{J\in\mathcal{J}} \mathbf{I}_{n_J}\otimes\mathcal{M}(d_J,\mathbb{C}).
\ee
\end{theorem}
This states that $\mathcal{A}$ reduces to a direct sum of irreducible complex matrix algebras $\mathcal{M}(d_J,\mathbb{C})$ of dimension $d_J$ with a multiplicity $n_J$. The Hilbert space on which they operate can be decomposed similarly as
\be\label{H_decomp}
\mathcal{H}=\bigoplus_{J\in\mathcal{J}} \mathcal{H}_{n_J}\otimes\mathcal{H}_{d_J}.
\ee 
From the theory of decoherence-free subspaces \cite{KBLW01}, the following result gives the condition for a subspace to be a DFS.
\begin{theorem}
The necessary and sufficient condition for a subspace $\mathcal{H}'$ to be a decoherence-free subspace of the dynamics is that it must lie in the degeneracy of a single one-dimensional irreducible representation.
\end{theorem}
This theorem states that in order to be a DFS, $\mathcal{H}'$ must lie in the $\mathbf{I}_{n_K}$ part of the decomposition for which $d_K=1$ i.e., using the decomposition in Eq. (\ref{H_decomp}), $\mathcal{H}' \subset \mathcal{H}_{n_K}$. When $d_K\neq 1$, we obtain the more general decoherence-free subsystems. To apply this to quantum walks, we are interested only in decoherence-free subspaces and so we only need one-dimensional irreducible subalgebras. Finally, to perform computations which preserve the DFS, the unitary operations must commute with the algebra inside the DFS. This means that the general form for the operations that preserve this DFS is
\be
\hat{U}\in \mathcal{M}(n_K,\mathbb{C})\oplus\mathcal{M}(d-n_K,\mathbb{C}) ,
\ee
where $d$ is the total dimension of the Hilbert space.

We now apply this to quantum walks with decoherence and find conditions under which the decoherence preserves the subspace of interest--$\mathcal{H}_H$. Firstly, using the condition of degeneracy in Eq. (\ref{DFS_degen}), we need that 
\be\label{DFS_orbit}
\hat{A}_i |\mathcal{O}_j\ra=c_i|\mathcal{O}_j\ra
\ee
where $\mathcal{O}_j$ is an orbit under the action of $H$ and is a basis vector in $\mathcal{H}_H$.
In order to find a condition in terms of the algebra generated by the symmetry operators, $\sigma(h)$, we first note that for the symmetry operators, if $\mathcal{H}'$ is a degenerate eigenspace i.e.,
\be\label{sigma_degen}
\sigma(g)|j\ra=c_g |j\ra,
\ee 
for all $|j\ra\in\mathcal{H}'$, then this subspace is preserved by the unitary evolution operator $\hat{U}$. This can be seen easily, for we have
\be
\sigma(g) \hat{U}^t |j\ra=\hat{U}^t\sigma(g)|j\ra=c_g\hat{U}^t|j\ra.
\ee
Now, the complex algebra generated by the group of symmetry operators (denote it by $\mathcal{A}_H$) is associative, $\dag$-closed and contains the identity. Using the above general theory, we can decompose this into a sum of irreducible subalgebras as
\be
\mathcal{A}_H\cong \bigoplus_{J\in \mathcal{J}'} \mathbf{I}_{n_J}\otimes\mathcal{M}(d_J,\mathbb{C}).
\ee
The condition Eq.(\ref{sigma_degen}) is equivalent to the condition that the subspace $\mathcal{H}_H$ must belong to the degeneracy of a one-dimensional irreducible subalgebra in order to be preserved by the evolution and the evolution operator $\hat{U}$ must lie in the algebra that commutes with $\mathcal{A}_H$ in the DFS, i.e., if for some $K'\in\mathcal{J}'$ for which $d_K'=1$, the subspace $\mathcal{H}_H$ lies in its degeneracy, then $\hat{U}$ must lie in the algebra $\mathcal{M}(n_K',\mathbb{C})\oplus\mathcal{M}(d-n_K',\mathbb{C})$. Combining the these conditions, we have that
\ber
\mathcal{A}& \cong& \mathbf{I}_{n_K}\otimes\mathbb{C} \oplus\bigoplus_{J\in\mathcal{J}-K} \mathbf{I}_{n_J}\otimes\mathcal{M}(d_J,\mathbb{C}) \nonumber \\
\mathcal{A}_H& \cong& \mathbf{I}_{n_K'}\otimes\mathbb{C} \oplus\bigoplus_{J\in\mathcal{J}'-K'} \mathbf{I}_{n_J}\otimes\mathcal{M}(d_J,\mathbb{C}) \nonumber \\
\hat{U} &\in& \mathcal{M}(n_K',\mathbb{C})\oplus\mathcal{M}(d-n_K',\mathbb{C}) \nonumber \\
&=& \mathcal{M}(n_K,\mathbb{C})\oplus\mathcal{M}(d-n_K,\mathbb{C}),
\eer
where we need $n_K'=n_K$. Now if $\mathcal{H}_H\subset\mathcal{H}_{n_K}$, then the quantum walk which starts inside $\mathcal{H}_H$ will be unperturbed by decoherence.

As an example, consider the discrete-time quantum walk on the hypercube. Consider a decoherence which has the following Kraus operators
\be
\hat{A}_i=\kappa_i \hat{S}_{ii+1}\otimes (|i\ra\la i+1| + |i+1\ra\la i|),
\ee
where $i\in\{1,2,\dots , n-1\}$. $\hat{S}_{ij}$ is the swap operation on the qubits numbered $i$ and $j$. It is given by
\be
\hat{S}_{ij}=\underbrace{\hat{I}\otimes\dots\otimes\hat{I}}_{i-1}\otimes\hat{S}_2\otimes\underbrace{\hat{I}\otimes\dots\otimes\hat{I}}_{n-i-1} ,
\ee
where
\be
\hat{S}_2 = |0\ra\la 0|\otimes |0\ra\la 0| + |1\ra\la 1|\otimes |1\ra\la 1|+ |0\ra\la 1|\otimes |1\ra\la 0| + |1\ra\la 0|\otimes |0\ra\la 1| .
\ee
The constants $\kappa_i$ need to satisfy
\be
\sum_{i=1}^{n-1} |\kappa_i|^2 =1
\ee
in order to have $\sum_i\hat{A}_i^\dag\hat{A}_i=\hat{I}$. This kind of decoherence might occur in implementations of quantum walks on the hypercube using $n$ spins. The decoherence is local in the sense that it only affects adjacent spins. This decoherence has the subspace $\mathcal{H}_H$ with basis vectors given above as a DFS. These basis vectors and Kraus operators satisfy the condition in Eq. (\ref{DFS_orbit}) since we have
\be
\hat{A}_i |\mathcal{O}_j\ra=\kappa_i(\hat{S}_{ii+1}\otimes (|i\ra\la i+1| + |i+1\ra\la i|)) |\mathcal{O}_j\ra=\kappa_i|\mathcal{O}_j\ra .
\ee 
Each $|\mathcal{O}_j\ra$ is an eigenvector of eigenvalue $1$ of the swap operators.

Continuous-time quantum walks evolve by an application of a Hamiltonian as defined in Eq. (\ref{Continuous_Ham}). Decoherence in this scenario would give rise to an evolution which can be described using the Lindblad semigroup master equation as
\be
\dot{\rho}(t)=[\hat{H}, \rho(t)] + \sum_i a_i(\hat{L}_i\rho(t)\hat{L}^\dag_i + \hat{L}^\dag_i\hat{L}_i\rho(t) + \rho(t)\hat{L}^\dag_i\hat{L}_i) .
\ee
The necessary and sufficient conditions for the existence of decoherence-free subspace for such an evolution is given in the following theorem \cite{LW03}.
\begin{theorem}
A subspace $\mathcal{H}'$ is a decoherence-free subspace of the above evolution if and only if the Lindblad operators are proportional to the identity on the subspace i.e., $\forall$ $|j\ra\in\mathcal{H}'$
\be\label{DFS_L}
\hat{L}_i |j\ra=d_i|j\ra .
\ee
\end{theorem}

An example of a decoherence model which has as its DFS the subspace $\mathcal{H}_H$ is where the Lindblad operators are
\be
\hat{L}_i = \kappa_i \hat{S}_{ij} ,
\ee
where $\hat{S}_{ij}$ has the same meaning as before. Indeed, as before one can derive a sufficient condition on the Lindblad operators which is that they must lie in the algebra generated by the group $H$. If this is true then the subspace $\mathcal{H}_H$ will be a DFS.

The decoherence operators (Kraus or Lindblad) need not lie in the algebra generated by the symmetry operators as in the examples. The above examples consider a kind of decoherence that might occur in a physical implementation of a quantum walk on the hypercube. But in general, the decoherence operators need not lie in the algebra generated by the symmetry operators. The decoherence operators just need to commute with the symmetry operators inside the DFS.

\section{Discussion}

Quantum walks can have invariant subspaces if the walk has sufficient symmetries. A walk can be endowed with enough symmetry such that there exists a suitable invariant subspace, which can be used to design an algorithm. The quantum-walk based search algorithms exploit such invariant subspaces. If such a walk is implemented using any quantum system, it will have decoherence affecting it. If the decoherence has symmetries i.e., if the decoherence operators have degeneracies, then we can have decoherence-free subspaces. We have determined conditions which the decoherence operators must satisfy in order to have a certain subspace of interest (from an algorithmic point of view) lies in a DFS of the dynamics. We presented examples of such a decoherence for a quantum walk on the hypercube.

In general, it is difficult to know the symmetries of the decoherence operators without looking at a specific implementation of quantum walks. Unfortunately, there are not many implementations of quantum walks and the few that exist are all for the walk on a line. Here we present the general conditions that the decoherence must satisfy in order to have the invariant subspace $\mathcal{H}_H$ as a DFS. For a specific implementation, one can check if the decoherence satisfies these conditions to know if that scheme can successfully implement a quantum walk algorithm which exploits invariant subspaces by preserving these subspaces even in the presence of decoherence.

\chapter{Conclusions}
Symmetry plays an important role in many areas of physics and especially in quantum mechanics. Many quantum systems have inherent symmetries which make it easier to analyze them. Since quantum computing involves harnessing the power of quantum mechanics to perform faster computations, we must take into account induced symmetries in order to make full use of quantum mechanics. Indeed one such example is in the theory of decoherence-free subspaces and subsystems. Based on the symmetries of the noise operators, we can determine subspaces of the Hilbert space where decoherence has no effect.

In view of this, it should not be surprising if symmetry plays a very important role in quantum algorithms. Indeed, symmetry often contributes in an implicit way. Consider the Grover search algorithm where the problem is to search for a marked node among many unmarked nodes (for simplicity we assume that there is only one marked node). In this abstract setting we can see that the labels of the unmarked nodes have no bearing on the problem. Hence it should have no bearing on the algorithm to search for the marked node. If the algorithm makes a distinction between unmarked nodes, it would mean that we are introducing structure where it does not exist and this cannot lead to an optimal algorithm. Thus if we permute the labels of all the unmarked nodes in any way, the problem must be the same. Therefore, the symmetry group of the problem is $S_{n-1}$ where $n$ is the number of nodes. Under the action of this symmetry group, we can easily see that there are two orbits: an equal superposition of all the unmarked nodes as one of them and the marked node as the other. Thus, symmetry arguments determine the subspace that should be the invariant subspace of the algorithm. Note that this does not give us an algorithm, but tells us the subspace that is important for the algorithm. However, in this case determining this invariant subspace is a very important step.

One can possibly extend this to other computational problems to determine the invariant subspaces relevant to them. In the case of Grover's search problem, there is no structure in the problem and hence every permutation of the nodes (except the marked node) is a symmetry. For problems with structure this will not be the case i.e., only certain permutations of the basis elements (which need to be defined appropriately) which preserve the structure of the problem are symmetries. This gives a group of symmetries-the structure group of the problem. We can then find the orbits under the action of this structure group and this gives us an invariant subspace of the problem: a space which respects all the symmetries. Any algorithm for this problem must also respect these symmetries and hence must lie in this invariant subspace. Explicitly considering the symmetries of a problem in this way and incorporating it into the solution would lead to more efficient algorithms.

\begin{singlespace}

\end{singlespace}
\end{document}